\pdfoutput=1

\documentclass[11pt]{article}

\usepackage[preprint]{acl}

\usepackage{times}
\usepackage{latexsym}
\usepackage{hyperref} 
\usepackage[T1]{fontenc}

\usepackage[utf8]{inputenc}

\usepackage{microtype}

\usepackage{inconsolata}

\usepackage{graphicx}

\usepackage{array,calc}

\usepackage{amssymb}     
\usepackage{pifont}      
\usepackage[T1]{fontenc} 

\newcommand{\xmark}{\ding{55}}  

\usepackage{amssymb}
\usepackage{amsbsy}
\usepackage{amsmath}
\usepackage{amsthm}
\usepackage{latexsym}
\usepackage{subcaption}
\usepackage{wrapfig}
\usepackage{float}
\usepackage{graphicx}
\usepackage{tabularx}
\usepackage{ulem}
\usepackage[font=small]{caption}
\usepackage{algorithm}
\usepackage{algorithmicx, algpseudocode}
\usepackage{mathtools}
\usepackage{arydshln}
\usepackage{titlesec}
\usepackage{lineno}

\usepackage{xcolor}         
\definecolor{myblue}{RGB}{33, 48, 163}
\definecolor{myred}{RGB}{192, 0, 0}
\usepackage{CJKutf8}
\usepackage{graphicx}

\usepackage{booktabs}
\usepackage{subcaption}

\usepackage{multirow}
\usepackage{booktabs}
\usepackage{array}

\title{VisFinEval: A Scenario-Driven Chinese Multimodal Benchmark for Holistic Financial Understanding}

\author{Zhaowei Liu\textsuperscript{1}$^{\dagger}$,
Xin Guo\textsuperscript{1}$^{\dagger}$,
Haotian Xia\textsuperscript{4}$^{\dagger}$,
Lingfeng Zeng\textsuperscript{1},
Fangqi Lou\textsuperscript{1}, 
Jinyi Niu\textsuperscript{5},\\
{\bf Mengping Li\textsuperscript{1},
Qi Qi\textsuperscript{1},
Jiahuan Li\textsuperscript{1},
Wei Zhang\textsuperscript{1},
Yinglong Wang\textsuperscript{6},
Weige Cai\textsuperscript{1},}\\
{\bf Weining Shen\textsuperscript{4}$^*$,
Liwen Zhang\textsuperscript{1,2,3}$^*$}\\
\textsuperscript{1}School of Statistics and Data Science, Shanghai University of Finance and Economics,\\
\textsuperscript{2}Shanghai Financial Intelligent
Engineering Technology Research Center, Shanghai\\ University of Finance and Economics,
\textsuperscript{3}Qinghai Provincial Key Laboratory of Big Data\\
in Finance and Artificial Intelligence Application Technology,
\textsuperscript{4}University of California, \\Irvine,
\textsuperscript{5}Fudan University, 
\textsuperscript{6}Johns Hopkins University\\
  \texttt{\small\{zhang.liwen\}@shufe.edu.cn},
  \texttt{\small\{weinings\}@uci.edu}
}

\begin{document}
\maketitle

\begin{abstract}
Multimodal large language models (MLLMs) hold great promise for automating complex financial analysis. To comprehensively evaluate their capabilities, we introduce \textbf{VisFinEval}, the first large‐scale Chinese benchmark that spans the full front‐middle‐back office lifecycle of financial tasks. VisFinEval comprises 15,848 annotated question–answer pairs drawn from eight common financial image modalities (e.g., K-line charts, financial statements, official seals), organized into three hierarchical scenario depths: Financial Knowledge \& Data Analysis, Financial Analysis \& Decision Support, and Financial Risk Control \& Asset Optimization. We evaluate 21 state-of-the-art MLLMs in a zero-shot setting. The top model, Qwen-VL-max, achieves an overall accuracy of 76.3\%, outperforming non-expert humans but trailing financial experts by over 14 percentage points. Our error analysis uncovers six recurring failure modes—including cross-modal misalignment, hallucinations, and lapses in business-process reasoning—that highlight critical avenues for future research. VisFinEval aims to accelerate the development of robust, domain-tailored MLLMs capable of seamlessly integrating textual and visual financial information. The data and the code are available at \url{https://github.com/SUFE-AIFLM-Lab/VisFinEval}.
\end{abstract}

\begin{figure*}[ht]   
    \centering
    \includegraphics[width=\textwidth]{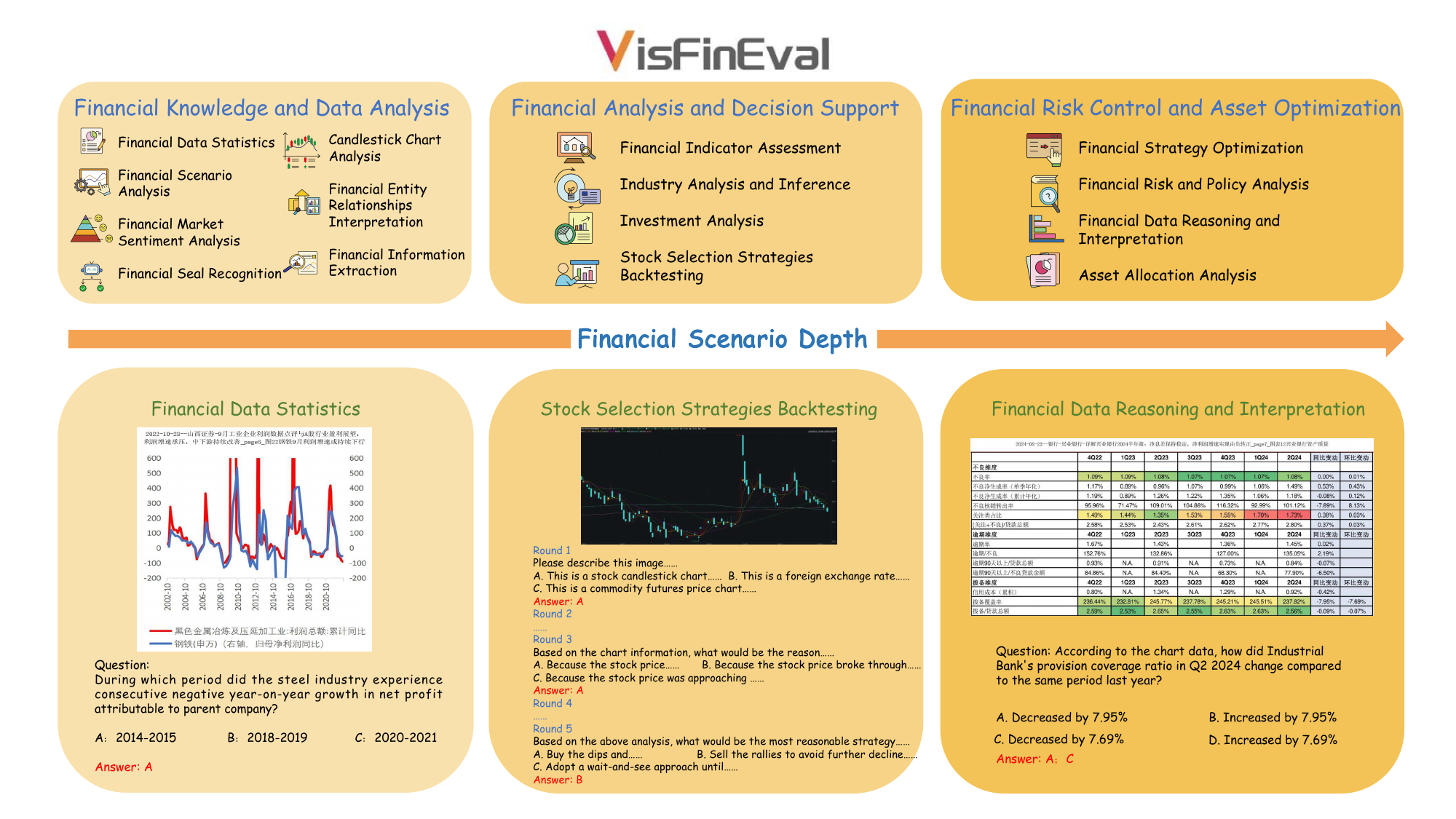}
    \caption{VisFinEval provides a multimodal evaluation framework for full-process financial operations. Starting from the perspective of business depth, it designs 3 major scenarios: Financial Knowledge and Data Analysis, Financial Analysis and Decision Support, and Financial Risk Control and Asset Optimization. Each major scenario corresponds to 7, 4, and 4 sub-scenarios, respectively, accurately reflecting the diverse business scenarios within the financial industry. Furthermore, it has constructed 15,848 multiple-choice and question-answering items based on 8 common types of images in the financial domain.The upper part of the image illustrates the overall structure of VisFinEval, where the business depth increases with the change in color. Concurrently, the demands on the model's understanding and analytical capabilities regarding financial business operations progressively increase. The lower part shows a specific example corresponding to the sub-scenarios.}
    \label{fig:frame}
    \vspace{-2pt}
\end{figure*}

\section{Introduction}
The advent of multimodal large language models (MLLMs) has dramatically broadened the scope of AI applications beyond pure text understanding to encompass tasks that require joint reasoning over images and text, including web navigation, sports analytics, and visual quality assessment \citep{deng2023musilingo,koh-etal-2024-visualwebarena,xia2024sportu,ku-etal-2024-viescore}. In the financial domain, practitioners routinely encounter richly formatted visual data—charts, tables, official seals—and yet existing benchmarks predominantly target textual comprehension, leaving a critical gap in the evaluation of MLLMs’ ability to integrate and reason over financial visuals. Text–only financial benchmarks such as FinEval \citep{guo2024fineval} and CFBenchmark \citep{zhu2024benchmarking} capture important language understanding skills but ignore chart- and document-based information that drives real-world decision making.  

Recent works, FinVQA \citep{bhatia2024fintral}, FIN-FACT \citep{zhang2024finagent}, MMMU \citep{wang2023finvis}, have begun to address multimodal finance, yet they suffer from limited scale, shallow question designs, or narrow coverage of business workflows. In practice, financial analysts progress through front-office data ingestion, mid-office analysis and decision support, and back-office policy and risk control. No existing benchmark systematically evaluates MLLMs across this full process, with tasks that range from basic chart reading to multi-step numerical calculations and counterfactual inferences under realistic perturbations.

To bridge these gaps, we present \textbf{VisFinEval}, the first large-scale Chinese benchmark for multimodal financial evaluation that mirrors end-to-end business scenarios. VisFinEval comprises 15,848 rigorously annotated QA pairs drawn from eight common financial image types (e.g., relationship graphs, K-line charts, official seals) and organized into three cascading scenario depths:  
\begin{itemize}
  \item \textbf{Financial Knowledge and Data Analysis (Front-Office)} tests foundational chart interpretation and basic numerical reasoning.  
  \item \textbf{Financial Analysis and Decision Support (Mid-Office)} challenges models with multi-image reasoning, metric computation, and investment backtesting.  
  \item \textbf{Financial Risk Control and Asset Optimization (Back-Office)} probes advanced capabilities in strategy optimization, policy impact analysis, and complex data extrapolation, including professional-exam-level questions.  
\end{itemize}
By simulating real-world document perturbations and multi-turn dialogues, VisFinEval captures the full complexity of financial workflows.

Our contributions are threefold:
\begin{itemize}
  \item \textbf{Comprehensive Multimodal Benchmark.} We construct VisFinEval with 15,848 QA pairs spanning eight types of financial images and three hierarchical scenario depths, filling a gap in financial MLLM evaluation.
  \item \textbf{Full-Process Business Workflow.} We align tasks with front-, mid-, and back-office functions, such as data perception, analytical decision support, and strategic optimization—thereby providing a practical, process-aware assessment framework.
  \item \textbf{Extensive Zero-Shot Evaluation.} We benchmark 21 state-of-the-art MLLMs in a zero-shot setting, analyze failure modes across six error categories, and compare model performance against non-expert and expert human baselines to highlight remaining challenges.
\end{itemize}
\begin{table*}[htbp]
\caption{Comparison of various benchmarks across multiple dimensions. Abbreviations in the header are: QT(Question Type), MC (Multiple-Choice questions), OE (Open-Ended questions), T/F (True/False questions), MLD(Multi‐level Difficulty), SD(Scenario Depth), RES (Realistic Environment Simulation), OS (Official Seal), FRG (Financial Relationship Graph), NoFFT (Number of Financial Figure Type), NoFS (Number of Financial Scenarios), NoQ (Number of Questions), and NoM (Number of Models). To better simulate real-world environments during the question-answering process, we introduced RES, which refers to simulating unexpected situations that may occur in real-world financial business scenarios.}
\label{tab:benchmarks-overview}
\renewcommand{\arraystretch}{1.3} 
\setlength{\tabcolsep}{14pt} 

\resizebox{\textwidth}{!}{%
\begin{tabular}{
  l 
  c 
  c 
  c 
  c 
  c 
  c 
  c 
  c 
  c 
  c 
}
\toprule[2pt]
\textbf{Benchmarks} & \textbf{QT} & \textbf{MLD} & \textbf{SD} & \textbf{RES} & \textbf{OS} & \textbf{FRG} & \textbf{NoFFT} & \textbf{NoFS} & \textbf{NoQ} & \textbf{NoM} \\
\midrule
\multicolumn{11}{c}{\textbf{Text}} \\
\midrule
FinDABench & OE & \checkmark & - & - & - & - & - & 5 & 2400 & 40 \\
SuperCLUE-Fin & OE & \xmark & - & - & - & - & - & 6 & 1000 & 11 \\
CFBenchmark & OE & \xmark & - & - & - & - & - & 8 & 3917 & 22 \\
FinEval & MC+OE & \xmark & - & - & - & - & - & 9 & 8351 & 19 \\
\midrule
\multicolumn{11}{c}{\textbf{MultiModal}} \\
\midrule
SEEDBENCH & MC & \xmark & \xmark & \xmark & \xmark & - & - & - & 19000 & 18 \\
MMMU & MC & \checkmark & \xmark & \xmark & \xmark & \xmark & - & - & 11500 & 30 \\
FinVQA & OE & \xmark & \xmark & \xmark & \xmark & \xmark & 2 & 2 & 1025 & 9 \\
FIN-FACT & T/F & \xmark & \xmark & \xmark & \xmark & \xmark & 2 & 5 & 3369 & 4 \\
FAMMA & MC+OE & \checkmark & \xmark & \xmark & \xmark & \xmark & 3 & 8 & 1758 & 4 \\
MME-Finance & OE & \checkmark & \xmark & \xmark & \xmark & \xmark & 6 & 11 & 2274 & 19 \\
\textbf{VisFinEval (Ours)} & MC+T/F+OE & \checkmark & \checkmark & \checkmark & \checkmark & \checkmark & 8 & 15 & 15848 & 21 \\
\bottomrule[2pt]
\end{tabular}%
}
\end{table*}

This paper is organized as follows. Section~\ref{s2} provides a review of related work in financial MLLMs and multimodal benchmarks. Section~\ref{s3} details the construction of VisFinEval, including data collection, question design, and quality control procedures. Section~\ref{s4} and Section~\ref{s5} presents our experimental setup and results across different difficulty levels, followed by error analysis. Finally, Section~\ref{s6} concludes the paper and discusses potential future directions in multimodal financial intelligence.

\section{Related Work}
\label{s2}


\noindent\textbf{Financial Scenario Analysis}\hspace{1em} Under the accelerating digital transformation in the financial sector, the groundbreaking advancements in Large Language Models (LLMs) have injected new momentum into the integration of artificial intelligence (AI) and finance. Early research primarily focused on unimodal technical applications, including text understanding ~\citep{masry2024longfin,wilson2024fin2sum}, sentiment analysis ~\citep{delgadillo2024finsosent,zhang2023instruct}, financial time-series forecasting ~\citep{li2024alphafin,li2024finreport,wang2024quantagent,mai2024stockgpt}, and decision support ~\citep{yu2024fincon,wang2023alpha,yu2024finmem,yang2023investlm,li2023multimodal}. However, these studies largely overlooked the critical value of chart-based data in financial contexts. Financial charts often encapsulate pivotal decision-making insights that penetrate beyond superficial data representations, only through accurate interpretation of such data can the core logic of financial decision-making be unveiled. This limitation was not alleviated until the emergence of multimodal large language models (MLLMs) ~\citep{bhatia2024fintral,zhang2024finagent,wang2023finvis} , which reconstructed MLLMs' comprehensive cognitive framework for the financial domain.

While existing benchmarks ~\citep{nie2024cfinbench,zhu2024benchmarking,koncel2023bizbench,zhu2021tat,zhao2024financemath,wang2024doctabqa,reddy2024docfinqa,chen2021finqa,chen2022convfinqa,chen2024fintextqa,chen2025mtbench} effectively evaluate models’ financial text comprehension capabilities, they remain inadequate for assessing models’ understanding of complex financial operations and multimodal chart data. Consequently, there is an urgent need to extend existing benchmarks to comprehensively evaluate models’ multimodal financial data comprehension and reasoning abilities, thereby more authentically reflecting their real-world applicability in financial scenarios.





\noindent\textbf{Multimodal Financial Benchmark}\hspace{1em} Up to today, the availability of dedicated benchmarks for multimodal financial scenarios remains limited. General multimodal benchmarks ~\citep{li2023seed,liu2024mmbench,guthaus2001mibench,yue2024mmmu,mathew2021docvqa} predominantly fail to adequately encompass domain-specific financial tasks, making it challenging to accurately assess models’ professional capabilities in financial contexts. Furthermore, existing studies on financial multimodal benchmarks are confined to knowledge-level validation and lack systematic evaluation of models’ operational depth and workflow integration in financial scenarios, thereby failing to holistically reflect their practical efficacy in real-world financial applications.

FAMMA ~\citep{xue2024famma} provides financial knowledge-related question-answering tasks, but its data primarily originates from university textbooks and examination questions, limiting its evaluation scope to knowledge verification rather than complex financial operational scenarios. FinTMMBench ~\cite{zhu2025fintmmbench} incorporates images of limited diversity, and its question design lacks explicit mapping to concrete financial business contexts. While MME-Finance~\citep{gan2024mme} addresses operational scenarios, its narrow business scope, limited question volume, and absence of task difficulty stratification aligned with real-world complexities restrict its generalizability and result in insufficient task depth, creating a disconnect from practical realities. Additionally, current research predominantly evaluates MLLMs’ performance in controlled environments while neglecting the inherent complexity of real-world financial scenarios, thereby impeding accurate assessment of large models’ true capabilities in financial applications.

To address these gaps in prior studies, we introduce VisFinEval, the first large - scale benchmark specifically designed for multimodal large language models in finance. This benchmark integrates diverse real-world financial scenarios and potential edge cases, employing hierarchical evaluation criteria to comprehensively cover tasks ranging from foundational knowledge to complex operational workflows. VisFinEval effectively bridges the capability gaps in existing evaluation frameworks, enabling rigorous and realistic assessment of models’ financial multimodal intelligence.

\section{VisFinEval Benchmark}
\label{s3}
\subsection{Overview}
We propose VisFinEval, a multimodal benchmark designed for the Chinese financial domain, which aims to evaluate the capabilities of MLLMs in processing and reasoning across the entire financial business workflow. 
As the first large-scale evaluation framework that deeply integrates multimodal tasks with end-to-end financial business scenarios, VisFinEval is constructed based on the actual operational flow of the financial industry. 
It establishes evaluation dimensions aligned with real-world needs, spanning from foundational front-office financial data perception, to mid-office analytical decision-making, and ultimately to high-level back-office strategic planning.
This structure reflects not only the high-frequency demands observed in practical financial contexts, but also follows a hierarchical and process-aware financial decision-making process. 
This enables the construction of a comprehensive evaluation framework that covers the entire financial business process. 
Therefore, VisFinEval provides a multimodal financial evaluation benchmark that is both professionally grounded and practically valuable. 
The overall framework is illustrated in Figure~\ref{fig:frame}.

VisFinEval is designed based on the front–mid–back office collaborative architecture commonly adopted in real-world financial systems, and establishes a three-tier evaluation framework that spans the entire financial business lifecycle. By integrating multimodal data with scenario-driven financial tasks, this benchmark systematically assesses the domain-specific capabilities of MLLMs within the vertical of financial scenarios.

This evaluation system is the first to achieve full-chain coverage of front-, mid-, and back-office financial functions. 
The front-office layer establishes a foundation of multimodal data perception, the mid-office layer constructs the core for analytical decision-making, and the back-office layer forms a closed loop for strategic optimization. 
Through modular decomposition and reorganization of financial workflows, VisFinEval ensures both the professional depth of evaluation tasks within each layer and the assessment of holistic model performance in cross-functional collaboration scenarios. 

The evaluation tasks adopt a wide range of objective formats, including single-choice, multiple-choice, true/false, and numerical reasoning questions, while also introducing dimensions such as multi-turn dialogue, counterfactual inference, multimodal consistency evaluation, and complex perturbation. By combining different question types and aligning them with specific sub-scenarios, the evaluation ensures a comprehensive assessment of MLLMs' capabilities in real-world financial tasks. Details on the task scenarios and dataset distributions are shown in Table~\ref{tab:visfineval_dist}, and representative examples of question types are provided in the Appendix~\ref{datasum}.

\subsection{Data Generation and Quality Control}
\label{quality}

During the data construction process of VisFinEval, most visual inputs are collected from PDF documents within the financial domain. These images are sourced from financial research reports, annual reports, and professional examinations such as the Chinese CPA and actuary exams. The dataset includes a diverse set of image types, including financial relationship graphs, line charts, histograms, candlestick (K-line) charts, pie charts, official seals, financial statements, and supporting data tables. Specifically, line charts, histograms, pie charts, and relationship graphs are primarily extracted from financial research reports; financial statements and supporting tables are collected from annual reports and exam questions; seal images are sourced from open-source datasets such as~\cite{trocr-seal-recognition}; and K-line charts are obtained from publicly accessible financial websites. All image materials are verified to be free from copyright restrictions.

The core of the data generation process lies in constructing scenario-specific prompts aligned with financial experts to ensure the domain relevance and consistency of question-answer (QA) pairs. These prompts guided the use of the Qwen-VL-Plus-latest~\cite{qwen2.5} to generate QA pairs based on the input images. The reliability of vision-language models in such generation tasks has been validated by prior work such as InstructBLIP~\cite{dai2023instructblip}. Furthermore, we used Qwen-max to classify the generated QA pairs into appropriate financial business scenarios spanning the full process. Detailed prompts used for the data generation and classification are provided in the Appendix~\ref{prompt}.

The QA data underwent a three-stage rigorous review process, including MLLM-based automated filtering based on multi-dimensional evaluation criteria, fine-grained annotation by trained and qualified undergraduate students, and cyclical validation by financial experts with ten years of work experience. This multi-layered processing pipeline ensures that we obtain high-quality QA pairs that meet standards such as accuracy, domain relevance, and consistency. Detailed review procedures and example prompts are provided in the Appendix~\ref{Details of Quality Review}.

Through this rigorous data generation and validation pipeline, VisFinEval offers a high-quality multimodal QA dataset tailored for evaluating the domain-specific capabilities of large multimodal models in the financial sector.

\subsection{VisFinEval Question Architecture}

VisFinEval, with financial business scenarios as its core starting point, has constructed a three-level hierarchical evaluation framework. This framework aims to systematically assess the comprehensive capability performance of Multimodal Large Language Models (MLLMs) in financial multimodal business, encompassing 15,848 high-quality QA pairs. It is further divided into the following three major real-world financial business scenarios based on scenario depth:

\noindent\textbf{Front-Office: Financial Knowledge and Data Analysis}\hspace{1em}Front-office operations in the financial domain are more oriented towards customer needs, focusing on the basic understanding of financial knowledge and data analysis-related capabilities. In this scenario, we have designed tasks covering financial cognition and data processing, primarily evaluating MLLMs' ability to understand customer needs and intentions in financial business and to process and analyze financial business data. This includes seven sub-scenarios: Financial Data Statistics, Candlestick Chart Analysis, 
Financial Indicator Assessment, Financial Entity Relationships Interpretation, Stock Selection Strategies Backtesting, Financial Information Extraction, and Financial Seal Recognition, corresponding to the real-world demands of financial front-office business activities.

\noindent\textbf{Mid-Office: Financial Analysis and Decision Support}\hspace{1em}The mid-office operations in the financial domain emphasize large-scale business facing the company or enterprise, requiring business personnel to have a deeper understanding and analysis of actual business, and to make clear and effective decisions on specific business issues under the influence of various factors in the real environment. Therefore, the design of this scenario aims to test the MLLM's comprehensive analysis and reasoning decision-making capabilities in a financial context. Tasks involve interpreting complex financial data structures and business logic, such as equity research and financial metric evaluation. Consequently, we have designed four core business sub-scenarios to evaluate the MLLM's information integration and systematic decision-making abilities: Financial Scenario Analysis, Industry Analysis and Inference, Investment Analysis, and Financial Market Sentiment Analysis, focusing on the analytical and decision-support functions typically undertaken by the mid-office.

\noindent\textbf{Back-Office: Financial Risk Control and Asset Optimization}\hspace{1em}
The back-office operations in the financial domain lean more towards strategic research and risk control. Business personnel need to possess strong domain expertise, mathematical calculation, and reasoning abilities to support front-office and mid-office operations through various internal decision-making processes, thereby ensuring the overall efficiency and effectiveness of financial business. To maintain the business authenticity of the evaluation, we have designed a series of highly complex financial tasks, including statistical inference, audit analysis, and expert-level reasoning, to assess whether MLLMs meet the requirements of actual business. Some of these tasks are adapted from challenging professional qualification exams, such as the Chinese CPA exam. Consequently, we have identified four sub-scenarios: Financial Strategy Optimization, Financial Risk and Policy Analysis, Financial Data Reasoning and Interpretation, and Asset Allocation Analysis, designed to simulate the strategic decision-making and optimization processes of the financial back office.

Meticulously designed based on the entire financial business process system, VisFinEval boasts advantages in terms of its systematic nature, practicality, and comprehensiveness. It can provide a professional and challenging benchmark for evaluating the real-world applicability of MLLMs in financial business.
\begin{table*}[ht]
\caption{Main Results. The higher the value in the table, the higher the accuracy of the surface model. The Financial Analysis and Decision Support assesses models with Financial Data Statistics (FDS), Candlestick Chart Analysis (CCA), Financial Indicator Assessment (FIA), Financial Entity Relationships Interpretation (FERI), Stock Selection Strategies Backtesting (SSSB), Financial Information Extraction (FIE), and Financial Seal Recognition (FSR). The Financial Analysis and Decision Support tests with Financial Scenario Analysis (FSA), Industry Analysis and Inference (IAI), Investment Analysis (IA), and Financial Market Sentiment Analysis (FMSA). The Financial Risk Control and Asset Optimization evaluates Financial Strategy Optimization (FSO), Financial Risk and Policy Analysis (FRPA), Financial Data Reasoning and Interpretation (FDRI), and Asset Allocation Analysis (AAA), concluding with the calculation of the Weighted Average (WA) score for each model. The table also indicates operational constraints encountered by certain models in multi-image tasks, such as Multi-image Limit and Context Window Limit.}
\label{tab:multimodal-results}
\vspace{12pt}
\resizebox{1.0\textwidth}{!}{
    \begin{tabular}{lcccccccccc@{\hspace{3.2em}}c@{\hspace{3.2em}}c@{\hspace{2em}}cc@{\hspace{3.2em}}c@{\hspace{3.2em}}c@{\hspace{2em}}cc}
    \toprule[2pt]
    \multirow{2.5}{*}{\textbf{Model}} & \multirow{2.5}{*}{\textbf{Size}} & \multirow{2.5}{*}{\textbf{Limit}} & \multicolumn{7}{c}{\textbf{Financial Knowledge and Data Analysis}} & \multicolumn{4}{c}{\textbf{Financial Analysis and Decision Support}} & \multicolumn{4}{c}{\textbf{Financial Risk Control and Asset Optimization}} & \multirow{2.5}{*}{\textbf{WA}}\\
    \cmidrule(lr){4-10} \cmidrule(lr){11-14} \cmidrule(lr){15-18}
    & & & \textbf{FDS} & \textbf{CCA} & \textbf{FIA} & \textbf{FERI} & \textbf{SSSB} & \textbf{FIE} & \textbf{FSR} & \textbf{FSA} & \textbf{IAI} & \textbf{IA} & \textbf{FMSA} & \textbf{FSO} & \textbf{FRPA} & \textbf{FDRI} & \textbf{AAA} \\
    \midrule[1pt]

    Qwen-VL-max & Unknown & / & \textbf{78.8} & \textbf{90.5} & \textbf{87.4} & \textbf{89.2} & \textbf{86.2} & 90.6 & 77.9 & \textbf{65.3} & \textbf{83.1} & 82.3 & 76.8 & \textbf{49.1} & \textbf{58.2} & \textbf{58.2} & 71.0 & \textbf{76.3} \\
    Qwen-VL-max-latest & Unknown & / & 76.0 & 84.5 & 86.1 & 87.1 &79.3 & 88.6 & 84.4 & 59.6 & 82.6 & \textbf{82.8} & \textbf{79.3} & 44.0 & 52.2 & 48.9 & 71.8 & 73.8 \\
    InternVL3-78B & 78B & / & 71.2 & 83.5 & 71.4 & 86.7 & 79.5 & 87.8 & 87.4 & 64.3 & 82.1 & 80.4 & 78.7 & \textbf{49.1} & 52.8 & 46.6 & 66.5 & 72.5 \\
    Doubao-1.5-vision-pro-32k & Unknown & / & 75.6 & 79.0 & 84.2 & 85.5 & 76.8 & \textbf{91.7} & 74.4 & 56.7 & 80.2 & 79.8 & 77.3 & 30.0 & 54.5 & 54.5 & \textbf{75.6} & 71.7 \\
    InternVL2.5-78B & 78B & / & 73.3 & 77.9 & 72.3 & 84.2 & 84.0 & 88.4 & 82.9 & 63.3 & 81.5 & 80.1 & 75.2 & 41.0 & 53.1 & 47.6 & 68.4 & 71.5 \\
    Qwen2.5-VL-72B & 72B & / & 75.9 & 77.0 & 72.8 & 85.4 & 81.5 & 88.3 & 80.4 & 57.4 & 82.4 & 80.3 & 74.5 & 41.4 & 53.4 & 42.6 & 71.9 & 71.0 \\
    GPT-4o-2024-11-20  & Unknown & / & 72.0 & 76.8 & 74.9 & 81.7 & 71.8 & 83.8 & 83.9 & 61.9 & 77.9 & 78.5 & 73.2 & 41.0 & 40.5 & 41.6 & 67.9 & 68.5 \\
    Step-1o-vision-32k & Unknown & / & 48.9 & 78.4 & 80.2 & 84.1 & 75.3 & 88.2 & \textbf{98.0} & 40.3 & 78.8 & 78.6 & 76.1 & 39.2 & 45.2 & 49.0 & 65.8 & 68.4 \\
    Moonshot-V1-32k-vision-preview & Unknown & / & 56.2 & 82.8 & 73.4 & 80.5 & 73.9 & 87.6 & 68.3 & 61.9 & 77.7 & 77.0 & 72.3 & 39.2 & 55.8 & 53.6 & 64.0 & 68.3 \\
    Qwen2.5-VL-7B & 7B & / & 71.4 & 75.9 & 69.2 & 80.9 & 74.0 & 85.5 & 69.9 & 53.4 & 79.7 & 76.5 & 70.7 & 37.2 & 37.6 & 35.4 & 63.2 & 65.4 \\
    InternVL3-8B & 8B & / & 68.2 & 78.0 & 62.8 & 87.0 & 74.1 & 84.0 & 77.4 & 56.5 & 76.1 & 76.8 & 71.7 & 29.7 & 46.2 & 36.8 & 55.3 & 65.4 \\
    Gemini-2.5-pro-exp-03-25 & Unknown & / & 73.6 & 76.7 & 72.6 & 81.0 & 73.0 & 89.4 & 87.4 & 53.2 & 72.4 & 70.8 & 75.5 & 28.4 & 38.0 & 41.5 & 37.7 & 64.7 \\
    Claude-3-7-Sonnet-20250219 & Unknown & / & 70.5 & 73.4 & 80.3 & 71.1 & 77.5 & 83.2 & 34.7 & 48.0 & 76.1 & 75.5 & 64.0 & 26.8 & 50.3 & 48.6 & 64.4 & 62.9 \\
    Qwen2.5-VL-3B & 3B & / & 69.5 & 81.1 & 65.9 & 76.6 & 73.6 & 83.4 & 72.4 & 50.0 & 75.4 & 74.7 & 66.6 & 22.9 & 34.8 & 35.9 & 53.8 & 62.4 \\
    MiniCPM-V-2.6 & 8B & / & 61.3 & 83.5 & 56.9 & 76.7 & 75.2 & 73.4 & 80.9 & 48.3 & 69.7 & 70.7 & 69.1 & 20.6 & 35.5 & 26.8 & 52.7 & 60.1 \\
    Llama-3.2-11B-Vision-Instruct & 11B & / & 56.9 & 40.8 & 59.3 & 63.9 & 62.9 & 73.1 & 70.4 & 45.3 & 69.7 & 67.1 & 63.4 & 18.0 & 22.1 & 19.9 & 31.1 & 50.9 \\
    Molmo-7B-D-0924 & 7B & / & 60.1 & 74.8 & 54.5 & 62.2 & 59.1 & 60.5 & 42.2 & 39.7 & 64.4 & 62.8 & 63.4 & 23.4 & 31.7 & 21.9 & 26.5 & 49.8 \\
    \midrule 
    GLM-4v-Plus-20250111 & Unknown & Multi-image Limit & 73.8 & 86.6 & 87.9 & 87.5 & 81.2 & 89.3 & 72.7 & 56.5 & 78.1 & 74.9 & 74.6 & 45.1 & 54.1 & 45.3 & 73.2 & 72.0 \\
    LLaVA-NEXT-34B & 34B & Context Window Limit & 55.3 & 79.8 & 92.3 & 63.2 & 87.8 & 55.0 & 58.8 & 54.3 & 88.2 & 88.1 & 66.9 & 13.1 & 17.5 & 12.7 & 7.7 & 56.0 \\
    LLaVA-v1.6-Mistral-7B & 7B & Context Window Limit & 54.6 & 73.4 & 65.9 & 62.1 & 47.4 & 47.0 & 62.3 & 42.3 & 58.3 & 56.4 & 63.7 & 10.2 & 16.3 & 35.9 & 21.1 & 47.8 \\
    LLaVA-NEXT-13B & 13B & Context Window Limit & 50.2 & 64.8 & 43.9 & 57.2 & 62.5 & 50.2 & 38.7 & 34.7 & 59.2 & 59.0 & 52.9 & 14.7 & 10.8 & 15.8 & 31.1 & 43.0 \\
    \bottomrule[2pt]
    \end{tabular}
}
\centering
\end{table*}

\section{Experiments Settings}

\label{s4}

\subsection{Models}
We tested 21 multimodal large language models , with close-source models accessed through their respective APIs and open-source models deployed locally. All inference tasks were run on NVIDIA A800 GPUs. For more details on the models please refer to Appendix \ref{modeloverview}.

\noindent\textbf{Closed-source models:} For close-source models, we evaluated 9 models, including Qwen-VL-max-lastest~\citep{qwen-vl-max}, Qwen-VL-max~\citep{qwen-vl-max}, Doubao-1.5-vision-pro-32k~\citep{Doubao-1.5-vision-pro-32k}, Step-1o-vision-32k~\citep{Step-1o-vision-32k}, Gemini-2.5-pro-exp-03-25~\citep{Gemini-2.5-pro-exp-03-25}, GPT-4o-2024-11-20~\citep{openai2024gpt4o}, Moonshot-V1-32k-vision-preview~\citep{moonshotai2024kimi}, Claude-3-7-Sonnet-20250219~\citep{cloude3.7sonnet} and GLM-4v-Plus-20250111~\citep{glm-4v-plus}.

\noindent\textbf{Open-source models:} For open-source models, we evaluated 12 models from several mainstream MLLMs, including Qwen2.5-VL-3B, Qwen2.5-VL-7B and Qwen2.5-VL-72B from the Qwen series~\citep{qwen2.5}; InternVL3-8B~\citep{InternVL3}, InternVL2.5-78B~\citep{InternVL} and InternVL3-78B~\citep{InternVL3} from the InternVL series~\citep{InternVL}; LLaVA-v1.6-Mistral-7B~\citep{llava}, LLaVA-NeXT-13B and LLaVA-NeXT-34B from the LLaVA series~\citep{llava-next}; as well as MiniCPM-V-2.6~\citep{MiniCPM-V-2_6}, Molmo-7B-D-0924~\citep{deitke2024molmo}, and Llama-3.2-11B-Vision-Instruct~\citep{touvron2023llama}.

\subsection{Evaluation Methods}


Despite our efforts to optimize prompts to improve model output, some models exhibit poor instruction following capabilities, making their output unsuitable for evaluation via rule-based extraction. To address this challenge, MMBench~\citep{xu2023mmbench} proposed leveraging LLMs as selection extractors, which significantly improved evaluation accuracy. Following a similar approach, we designed specific prompts and employed Qwen-max-latest as the judge model to evaluate the outputs of various models. To validate the judge model's evaluations, we conducted a manual review of all the results it provided for each model and task. The review showed that the accuracy of the judge model's judgments exceeded 98\%.
\section{Results}
\label{s5}
\subsection{Main Results}

We evaluated 21 mainstream MLLMs, as shown in Table \ref{tab:multimodal-results}. Due to a few limitations such as context length or multi-image support, certain questions were excluded from evaluation for some models; their results are provided separately for reference.

Among all the results, Qwen-VL-max achieved the best overall performance, with an average accuarcy of 76.3\%. It ranked highest among all evaluated models in 10 out of 15 sub-scenarios, strongly indicating Qwen-VL-max’s stable and powerful capabilities across diverse and in-depth multimodal financial scenarios. Closely following was Qwen-VL-max-latest, with only a 2.5\% difference, also demonstrating outstanding performance in FMSA and IA. Together, these results highlight the Qwen series’ excellence in the financial multimodal domain. Ranked third to sixth were InternVL3-78B, Doubao-1.5-vision-pro-32k, InternVL2.5-78B, and Qwen2.5-VL-72B, with relatively close scores. InternVL3-78B tied for the top score in FSO with Qwen-VL-max, reflecting its ability to optimize strategies in response to various challenges in financial business. Doubao-1.5-vision-pro-32k performed well in FIE , demonstrating strong visual information extraction capabilities in multimodal settings, and its high score in AAA further underscores its competence in asset allocation and financial analysis tasks. It is worth noting that Step-1o-vision-32k achieved an exceptionally high accuracy of 98.0\% in FSR significantly outperforming all other models. This suggests strong capabilities in this sub-scenario. In stark contrast, Claude-3-7-Sonnet-20250219 scored only 34.7\% in the same scenario, often failing to recognize seals correctly and sometimes even producing incorrect responses despite correct recognition. We attribute this primarily to its poor semantic alignment with Chinese, resulting in hallucinations.

From a pattern perspective, the performance gap between the open-source model InternVL3-78B and the closed-source model Qwen-VL-max is only 3.8\%, suggesting that as MLLMs continue to evolve, the performance disparity among top-tier models in financial tasks will gradually diminish. Regarding model size, both the Qwen and InternVL series show a clear trend where larger parameter models exhibit stronger capabilities. However, this trend is not observed in the LLaVA series, likely due to its origin from a startup organization, which may lack the training stability and iterative refinement seen in models backed by large internet companies like the Qwen series. A similar phenomenon is also observed in Molmo-7B-D-0924, a topic further explored in detail within FinEval \citep{guo2024fineval}. As task complexity increases, all models exhibit a noticeable decline in performance. This also demonstrates that VisFinEval effectively tests the boundaries of MLLMs' financial business capabilities, reflecting its authenticity and effectiveness.

\begin{figure*}[ht]   
    \centering
    \includegraphics[width=0.8\textwidth]{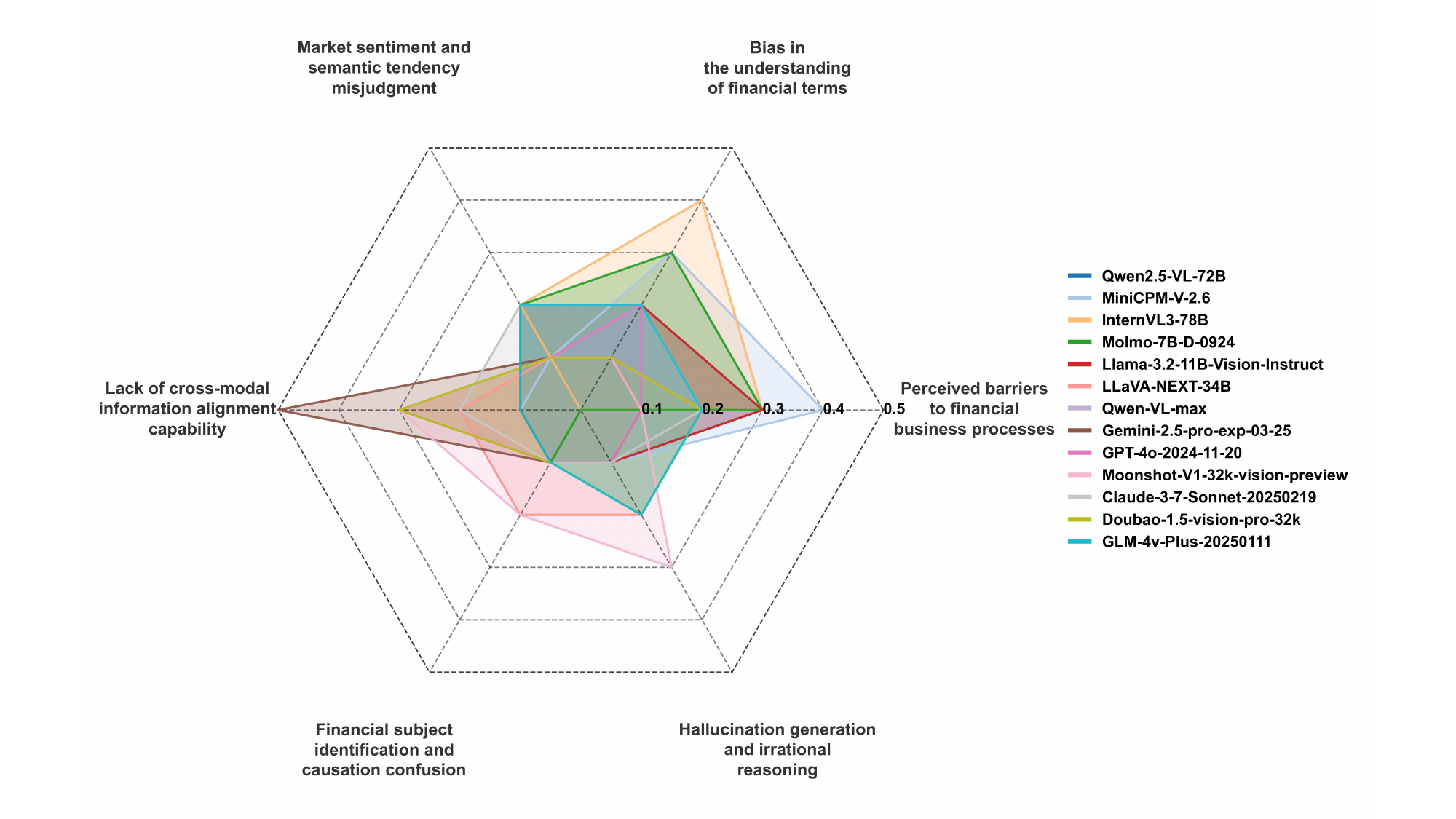}
    \caption{Error types' distribution across different MLLMs in VisFinEval tasks.}
    \label{erroranalysis}
    \vspace{-2pt}
\end{figure*}

\subsection{Comparative Analysis}
To better compare the capabilities of MLLMs and make a meaningful contribution to model research, we randomly selected 2\% of the questions from VisFinEval (approximately 300 questions) to conduct a competition among models, non-experts, and financial experts. Considering differences in domain knowledge and in order to better reflect the current stage of MLLM development, we selected the top two performing models from both open-source and closed-source categories for comparison. On the human side, we included a undergraduate students with no background in finance as representatives of non-experts, while the financial expert role was filled by a PhD candidate majoring in finance. All participants were uninvolved in any data annotation or review processes related to this study, and all responses were collected under closed-book conditions.

Unlike Table \ref{tab:multimodal-results}, here we calculate the average results for each of the three major scenarios as well as the overall average result to compare with human performance. As shown in Table \ref{tab:top}, the top-performing MLLMs have already outperformed the non-experts in all three major scenarios as well as in terms of overall average score. However, there remains a performance gap of over 14\% between the MLLMs and the financial expert, indicating that further iterations and improvements are still needed. A detailed analysis of the three major scenarios can be found in Appendix \ref{Details of Evaluation Results}.

\begin{table}[ht]
\caption{Performance comparison across non-experts, experts and MLLMs. FKDA represented Financial Knowledge and Data Analysis. FADS refers to Financial Analysis and Decision Support. FRCAO means Financial Risk Control and Asset Optimization. }
\label{tab:top}
\vspace{5pt}
\resizebox{0.5\textwidth}{!}{
    \begin{tabular}{lccccc}
    \toprule[2pt]
    \textbf{Source} & \textbf{Category} & \textbf{FKDA} &\textbf{FADS} & \textbf{FRCAO} & \textbf{Average} \\
    \midrule[1pt]
    Human & Non-experts & 72.1 & 57.0 & 40.1 & 56.4 \\
        & Experts & 93.3 & 88.0 & 82.8 & 88.0 \\
    \midrule
    Closed-Source & Qwen-VL-max & 85.8 & 76.9 & 59.1 & 73.9 \\
             & GPT-4o-2024-11-20  & 77.8 & 72.9 & 47.7  &66.1 \\
    \midrule
    Open-Source & InternVL3-78B & 81.1 & 76.4 & 53.7 & 70.4 \\
            & Qwen2.5-VL-72B & 80.2 & 73.6 & 52.3 & 68.7 \\
    \bottomrule
    \end{tabular}
}
\end{table}

\subsection{Error Analysis}


In all the incorrect answers from the evaluated MLLMs, we conducted a stratified sampling of 10\% of the questions for error analysis to investigate the issues MLLMs face in financial domain business capabilities. Based on the various types of errors made by MLLMs, we summarized six major problems in the financial domain: Lack of cross-modal information alignment capability, Market sentiment and semantic tendency misjudgment, Bias in the understanding of financial terms, Perceived barriers to financial business processes, Hallucination generation and irrational reasoning, Financial subject identification and causation confusion. These issues limit the MLLMs' performance in specific financial scenarios.

From Figure \ref{erroranalysis}, most MLLMs exhibit a relatively even distribution of errors, such as Qwen-VL-max and MoonShot-V1-32k-vision-preview, while a few models, such as Gemini-2.5-pro-exp-03-25, InternVL3-78B, and MiniCPM-V-2.6, show a higher concentration of errors in areas such as cross-modal consistency, understanding of financial terms, and financial business process, respectively. For more detailed analysis and related error examples, please refer to Appendix \ref{Examples for Error analysis}.

\section{Conclusion}
\label{s6}
This paper introduces VisFinEval, a benchmark designed to evaluate MLLMs' based on the full-process business system of the financial domain. t assesses MLLMs’ understanding and application abilities in real-world financial tasks through three major business scenarios, which together comprise fifteen sub-scenarios. Through comprehensive analysis of these scenarios, VisFinEval identifies eight commonly used chart types that cover a wide range of financial applications, enabling a performance evaluation grounded in actual business competencies. The results indicate that Qwen-VL-max performs the best overall; however, it still experiences some performance degradation in scenarios with the highest business complexity. Compared to humans, most current MLLMs have already outperformed non-expert individuals without a financial background, but a substantial gap remains when compared to financial experts. In addition, the error analysis highlights six major capability deficiencies that MLLMs exhibit in real-world financial applications. 
As a benchmark rooted in the full-process business workflows of the financial industry, VisFinEval provides a structured framework for measuring MLLMs’ practical capabilities in finance. We hope it will drive progress in MLLM research and contribute to enabling a deeper understanding of real-world financial scenarios.

\section*{Limitations}
While VisFinEval has made significant progress in evaluating multimodal large language models in the financial domain, it still has some limitations. Although VisFinEval includes some analysis of dynamic trend changes, it lacks in-depth research on the more dynamic micro and macro financial markets, which are closely related to time. Future work needs to consider designing an evaluation framework that can better assess the performance of MLLMs in more dynamic scenarios. The current work's evaluation is mainly focused on zero-shot performance, and it is necessary to further consider the potential of MLLMs to adapt through few-shot learning. Finally, although VisFinEval includes various financial image types, the distribution of these types and their relative importance in real-world financial analysis could be further refined, as the importance varies across different business scenarios. It is necessary to design more appropriate business scenario weights to evaluate the true performance of MLLMs in the financial domain.

\section*{Acknowledgments}
This work was supported jointly by the National Social Science Fund of China (Grant No. 22BTJ031, Liwen Zhang) and the Qinghai Provincial Key Laboratory of Big Data in Finance and Artificial Intelligence Application Technology.

\bibliography{VisFinEval}
\newpage
\appendix

\section{Details of VisFinEval}
\label{datasum}

\subsection{Design and Examples of Financial Business Scenarios }

We list the detailed information of VisFinEval data in Table \ref{tab:visfineval_dist}. 
Since we adopt a three-tier structure of financial business scenarios, comprising front-office, mid-office, and back-office. The detailed information of the financial business cenarios are presented below. 

Financial Knowledge and Data Analysis includes the following seven financial business scenarios:

\noindent\textbf{Financial Data Statistics}\hspace{1em}Organizing and analyzing enterprise or market financial data to support subsequent modeling and decision-making.

\noindent\textbf{Financial Information Extraction}\hspace{1em}Extracting key information from structured or unstructured data using NLP and computer vision techniques.

\noindent\textbf{Financial Indicator Assessment}\hspace{1em}Interpreting key financial indicators to assess operational capabilities and profitability of enterprises.

\noindent\textbf{Financial Entity Relationships Interpretation}\hspace{1em}Analyzing the logical and business relationships among institutions, individuals, and products presented in texts or images.

\noindent\textbf{Stock Selection Strategies Backtesting}\hspace{1em}Designing and backtesting quantitative stock selection strategies to evaluate historical performance and robustness.

\noindent\textbf{Candlestick Chart Analysis}\hspace{1em}Interpreting price trends and technical signals embedded in candlestick charts to support technical trading analysis.

\noindent\textbf{Financial Seal Recognition}\hspace{1em}Automatically detecting and verifying seals in financial documents (e.g., bills, contracts) to assist in compliance auditing.

Financial Analysis and Decision includes four core business scenarios:

\noindent\textbf{Industry Analysis and Inference}\hspace{1em}Leveraging industry data and trends to analyze industrial chain structures, competitive landscapes, and development trajectories.

\noindent\textbf{Investment Analysis}\hspace{1em}Evaluating asset allocation, valuation levels, and market outlooks to support investment decisions with quantitative insights.

\noindent\textbf{Financial Market Sentiment Analysis}\hspace{1em}Mining sentiment signals from sources such as news and social media to inform market forecasting and risk control.

\noindent\textbf{Financial Scenario Analysis}\hspace{1em}Identifying and modeling typical financial events, market behaviors, or trading contexts to assist in complex scene understanding.

Financial Risk Control and Asset Optimization includes four high-level financial task scenarios:

\noindent\textbf{Financial Strategy Optimization}\hspace{1em}Optimizing trading, investment, or risk management strategies under given constraints to improve the risk-return profile.

\noindent\textbf{Financial Risk and Policy Analysis}\hspace{1em}Identifying and quantifying systemic financial risks, and assessing the impact of macroeconomic and regulatory policy shifts on market stability and institutional behavior.

\noindent\textbf{Financial Data Reasoning and Interpretation}\hspace{1em}Building predictive models and causal inference frameworks from multi-source financial data to support strategic decision-making.

\noindent\textbf{Asset Allocation Analysis}\hspace{1em}Optimizing allocations across multiple asset classes to balance risk and return, aiming to construct optimal investment portfolios.

\begin{figure*}[ht]
    \centering
\includegraphics[width=1\textwidth]{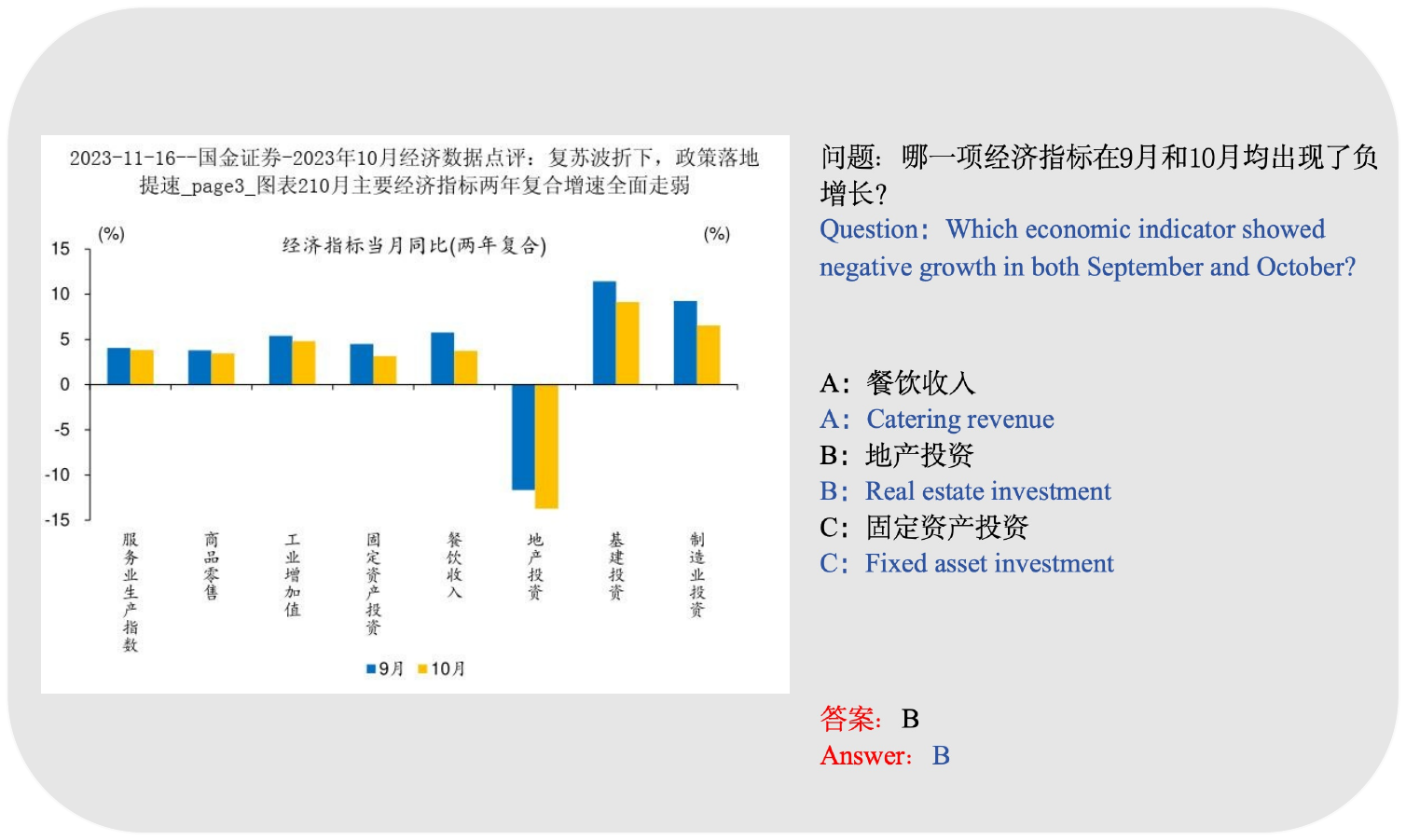}
    \caption{This is a three-option single-choice question related to Financial Data Statistics scenario. To answer this question accurately, the model must read the chart, identify the monthly growth rates of various economic indicators, and determine whether they showed negative growth in both September and October. This requires cross-temporal comparison and judgment of negative trends. The question assesses the model’s ability to extract consistent temporal trends from time series indicators, testing its precision in structured understanding and numerical reasoning over financial data.
    }
    \label{fig:L1a}
\end{figure*}

\begin{figure*}[ht]
    \centering
\includegraphics[width=1\textwidth]{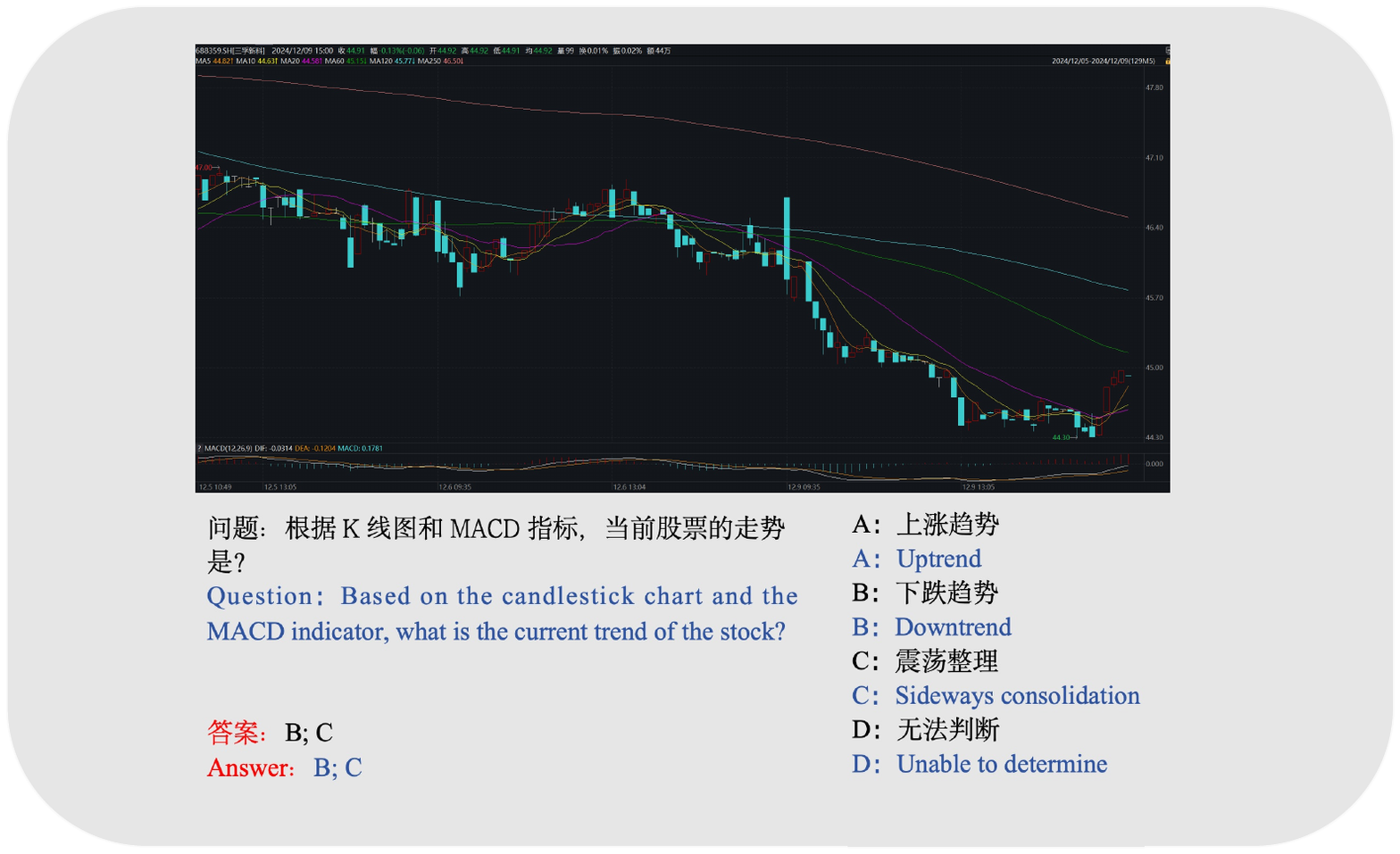}
    \caption{This is a four-option multiple-choice question focused on Candlestick Chart Analysis scenario. By identifying the price patterns in the candlestick chart and determining whether the MACD indicator has formed a "death cross" or "golden cross," the model must judge whether the stock is in an upward, downward, or sideways trend. Accurate answering requires interpretation of price action and understanding of MACD crossovers. The question evaluates the model’s ability to jointly reason over technical charts and financial indicators, testing its multimodal chart understanding and trend analysis capabilities in financial contexts.
    }
    \label{fig:L1b}
\end{figure*}

\begin{figure*}[ht]
    \centering
\includegraphics[width=1\textwidth]{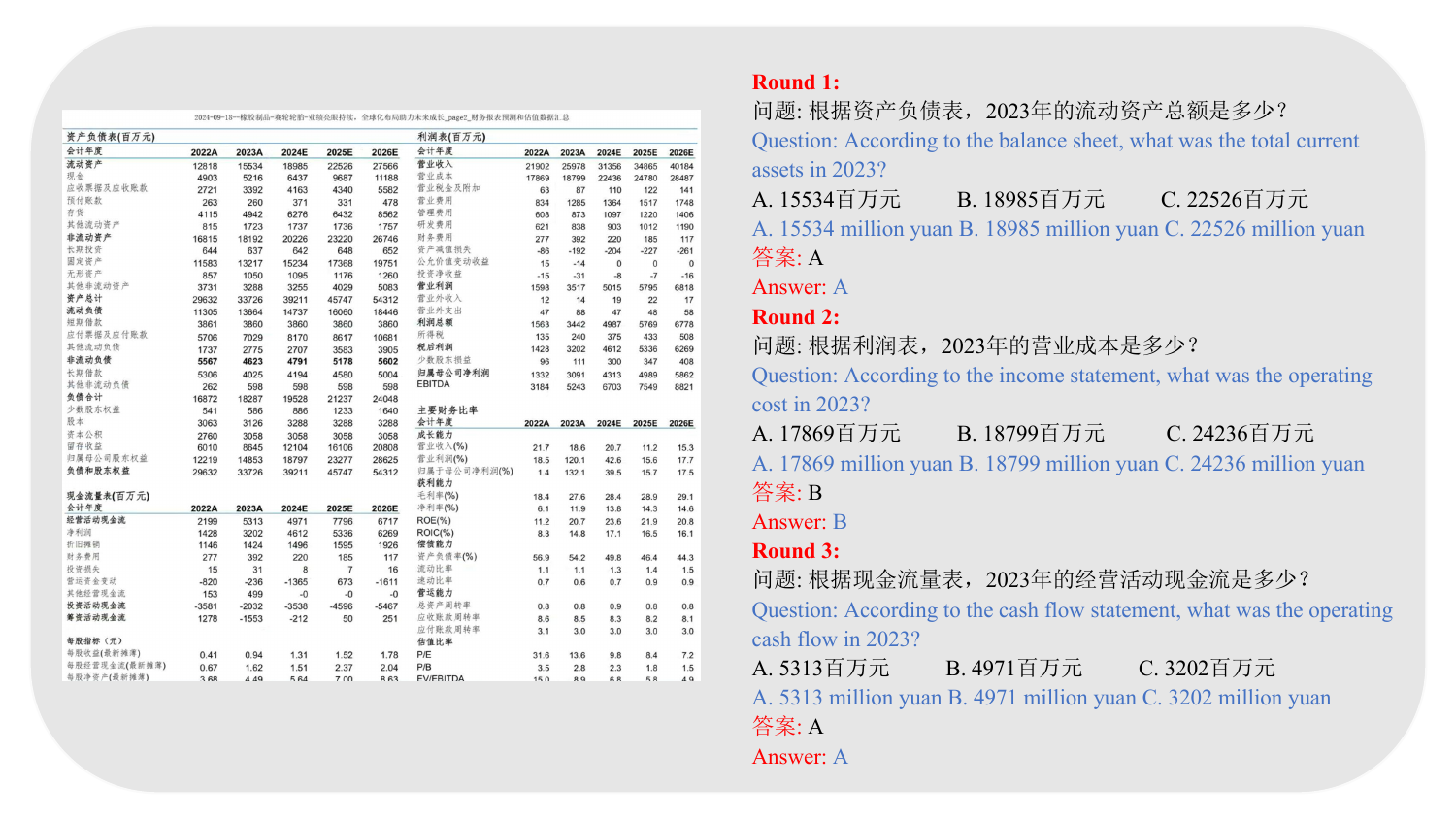}
    \caption{This is a single-choice question involving Financial Indicator Assessment, serving as an example of such analysis. Accurately answering this question requires the large model to identify the specified data year and specific accounting items in the question, locate the corresponding line items in the financial statements on the left, and then verify the data units and numerical precision while excluding distractors in the options. The accurately extracted financial data must then be compared with each option one by one. By examining key data from the three core financial statements (balance sheet, income statement, and cash flow statement), the question assesses the large model's fundamental ability to interpret a company's financial condition.For better readability, the English translation is displayed below the corresponding Chinese text.
    }
    \label{fig:L2b}
\end{figure*}

\begin{figure*}[ht]
    \centering
\includegraphics[width=1\textwidth]{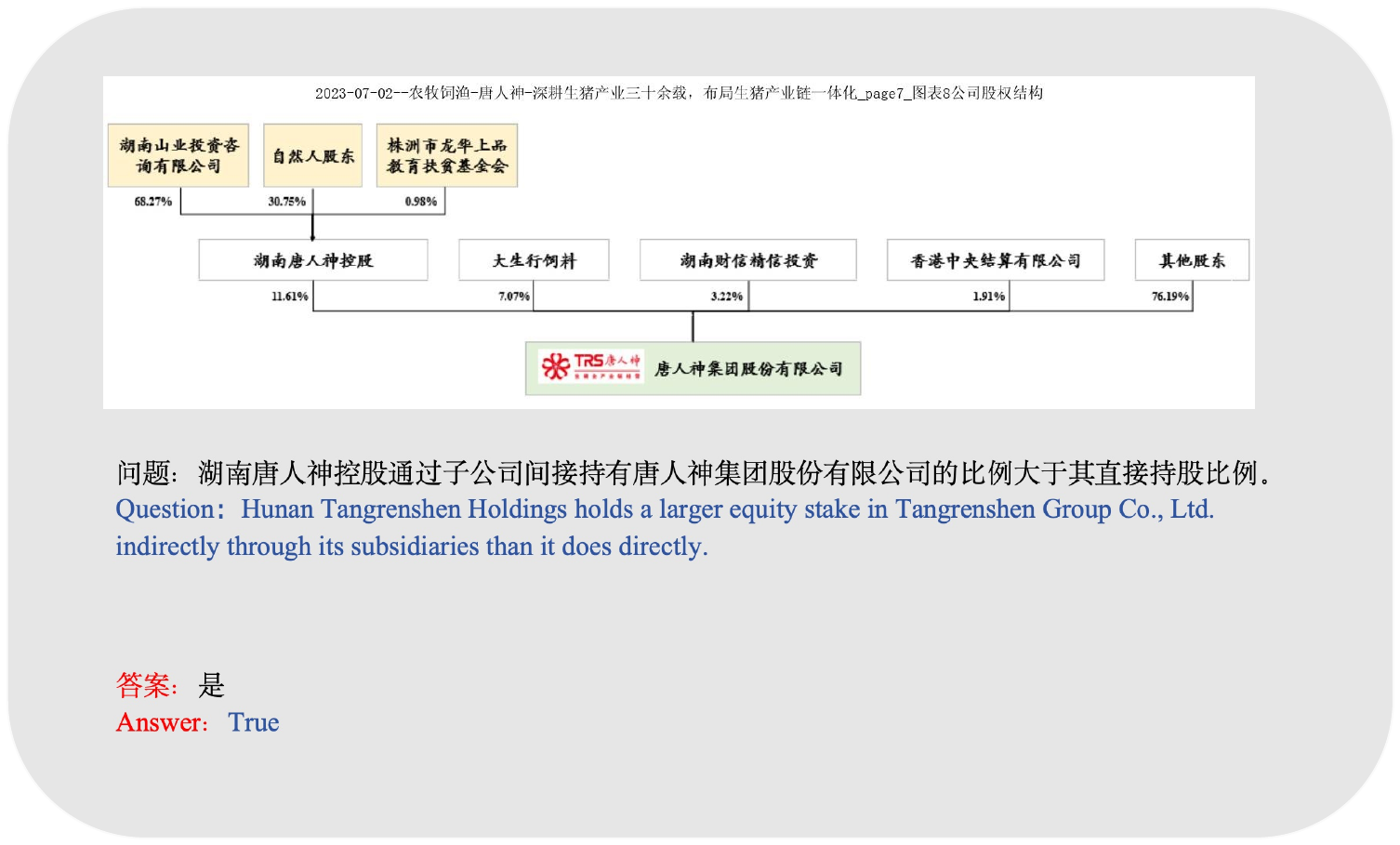}
    \caption{This is a true/false question in the scenario of Financial Entity Relationships Interpretation. To answer correctly, the model needs to analyze a corporate structure diagram or table, extract both direct and indirect shareholding paths, and compute the aggregate stake to determine whether indirect holdings exceed direct ones. The main challenge lies in multi-layered structural parsing and path aggregation. This question assesses the model’s ability to parse corporate ownership structures and reason over control paths, testing its accuracy in abstracting and comparing hierarchical entity relationships.
    }
    \label{fig:L1d}
\end{figure*}

\begin{figure*}[ht]
    \centering
\includegraphics[width=1\textwidth]{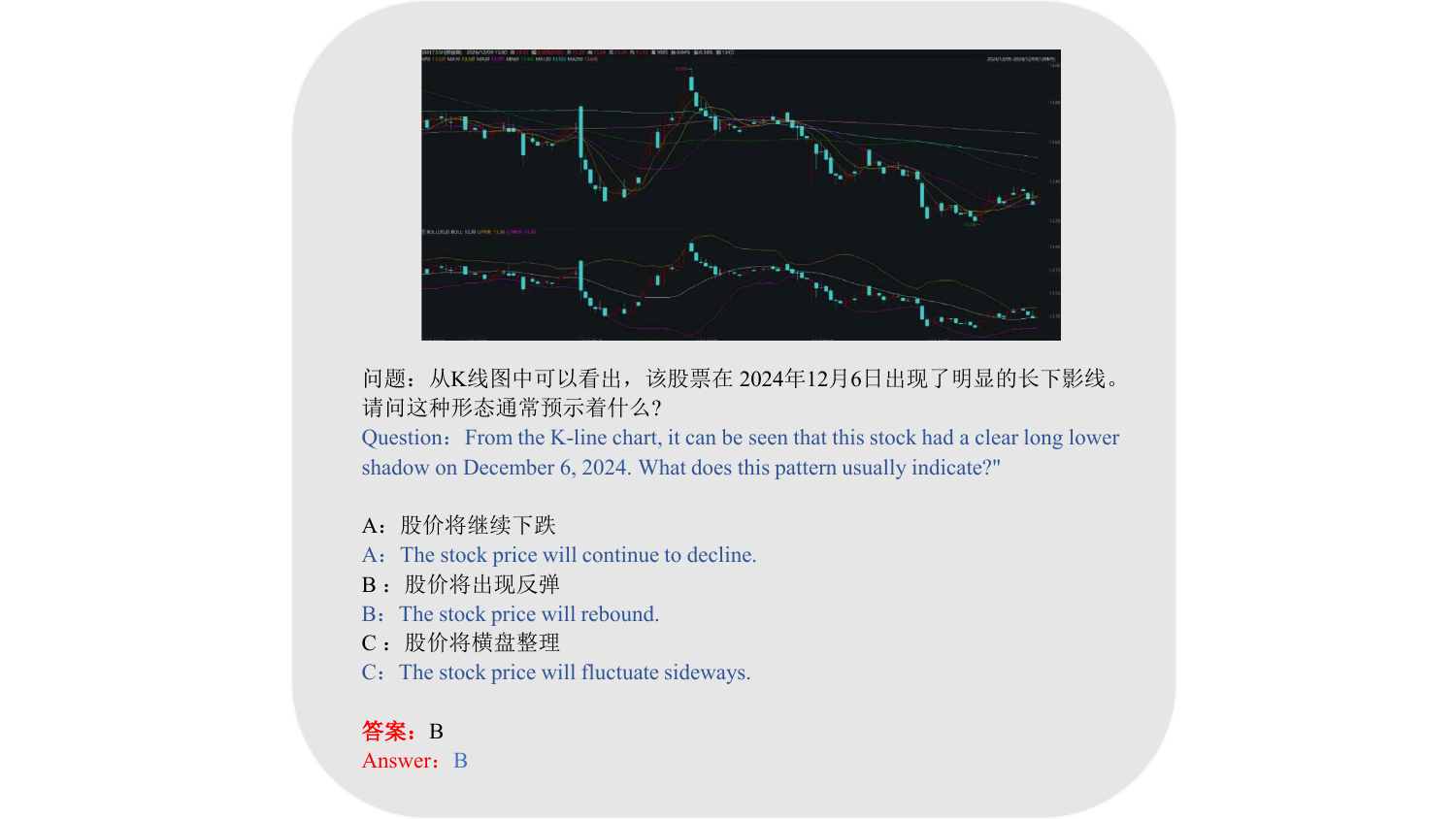}
    \caption{This is an example of Stock Selection Strategies Backtesting. Accurately answering this question requires the large model to identify the characteristics of the candlestick pattern and analyze the market implications of a long lower shadow. By recognizing the candlestick pattern (long lower shadow), this question tests the large model's quantitative application ability regarding technical analysis indicators.For better readability, the English translation is displayed below the corresponding Chinese text.
    }
    \label{fig:L2c}
\end{figure*}

\begin{figure*}[ht]
    \centering
\includegraphics[width=1\textwidth]{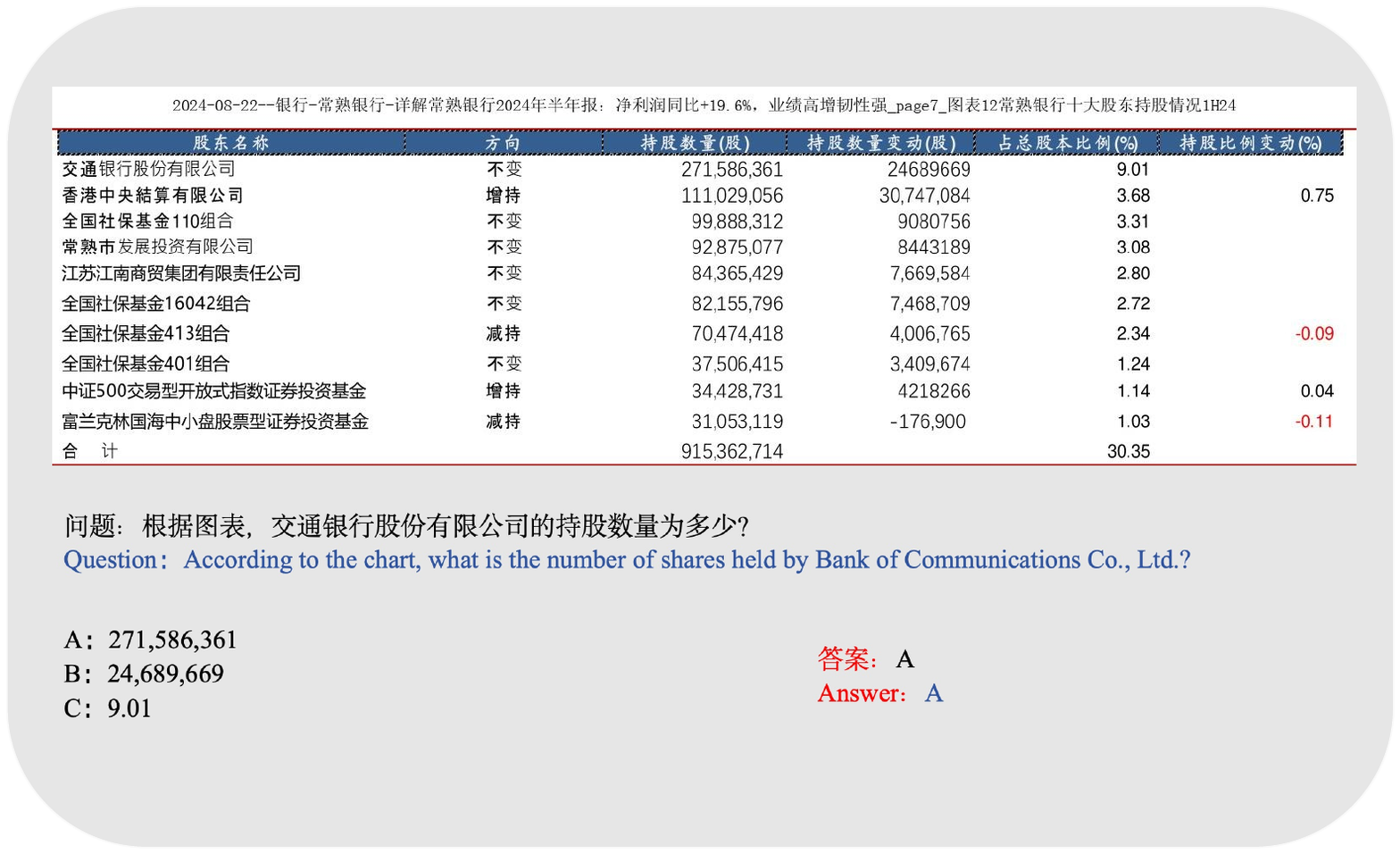}
    \caption{This is a three-option single-choice question in the scenario of Financial Information Extraction. To answer correctly, the model must locate the relevant row and column associated with the Bank of Communications in a tabular or graphical chart and extract the corresponding numerical value. The key lies in precise visual localization and accurate data extraction. This question tests the model’s ability to locate and extract key information from structured visual content, evaluating its accuracy in structured vision-language understanding and entity-value alignment.
    }
    \label{fig:L1f}
\end{figure*}

\begin{figure*}[ht]
    \centering
\includegraphics[width=1\textwidth]{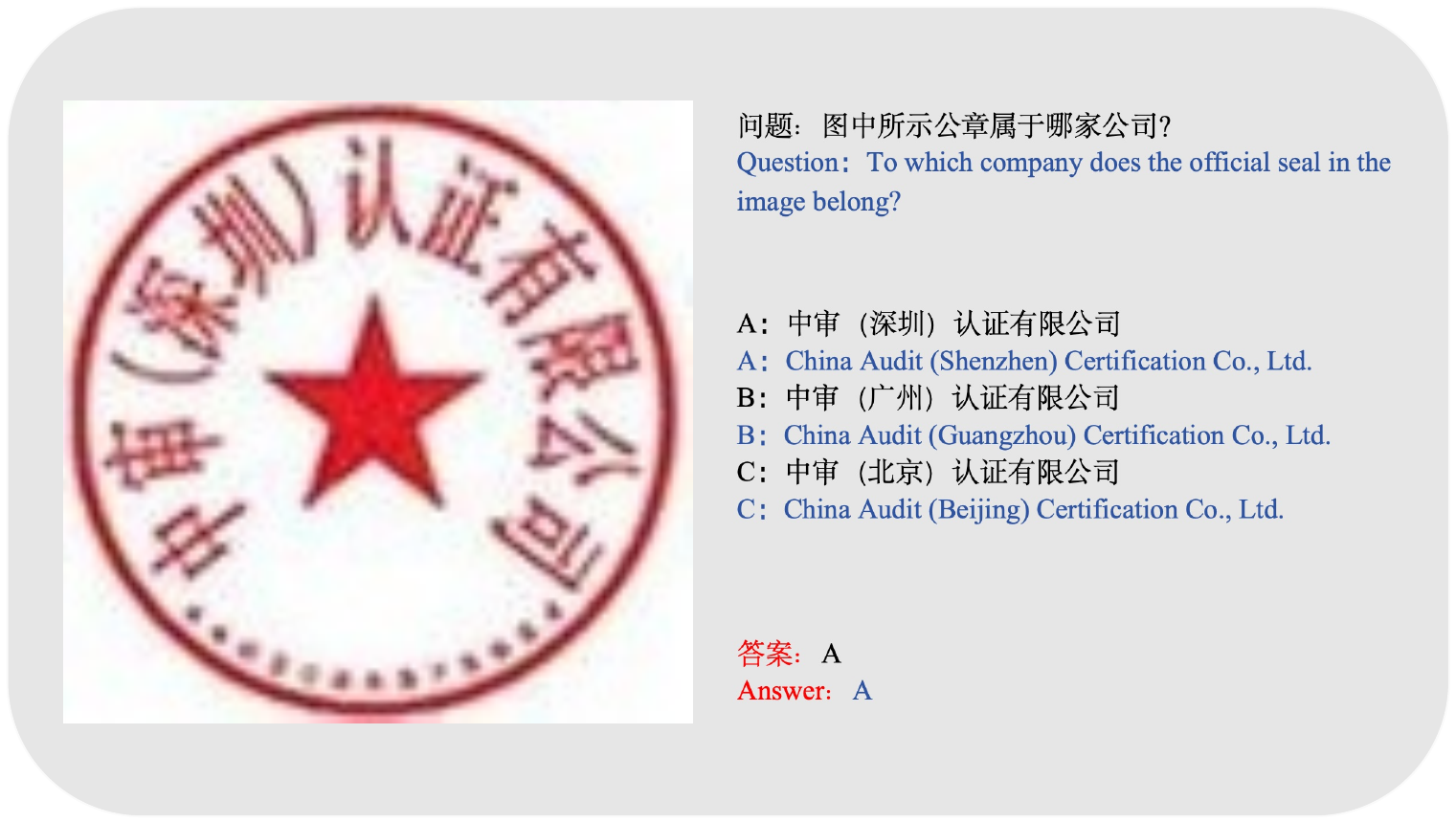}
    \caption{This is a three-option single-choice question on Intelligent Seal Recognition. To answer this question accurately, the model must analyze the textual structure and formatting of the seal shown in the image and compare it against the names of candidate institutions. The focus is on recognizing key terms in the seal and determining hierarchical or departmental alignment. This question assesses the model’s capacity to understand textual structures within stamp images and judge visual-semantic similarity, testing its fine-grained multimodal entity recognition and image-text alignment skills.
    }
    \label{fig:L1g}
\end{figure*}

\begin{figure*}[ht]
    \centering
\includegraphics[width=1\textwidth]{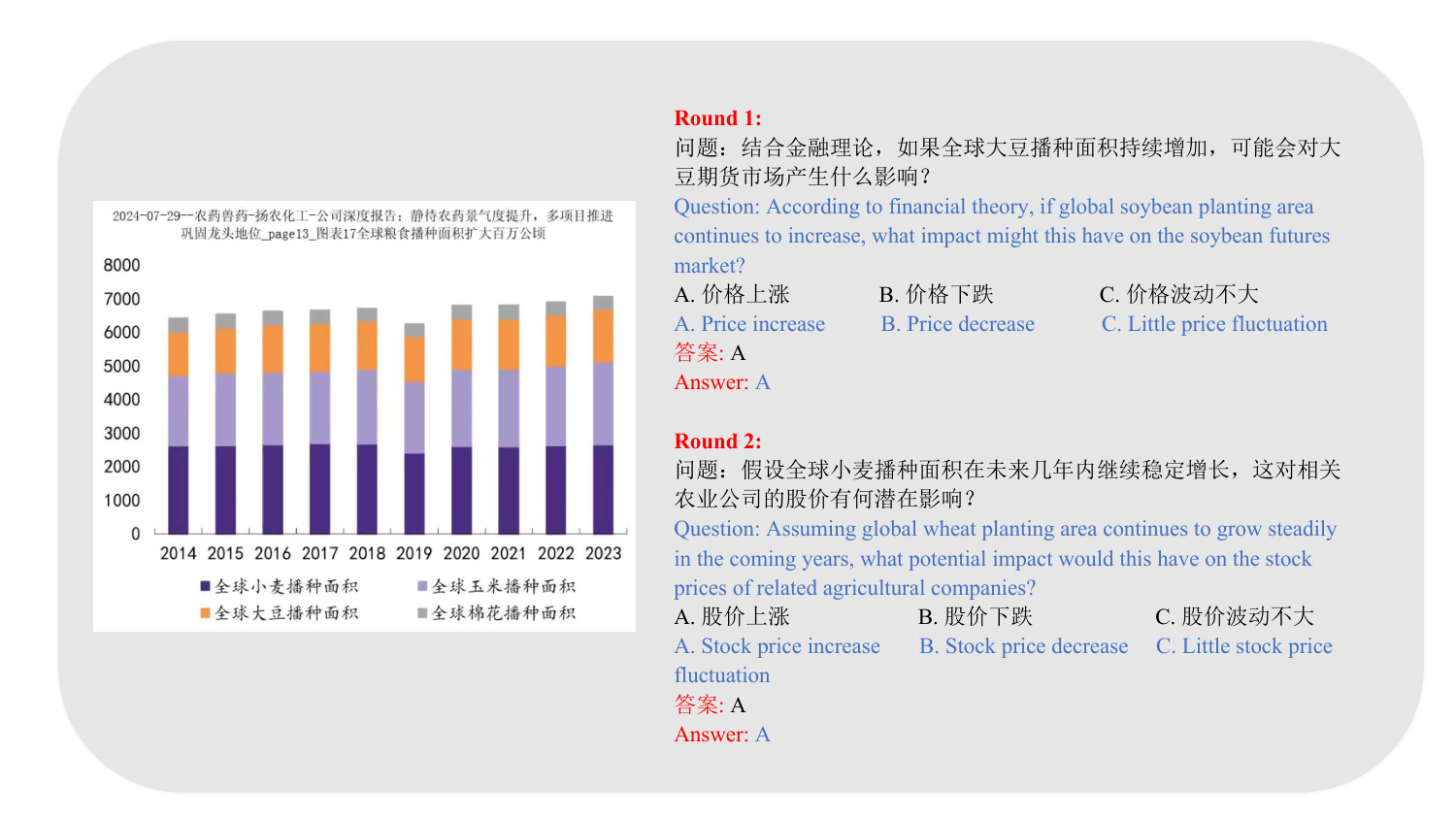}
    \caption{This is a single-choice question involving Industry Analysis and Inference. Accurately answering this question requires the large model to analyze the trend of chart data, observe the changes in the bar chart of global major crop planting areas from 2014 to 2023, and combine the question to interpret the market impact mechanism and summarize the transmission logic. By examining the transmission impact of changes in global crop planting areas on futures markets and the stock prices of listed companies, the question tests the large model's comprehensive analytical capabilities regarding the supply-demand relationship in the agricultural industry chain and investment logic.For better readability, the English translation is displayed below the corresponding Chinese text.
    }
    \label{fig:L2a}
\end{figure*}

\begin{figure*}[ht]
    \centering
\includegraphics[width=1\textwidth]{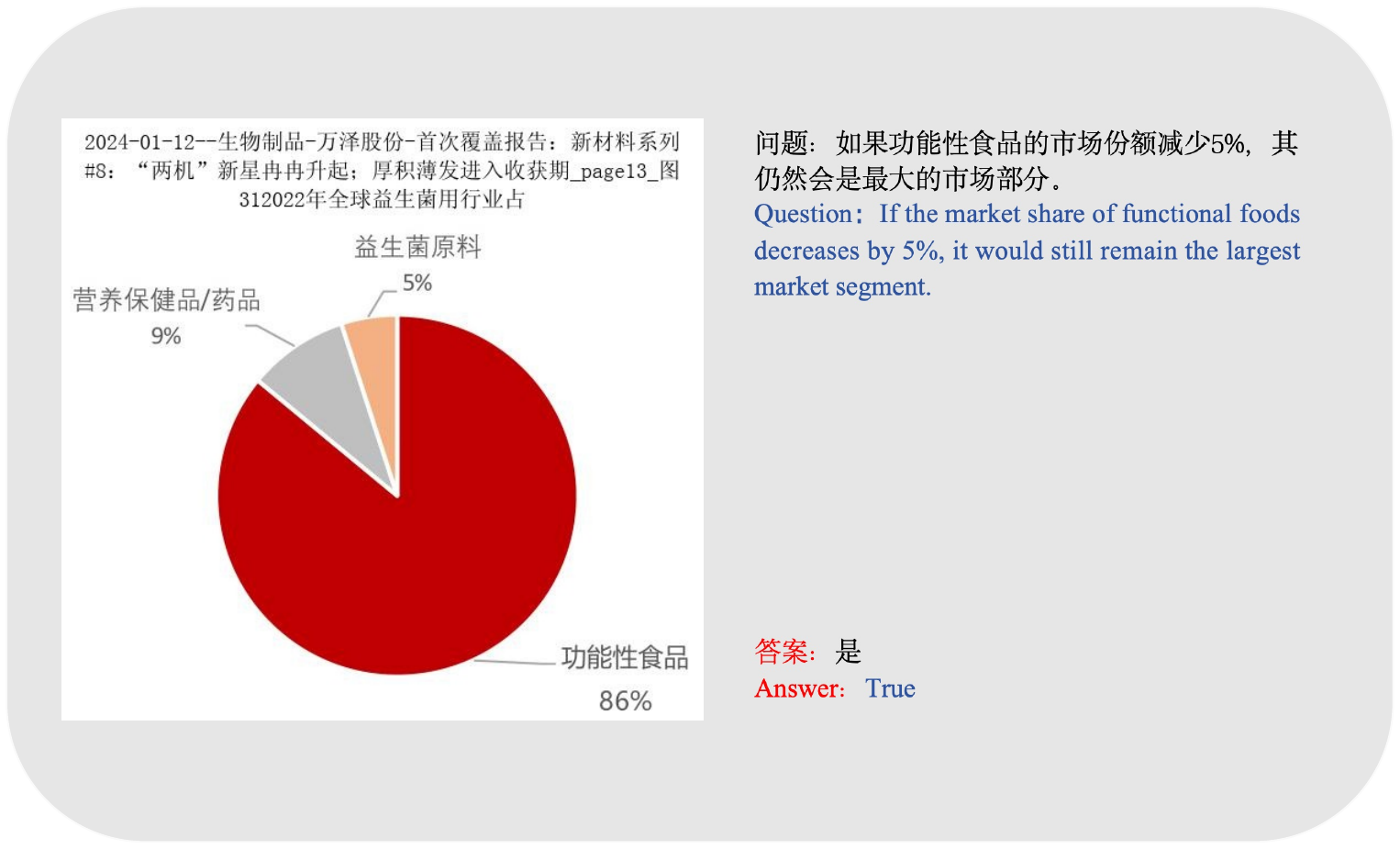}
    \caption{This is a counterfactual inference question within the Financial Scenario Analysis. A correct answer requires the model to perform hypothetical adjustments to the original market share data and determine whether functional foods would still hold the largest market share after a 5\% decrease. The key lies in constructing a counterfactual scenario and comparing adjusted values. The question examines the model’s sensitivity to causal changes among variables and the rigor of its reasoning process, testing its ability in numerical inference and logical reasoning under hypothetical financial settings.
    }
    \label{fig:L1c}
\end{figure*}

\begin{figure*}[ht]
    \centering
\includegraphics[width=1\textwidth]{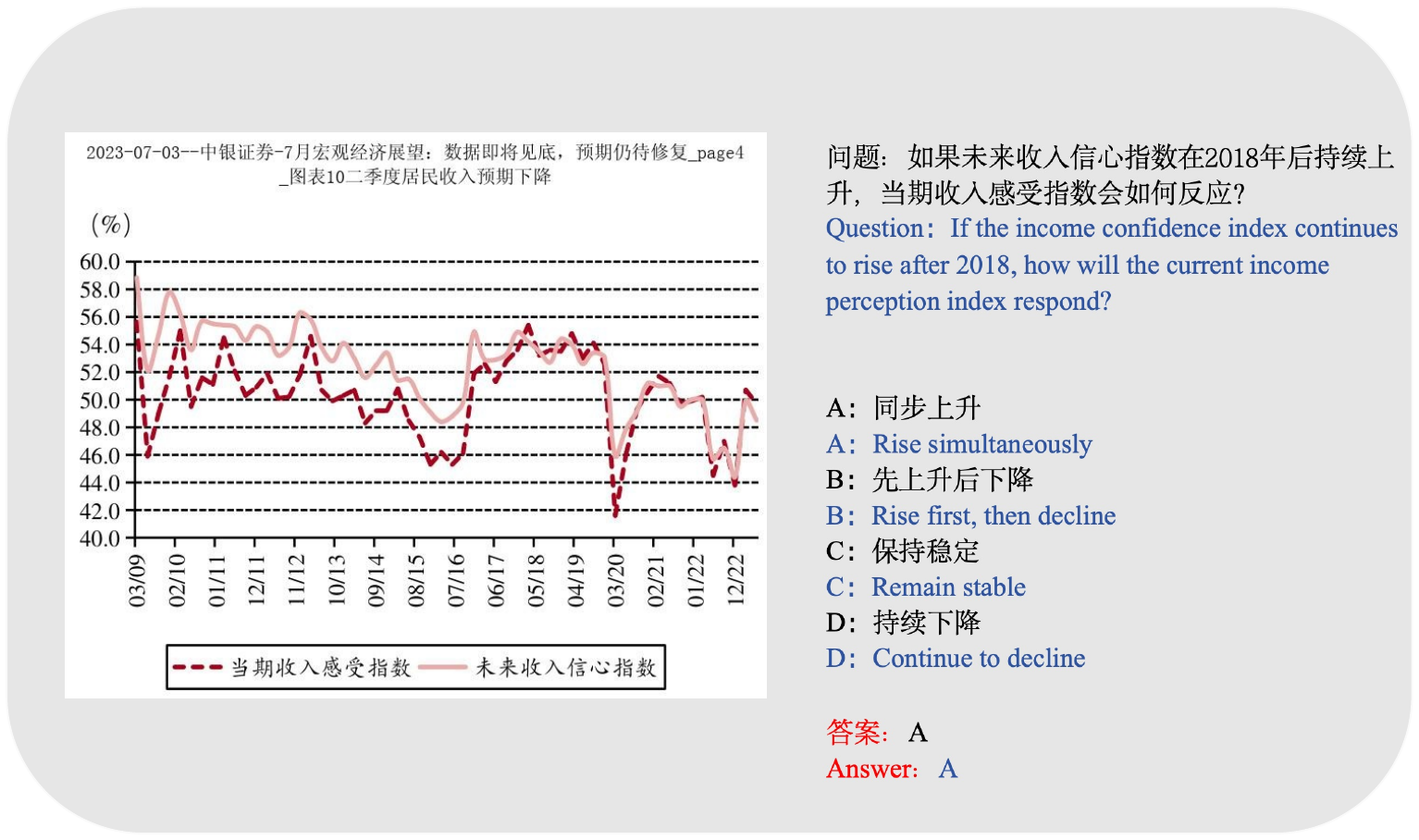}
    \caption{This is a example for Financial Market Sentiment Analysis. To answer this question accurately, the model must understand the logical relationship between the income confidence index and the income perception index, typically assuming that rising confidence leads to a rise in perception. The key is grasping trend co-movement and the economic implications of sentiment indicators. This question evaluates the model’s understanding of dynamic relationships among macro sentiment variables, testing its capacity in predictive reasoning and sentiment-driven analyzing in financial psychology contexts.
    }
    \label{fig:L1e}
\end{figure*}

\begin{figure*}[ht]
    \centering
\includegraphics[width=1\textwidth]{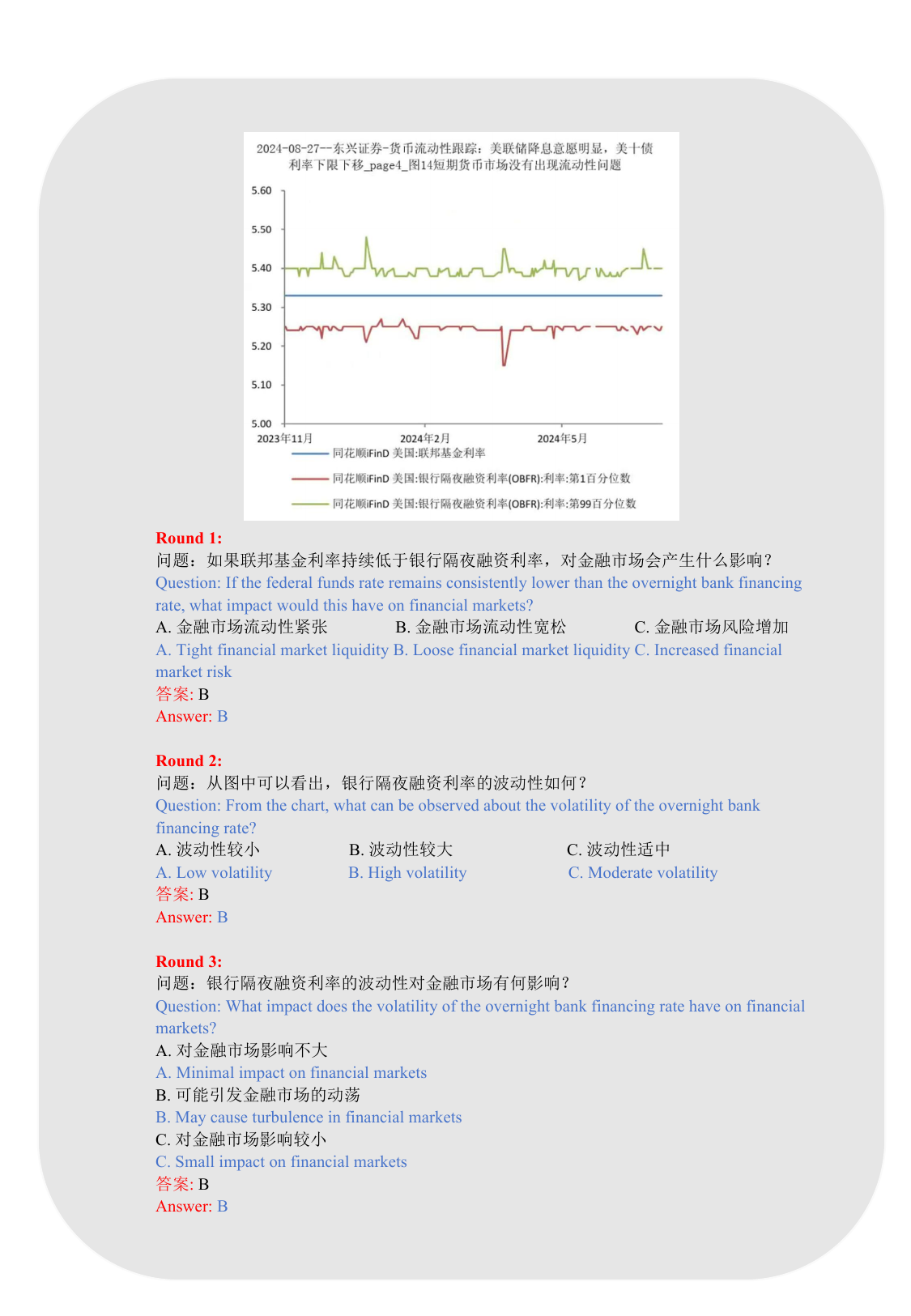}
    \caption{This is a standard Investment Analysis question. To respond accurately, the model must analyze the trends of three interest rate curves in the chart, interpret liquidity easing signals based on the question's context, and assess OBFR rate volatility's impact on financial markets to evaluate short-term money market liquidity risks and stability. The question tests the model's ability to analyze monetary policy rate differentials, market rate volatility, and their transmission effects, assessing its comprehensive understanding of money market liquidity and systemic risk mechanisms.}
    \label{fig:L2d}
\end{figure*}

\begin{figure*}[ht]
    \centering
\includegraphics[width=1\textwidth]{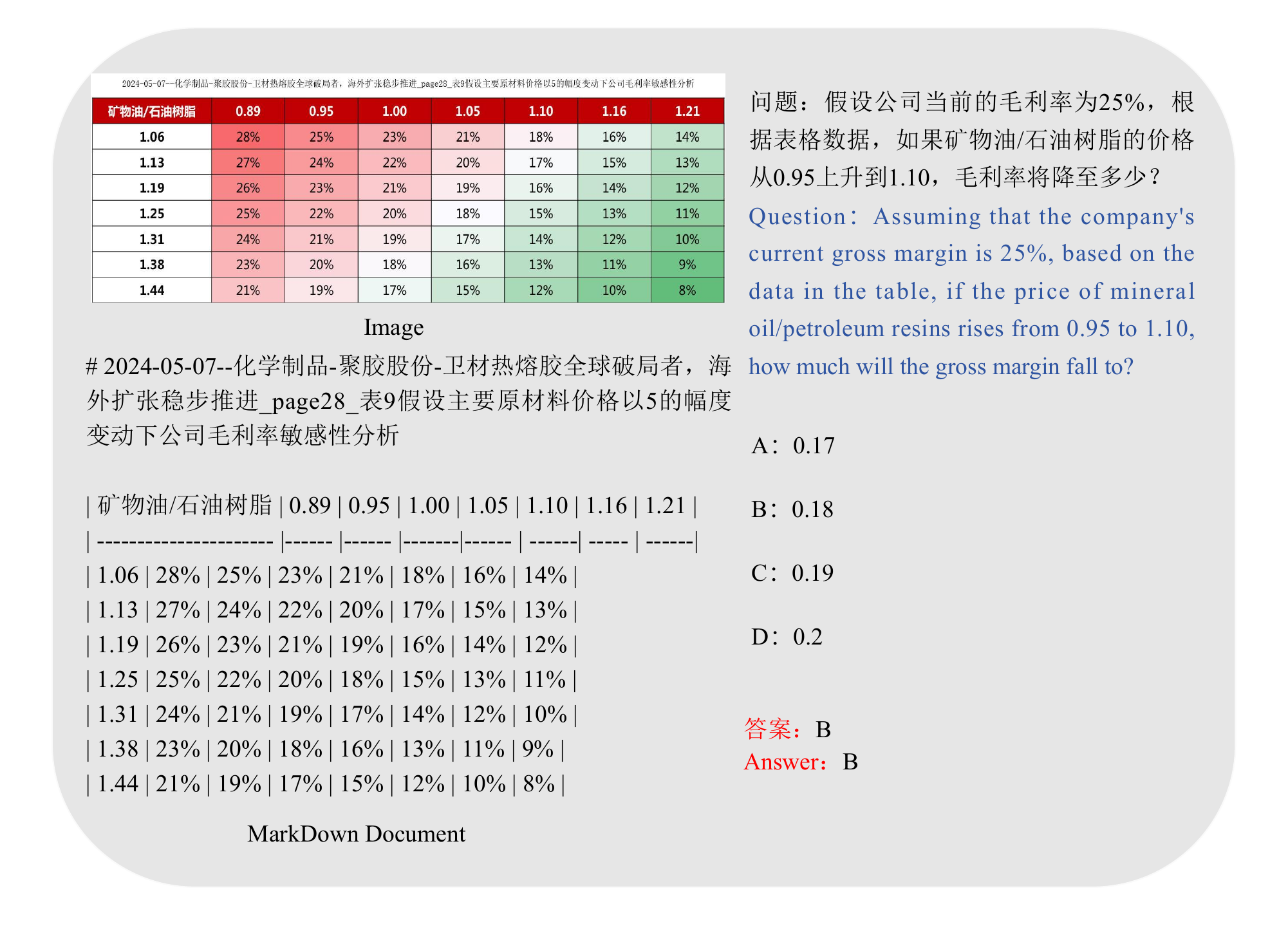}
    \caption{This is an example of for Financial Strategy Optimization. First, the large model needs to locate the row and column in the table where the price of mineral oil/petroleum resin is 0.95 and the corresponding gross margin is 25\%, then find the corresponding gross margin value when the price rises to 1.10 in the row, and finally match the value with the options. This question tests the model's ability to accurately find and locate the data in the table and analyze it according to the correlation between the data in financial scenarios. In addition, it also verifies the model's ability to extract consistent key data from images (visual) and structured text (Markdown table) to get the correct answer.}
    \label{fig:L3a}
\end{figure*}

\begin{figure*}[ht]
    \centering
\includegraphics[width=1\textwidth]{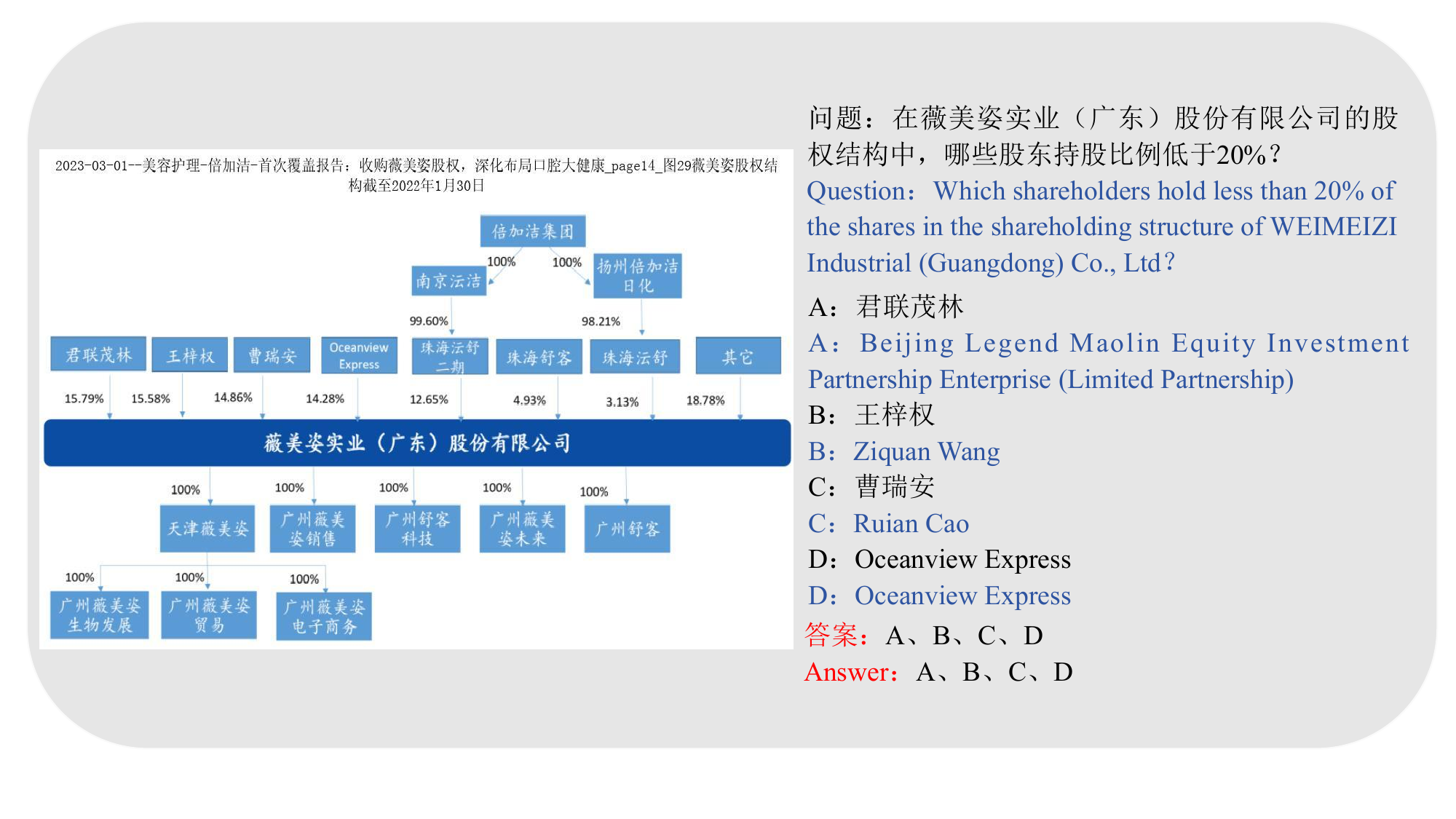}
    \caption{This is a question for Asset Allocation Analysis. Answering this question requires extracting information about the identity of shareholders in the equity structure diagram, obtaining their shareholdings, and comparing the 20\% threshold. The question tests the large model's ability to understand and extract data from the mapping of financial relationships, equity penetration analysis, and the ability to make judgments about specific conditions.}
    \label{fig:L3b}
\end{figure*}

\begin{figure*}[ht]
    \centering
\includegraphics[width=1\textwidth]{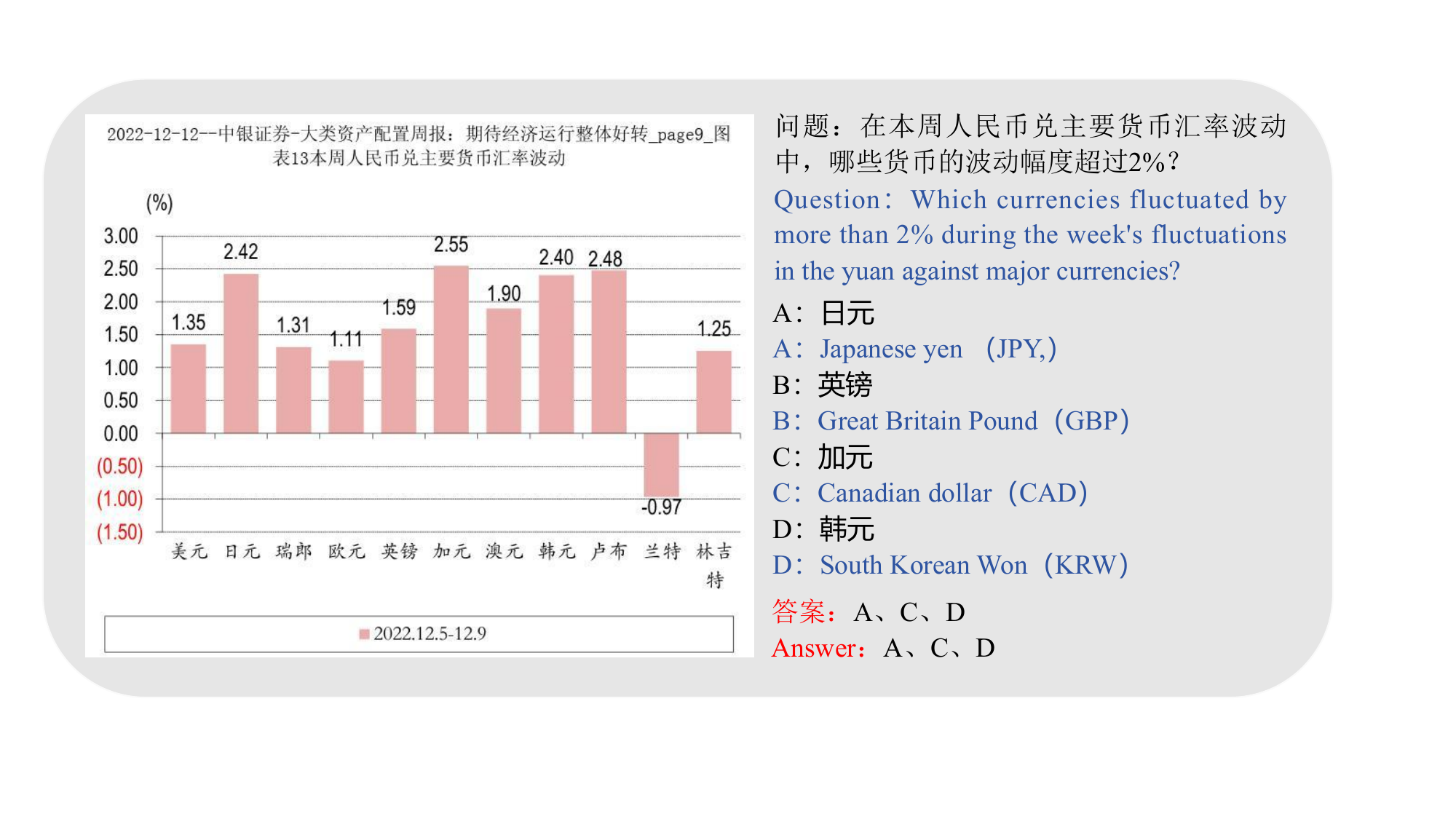}
    \caption{This question is a multiple‐choice item centered on Financial Risk and Policy Analysis. The critical task is to accurately extract the percentage fluctuations in the exchange‐rate chart and benchmark them against a 2 percent threshold. The item evaluates the model’s proficiency in recognizing financial data, assessing risk thresholds, and providing decision support.
    }
    \label{fig:L3c}
\end{figure*}

\begin{figure*}[ht]
    \centering
\includegraphics[width=1\textwidth]{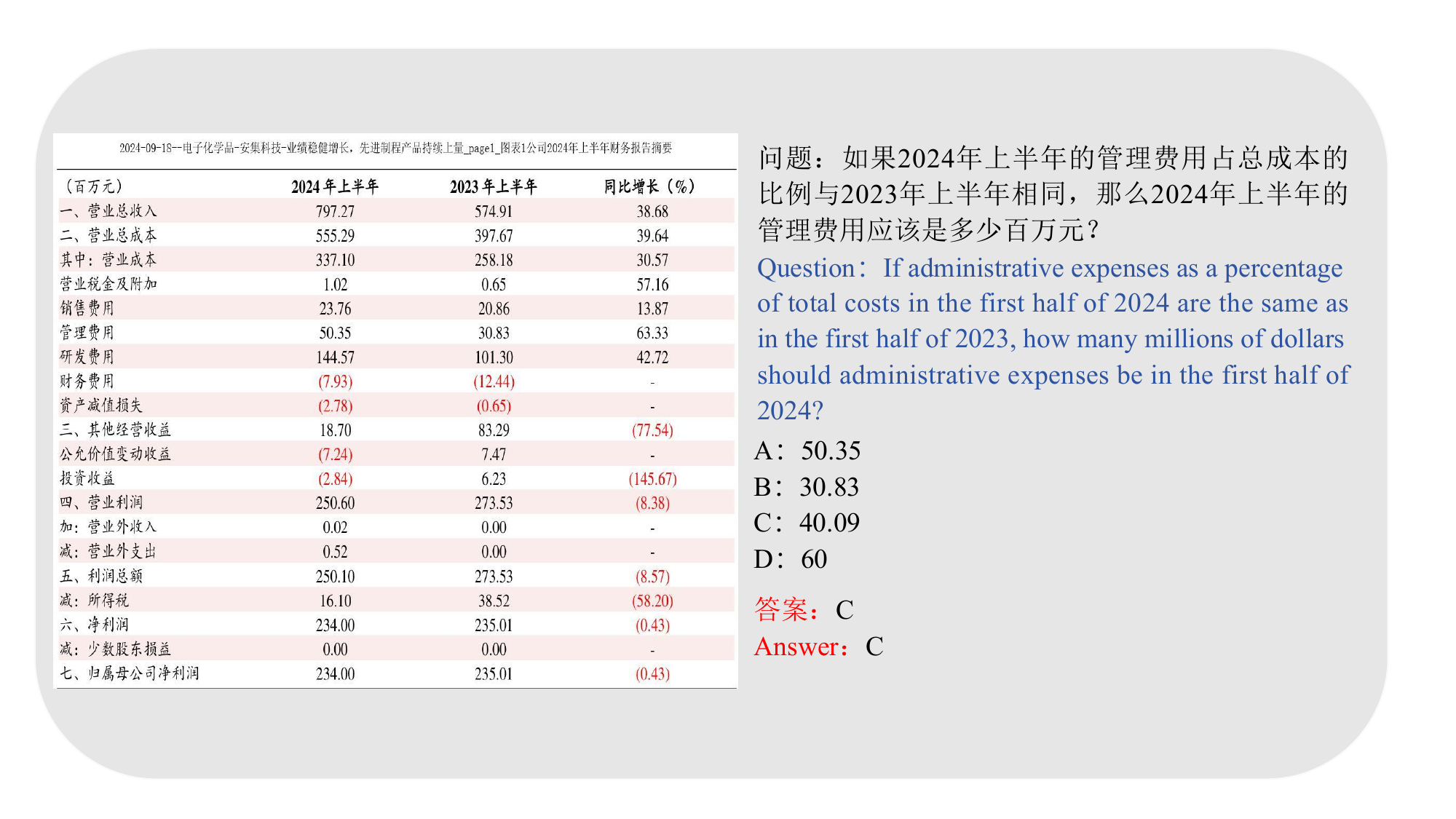}
    \caption{This is a multiple choice question involving Financial Data Reasoning and Interpretation. First, the large model needs to identify the overhead and total cost data extracted from the first half of 2023 and calculate its ratio; use this ratio to extrapolate with the total cost data obtained in 2024, and finally calculate the theoretical value of overhead in 2024. The question tests the ability of the large model to extract, calculate, and logically extrapolate financial statement data.}
    \label{fig:L3d}
\end{figure*}

Figure \ref{fig:L1a}, \ref{fig:L1b}, \ref{fig:L2b}, \ref{fig:L1d}, \ref{fig:L2c}, \ref{fig:L1f}, \ref{fig:L1g} are the examples of Financial Knowledge and Data Analysis. 
Figure \ref{fig:L2a}, \ref{fig:L1c}, \ref{fig:L1e}, \ref{fig:L2d} are the examples of Financial Analysis and Decision Support. 
Figure \ref{fig:L3a}, \ref{fig:L3b}, \ref{fig:L3c}, \ref{fig:L3d} are the examples of Financial Risk Control and Asset Optimization.

\subsection{Details of Quality Control}
\label{Details of Quality Review}
\begin{figure*}[ht]
    \centering
    \includegraphics[width=1\linewidth]{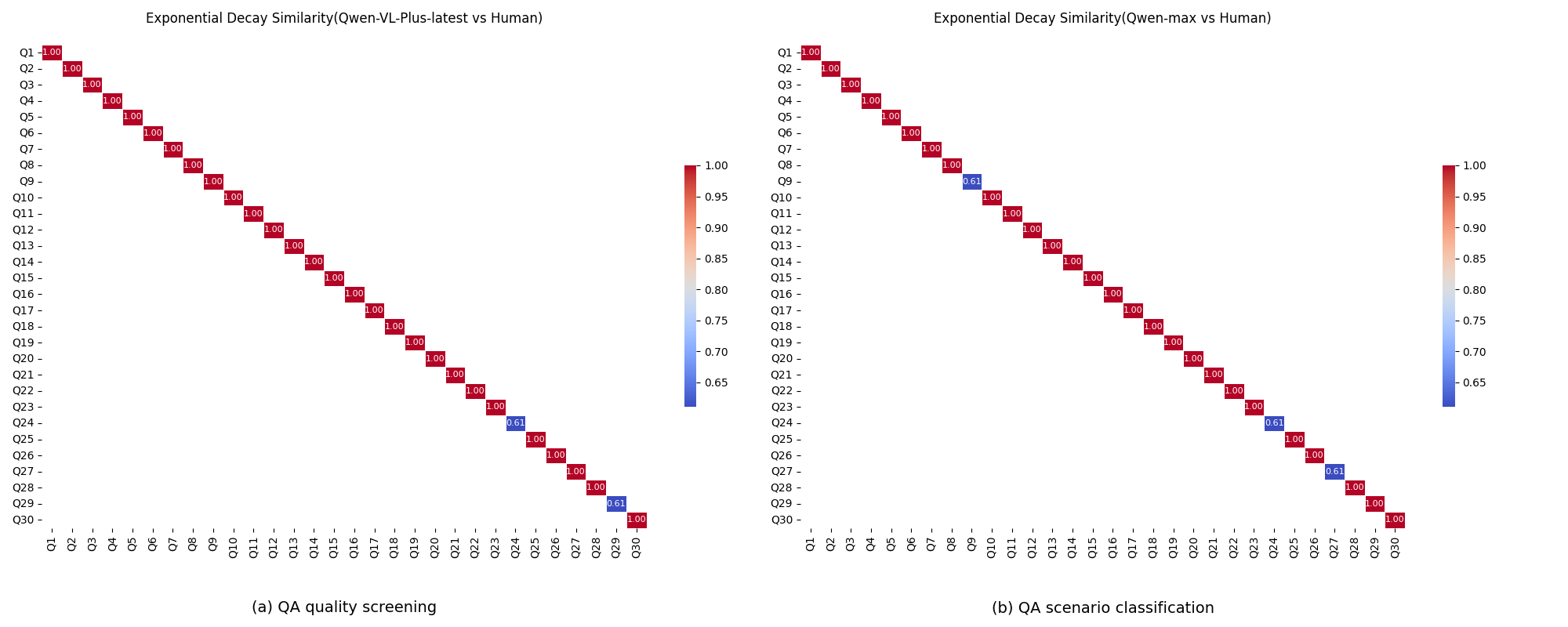}
    \caption{The graphs show the exponential decay similarity between LLMs and human evaluation. (a) depicts the similarity between LLM used for quality screening (Qwen-VL-Plus-latest) and human evaluation, where a value of 1 indicates complete similarity, and a value of 0.61 represents a non-ideal match. (b) illustrates the similarity between LLM used as a classifier (Qwen-max) and human evaluation, with the same similarity scale: 1 for complete similarity and 0.61 for a lower match. }
    \label{fig:fenlei-label}
\end{figure*}

The similarity heatmap for scene classification of the QA pairs is shown in the figure~\ref{fig:fenlei-label}, with (b) corresponding to this task. The results show that the similarity between Qwen-max's performance on this task and human performance is relatively high.
\\
The QA data underwent a three-stage quality control process to ensure accuracy and domain relevance:

\noindent\textbf{Automated Filtering Based on Multi-Dimensional Evaluation Metrics}\hspace{1em}We develop an automatic screening process driven by a set of prompts and scoring criteria, including image information density, semantic validity of the QAs, data diversity, objectivity, and computational complexity. Qwen-VL-Plus-latest is employed to score and filter the generated QA pairs. The similarity heatmap for quality filtering of the QA pairs is shown in (a) of Figure~\ref{fig:fenlei-label}. This stage focuses on removing incorrect answers, highly ambiguous data, and constructing a clean base dataset. Prompt examples used in this phase are shown in Table~\ref{promptquality1} and Table~\ref{promptquality2}.

\noindent\textbf{Manual Annotation}\hspace{1em}In the second stage, all QA pairs are manually annotated by six trained undergraduate students majoring in finance. These annotators have a strong background in financial knowledge and undergo a rigorous training process, including tests to ensure their competence in evaluating the correctness, domain specificity, verifiability of answers, completeness of visual elements, contextual alignment with financial scenarios, and logical coherence of question design. Only after passing the evaluation phase are annotators allowed to proceed with batch annotation, ensuring the overall consistency and accuracy of the labeled data.    

\noindent\textbf{Final Review by Financial Experts}\hspace{1em}The third stage involves a comprehensive review by three financial experts, each with over a decade of experience in finance (the same domain experts who contribute to prompt design). The review focuses on several critical aspects, including logical rigor, policy compliance, decisional determinacy, accuracy of terminology, and completeness of scenario coverage. Each QA pair had to be unanimously approved by all three experts to pass this stage. This final review guarantees that each QA item is well-designed, has a unique correct answer, and faithfully reflects real-world financial business logic. 

\subsection{Examples of Environmental Perturbation Simulations}





\begin{figure*}[ht]
    \centering
\includegraphics[width=1\textwidth]{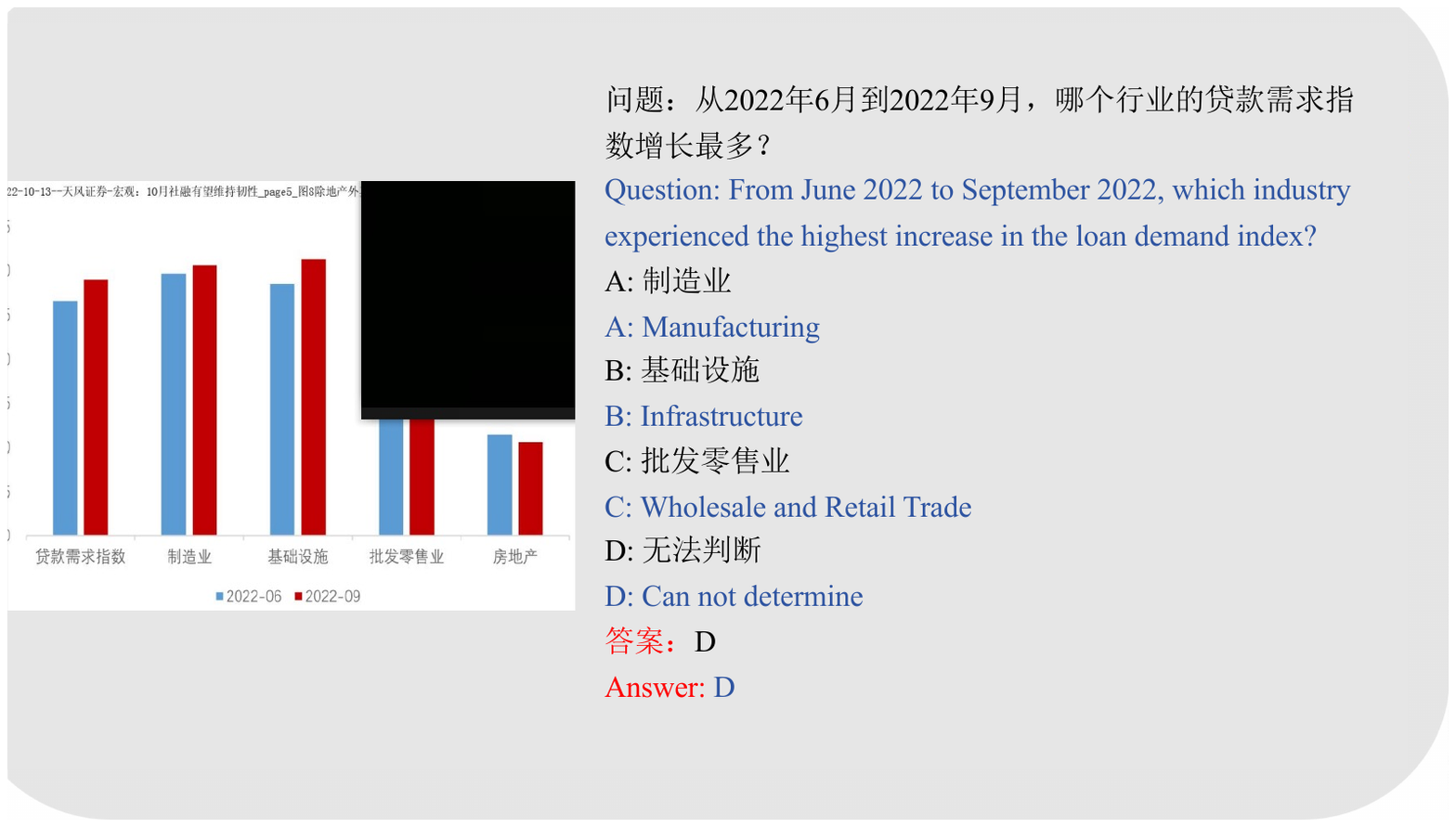}
    \caption{This is an example of Key Information Occlusion. The upper-right corner of the histogram contains loan demand index data for the wholesale and retail sector, as well as the real estate industry. However, this critical region is obscured, preventing the model from accurately extracting the necessary information. To generate correct responses under such conditions, the model must have the ability to detect occlusion or missing information.}
    \label{errorexamples:Occlusion}
\end{figure*}

In real-world financial applications, environmental perturbations often arise from practical factors such as document quality degradation, scanning errors, complex layout structures, or missing information. To simulate these common sources of noise and disturbance in financial document processing, VisFinEval incorporates a set of environmental perturbation tasks. These simulations provide a more realistic assessment of model performance under non-ideal visual conditions.
We categorize four representative types of perturbations as follows:

\noindent\textbf{Key Information Occlusion}
Critical regions of the image—such as data tables, axis labels, or seal texts—are partially obscured or blurred.
Figure~\ref{errorexamples:Occlusion} presents an example of this perturbation.

\begin{figure*}[ht]
    \centering
\includegraphics[width=1\textwidth]{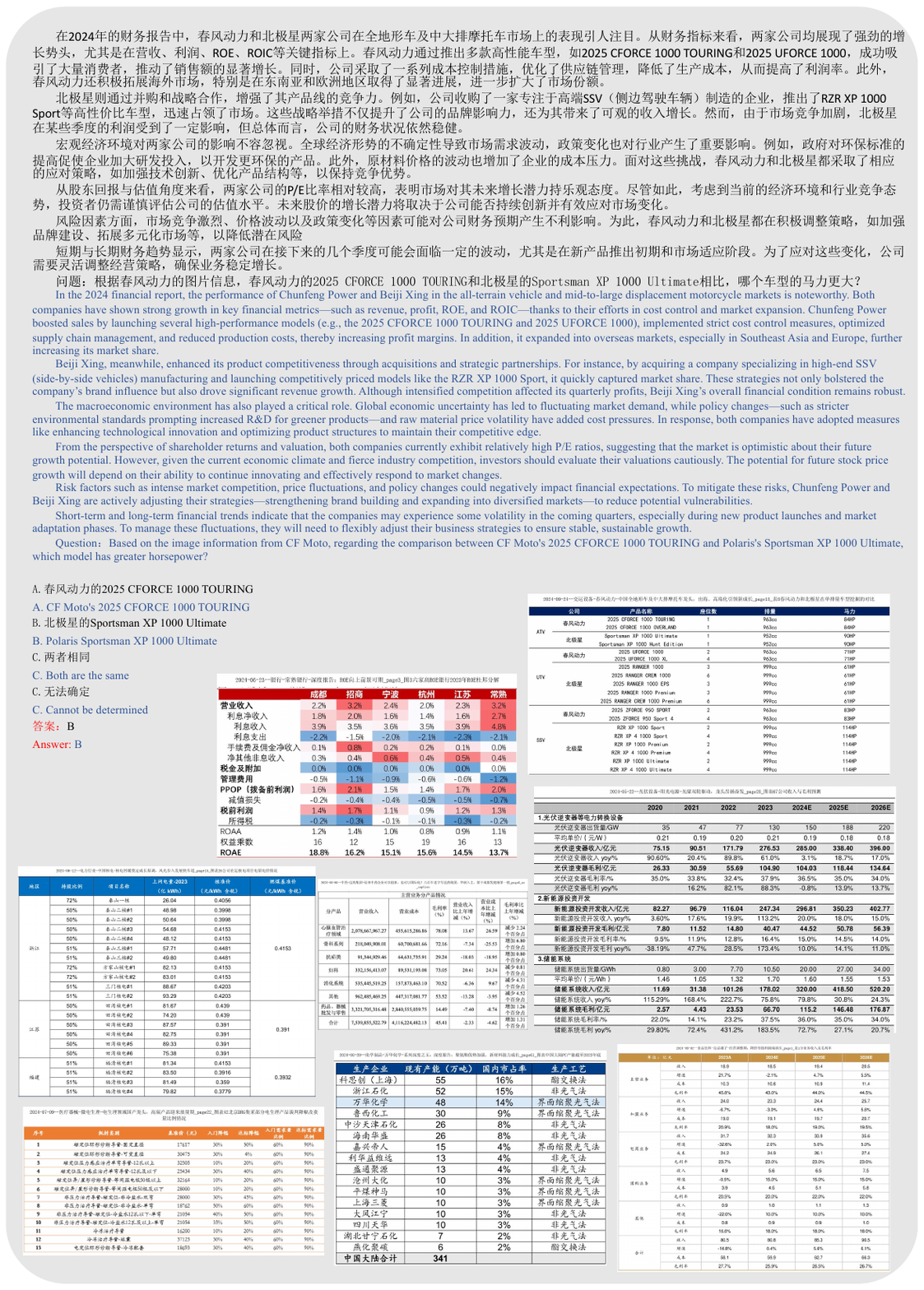}
    \caption{This is an example of Redundant Image Perturbation. As multiple images are used for perturbation in this case, the resolution has been compressed for display purposes. This example is for illustration only; the resolution of the original images remains unchanged in the actual dataset. The image contains multiple financial tables with similar formats, most of which are unrelated to the question. The model must possess effective vision-language alignment and contextual matching capabilities to accurately locate the one table that is relevant to the question in order to answer correctly.}
    \label{errorexamples:Image}
\end{figure*}

\noindent\textbf{Redundant Image Perturbation}
The original image is overlaid or mixed with visually similar but irrelevant graphical content, such as unrelated charts.
Figure~\ref{errorexamples:Image} illustrates this type of perturbation.

\begin{figure*}[ht]
    \centering
\includegraphics[width=1\textwidth]{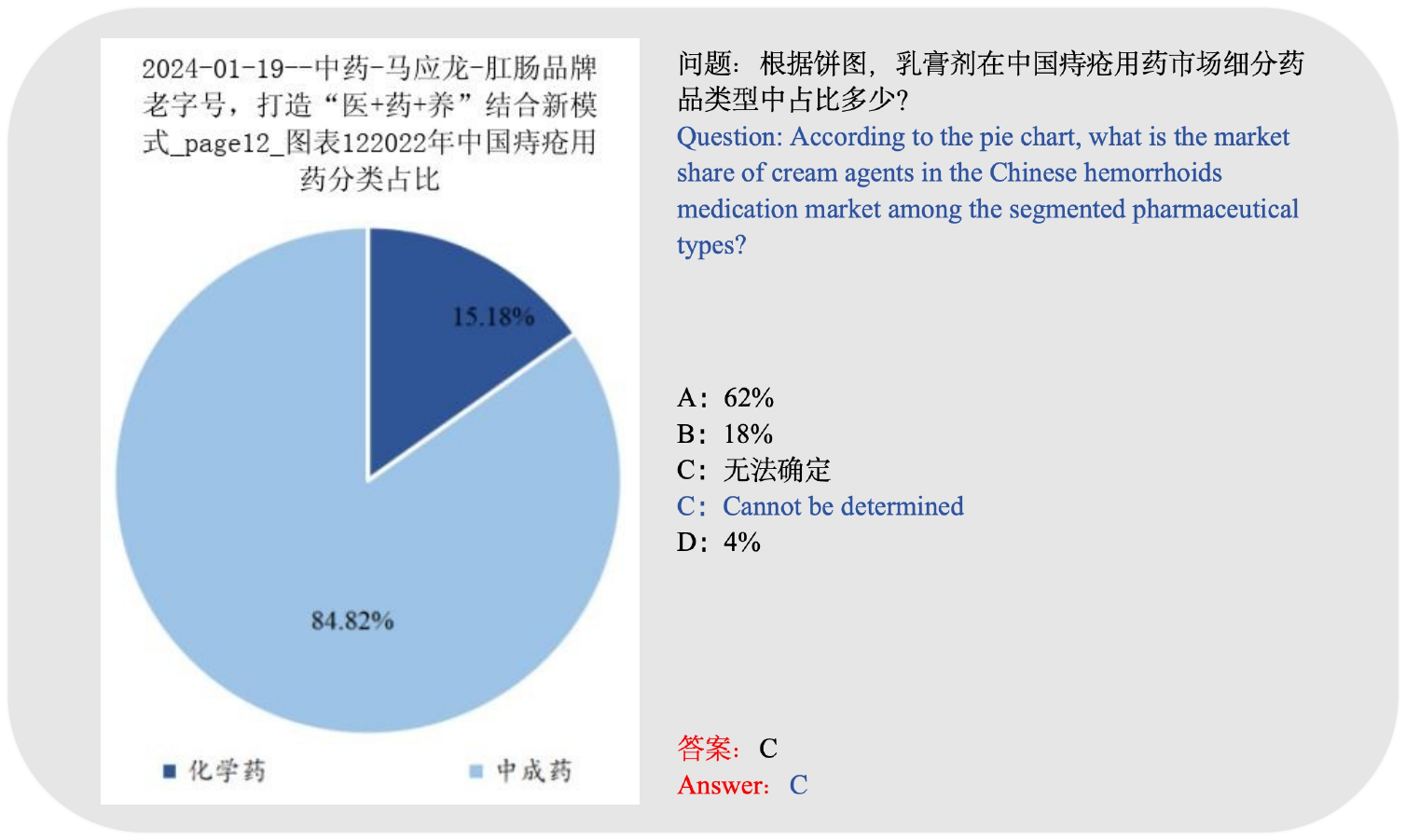}
    \caption{This is an example of Missing Relevant Visual Information. These types of questions often ask about content that doesn't exist in the image. Therefore, the model needs to determine whether relevant information is present in the image. This scenario simulates situations where business personnel might encounter incomplete customer information in real-world operations.}
    \label{errorexamples:HE}
\end{figure*}

\noindent\textbf{Missing Relevant Information}
The image lacks the information explicitly referenced in the question, simulating scenarios where relevant content is omitted due to formatting issues or cross-page references.
Figure~\ref{errorexamples:HE} provides an example of this case.

\begin{figure*}[ht]
    \centering
\includegraphics[width=1\textwidth]{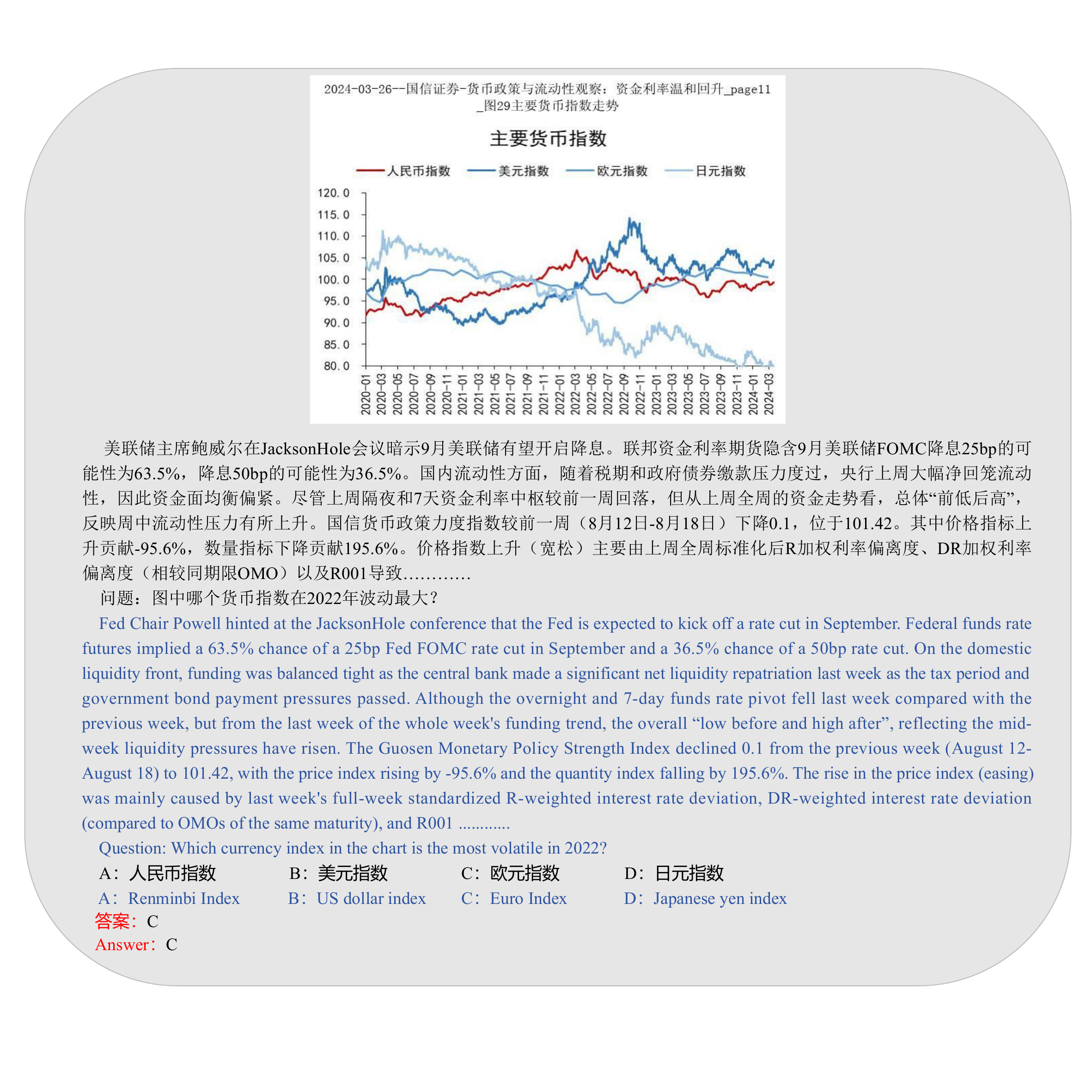}
    \caption{This is an example of Irrelevant Information Perturbation. These types of questions add text that is similar to the question during input but actually provides no assistance for answering. This introduces semantic noise, thereby increasing the difficulty for the model to answer and simulating the real-world scenario where business personnel need to conduct business operations under the interference of a large amount of irrelevant information.}
    \label{errorexamples:Content}
\end{figure*}

\noindent\textbf{Irrelevant Information Perturbation}
Unrelated content is added to the image without removing the original task-relevant information, resulting in semantic interference.
Figure~\ref{errorexamples:Content} demonstrates this perturbation type.

These four types of perturbations reflect common visual complexities in real financial scenarios and enable a systematic evaluation of multimodal large language models under environmentally degraded conditions.

\section{Details of MLLMs}
\label{modeloverview}
We list details of the MLLMs evaluated using VisFinEval in Table \ref{tab:model-overview}.

\subsection{Details of Evaluation Results}
\label{Details of Evaluation Results}
In our experiments, we conducted an in-depth comparative analysis of the performance of different models in specific financial business scenarios. The experimental results show that the models exhibited significant performance differentiation across scenarios of varying difficulty levels. Notably, the Qwen-VL-max ranked first across all three difficulty levels, demonstrating outstanding performance and strong adaptability.

\noindent\textbf{Financial Knowledge and Data Analysis} The comparative results of different models under the Financial Knowledge and Data Analysis scenario are detailed in Table \ref{tab:FKDA}, Qwen-VL-max secured first place in multiple tasks, ultimately achieving the top accuracy of 85.8. It is noteworthy that Moonshot-V1-32k-vision-preview far outperformed other models in the FSR(Financial Seal Recognition) task with the accuracy of 98.0, only 2 points behind human experts. 

\noindent\textbf{Financial Analysis and Decision Support} Table \ref{tab:FADS} shows how each model performed in the Financial Analysis and Decision Support scenario, InternVL3-78B, which ranked first in the FSA (Financial Scenario Analysis) task, secured the second position overall with the accuracy of 76.4. This surpassed Qwen-VL-max-latest, which performed excellently in the IA (Investment Analysis) and FMASA (Financial Market Anomaly Sentiment Analysis) tasks. Furthermore, InternVL3-78B was only 0.5 points behind the top-ranked Qwen-VL-max, making it the most powerful open-source model. 

\noindent\textbf{Financial Risk Control and Asset Optimization} Table \ref{tab:FRCAO} summarizes the performance of the models in the more complex scenario, there was a considerable gap between the models' performance and that of human experts. Although the Qwen-VL-max model ranked first with the accuracy of 59.1, it was still 23.7 points lower than human experts. This indicates that there is still significant room for improvement in model performance in complex financial business scenarios. 

Overall, while some models can approach human experts in specific scenarios, a significant disparity remains between models and human experts in tasks of higher complexity and difficulty.

\subsection{Examples for Error Analysis}
\label{Examples for Error analysis}
In this section, we explain in detail the meaning of six types of errors of MLLMs in financial business scenarios and provide examples and related error analysis.

\begin{figure*}[ht]
    \centering
\includegraphics[width=1\textwidth]{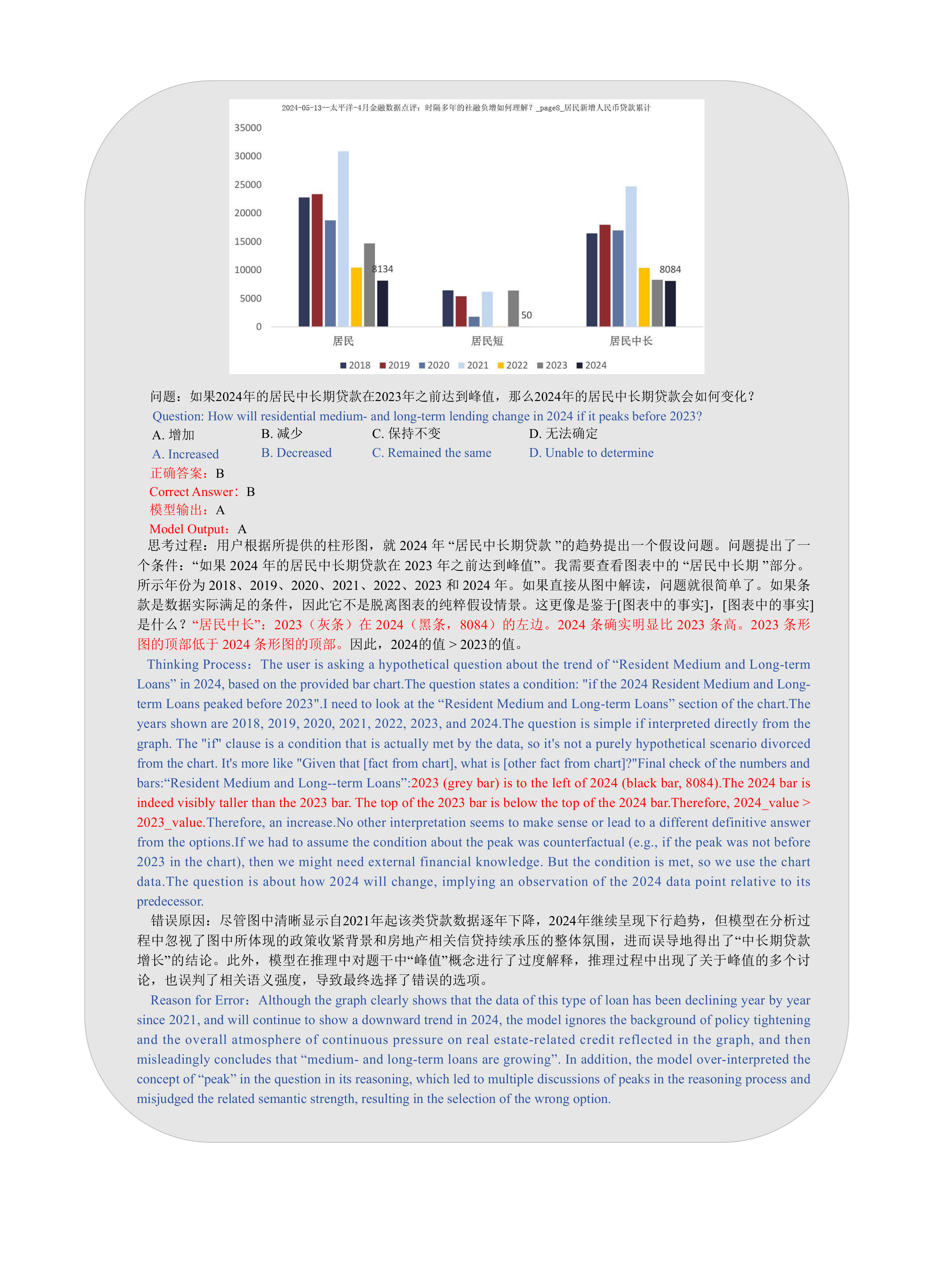}
    \caption{An example of Errors in misjudging Market Sentiment and Semantic Tendency Misjudgment. The large model over-interpreted the concept of “peak” in the question stem and ignored the overall atmosphere of continued pressure on real estate-related credit presented by the data trends in the chart. This reflects the fact that the model was too obsessed with a particular detail in the question and answer and lacked a big-picture view of the overall trend behind the data.}
    \label{errorexamples:MS}
\end{figure*}

\noindent\textbf{Market sentiment and semantic tendency misjudgment} The model is unable to accurately capture the front and back logic and key nodes of business operations, resulting in output results that are out of sync with real business processes or do not conform to real business thinking logic. An example of this can be seen in Figure \ref{errorexamples:MS}.

\begin{figure*}[ht]
    \centering
\includegraphics[width=0.9\textwidth]{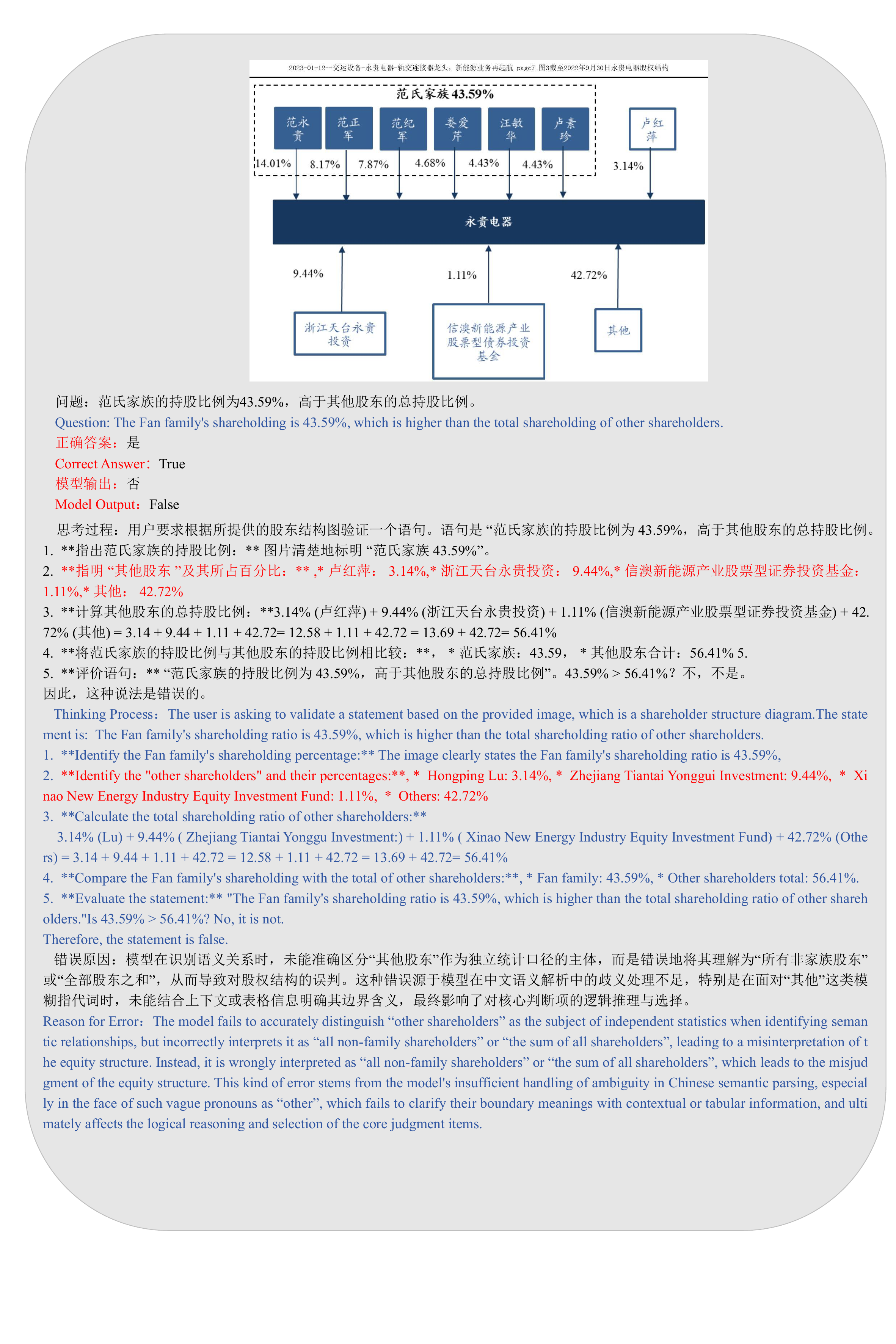}
    \caption{An example of Financial subject identification and causation confusion. Failure to accurately distinguish “other shareholders” as a subject of independent statistical caliber resulted in an incorrect answer. This reflects the model's lack of ability to handle ambiguity in Chinese semantic parsing.}
    \label{errorexamples:FS}
\end{figure*}

\noindent\textbf{Financial subject identification and causation confusion} The model has insufficient understanding of the definitions and calculation logic of specialized terms or financial indicators, which may easily lead to conceptual confusion or numerical calculation errors. An example of this can be seen in Figure \ref{errorexamples:FS}.

\begin{figure*}[ht]
    \centering
\includegraphics[width=0.9\textwidth]{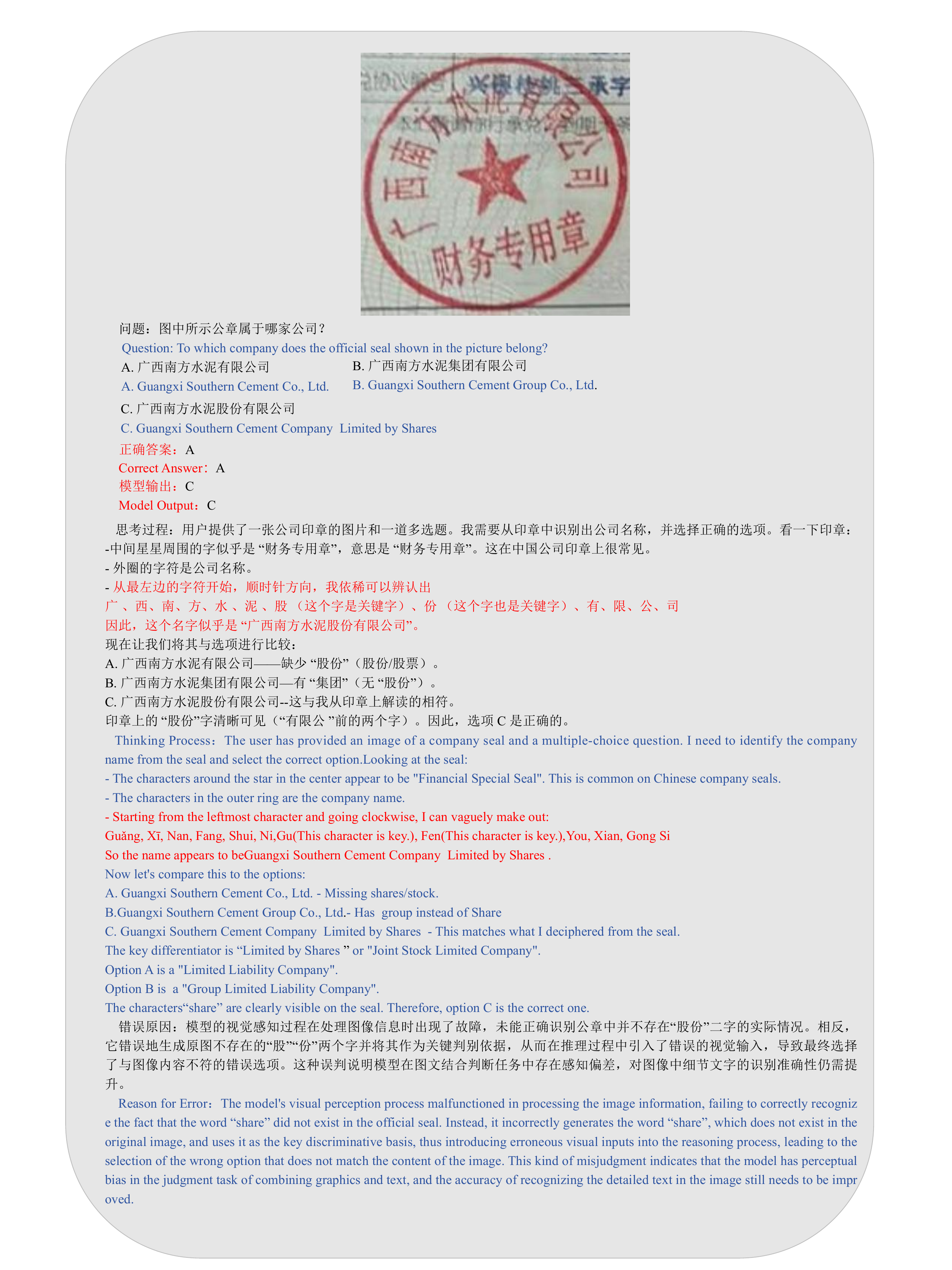}
    \caption{An example of Hallucination generation and irrational reasoning. Although the model has correctly recognized the words on the official seal, there is no information or words about “shares” on the official seal, so the model creates an illusion in the output process and uses the shares as the basis for judgment, resulting in an incorrect answer.}
    \label{errorexamples:HG}
\end{figure*}

\noindent\textbf{Hallucination generation and irrational reasoning} The model in the parsing of financial texts, public opinion, research reports, etc., misjudges emotional tendency and semantic intensity, ignoring or misinterpreting the policy signals and industry atmosphere. An example of this can be seen in Figure \ref{errorexamples:HG}.

\begin{figure*}[ht]
    \centering
\includegraphics[width=0.9\textwidth]{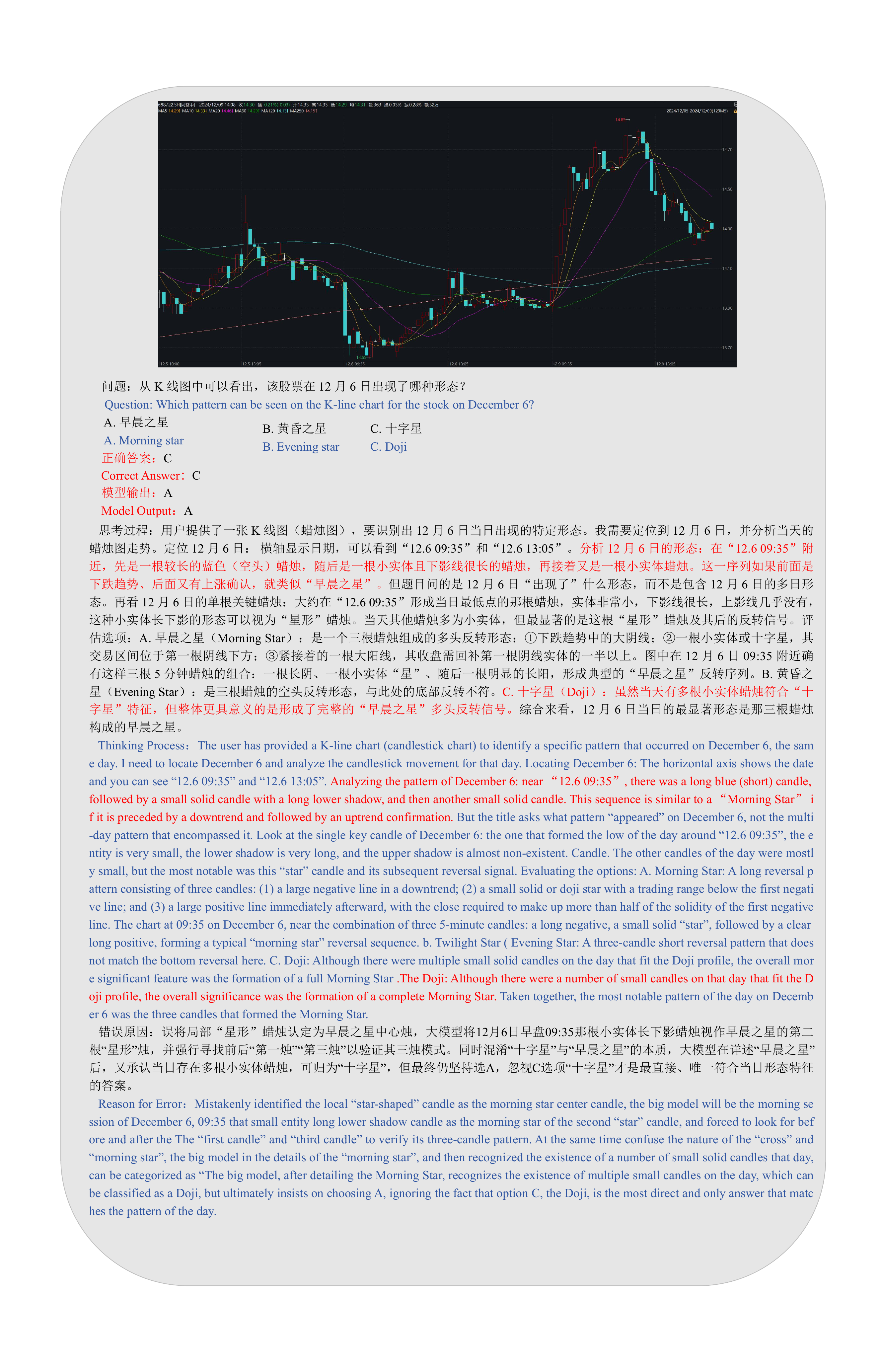}
    \caption{An example of Bias in the understanding of financial terms and indicators. Although the Model correctly extracted information such as the time point and the shape of the K-line chart, it confused the difference between different K-line patterns and eventually chose the wrong answer. This reflects the Big Model's lack of ability to discriminate between financial terminology and indicators.}
    \label{errorexamples:Bias}
\end{figure*}

\noindent\textbf{Bias in the understanding of financial terms and indicators} Difficulty for models to effectively integrate charts, tables and contextual textual information, leading to biased understanding of trends, data correlations, or visualization content. An example of this can be seen in Figure \ref{errorexamples:Bias}.

\begin{figure*}[ht]
    \centering
\includegraphics[width=1\textwidth]{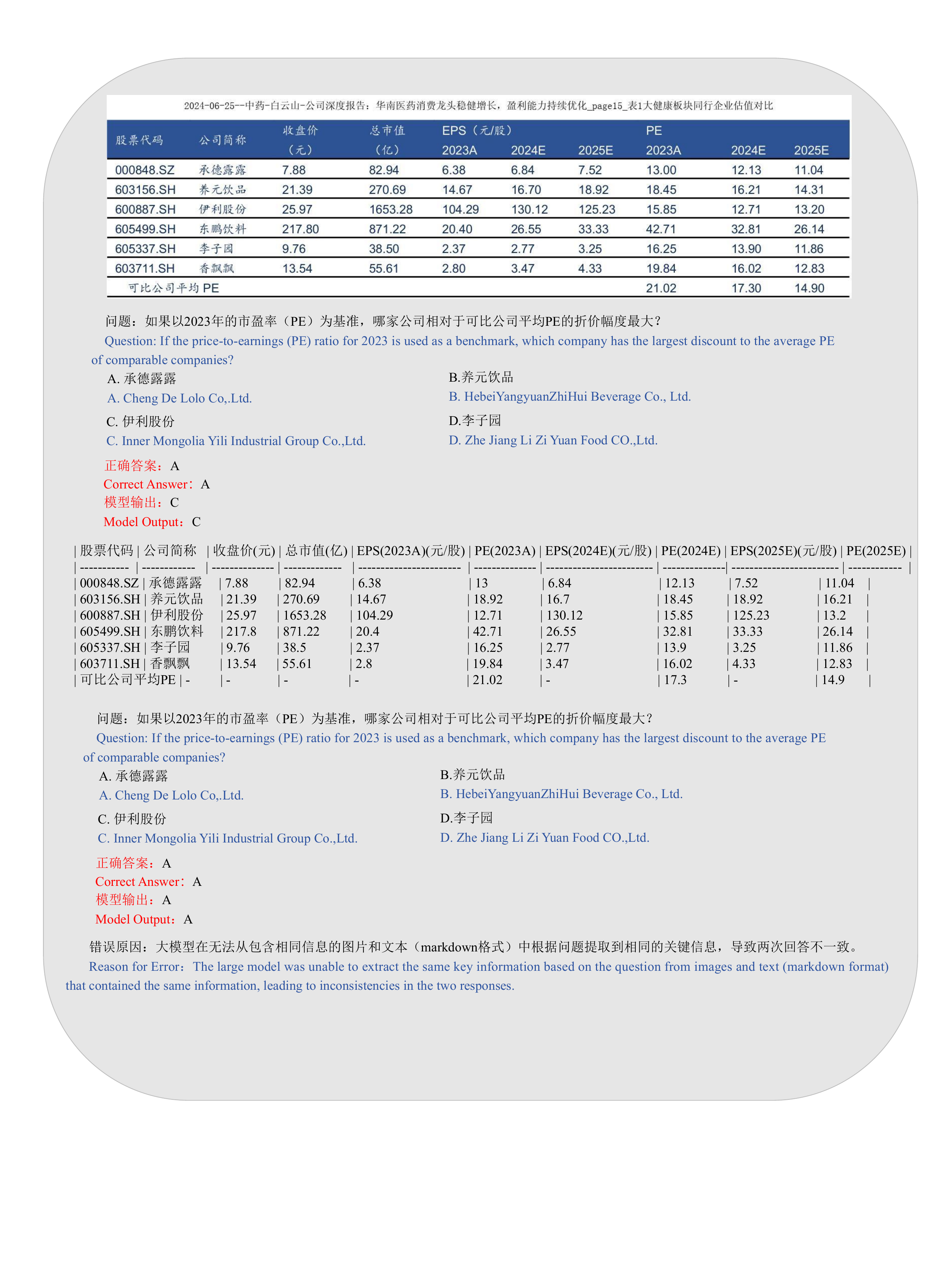}
    \caption{An example of Lack of cross-modal information alignment capability. This error type is generally a failure of the model to effectively combine image, chart, table, and text semantics, resulting in an incorrect trend determination or a numerical reading error. Although the large model found the key information and answered the question correctly in the text format of markdown, it answered the question incorrectly in the image format, containing the same information, which reflects the large model's lack of cross-modal information alignment ability.}
    \label{errorexamples:LAC}
\end{figure*}

\noindent\textbf{Lack of cross-modal information alignment capability} The model is unable to accurately discern the actual causal relationship between companies, industries, or indicators, and is prone to treating correlation as causation or confusing the roles of different subjects. An example of this can be seen in Figure \ref{errorexamples:LAC}.

\begin{figure*}[ht]
    \centering
\includegraphics[width=0.9\textwidth]{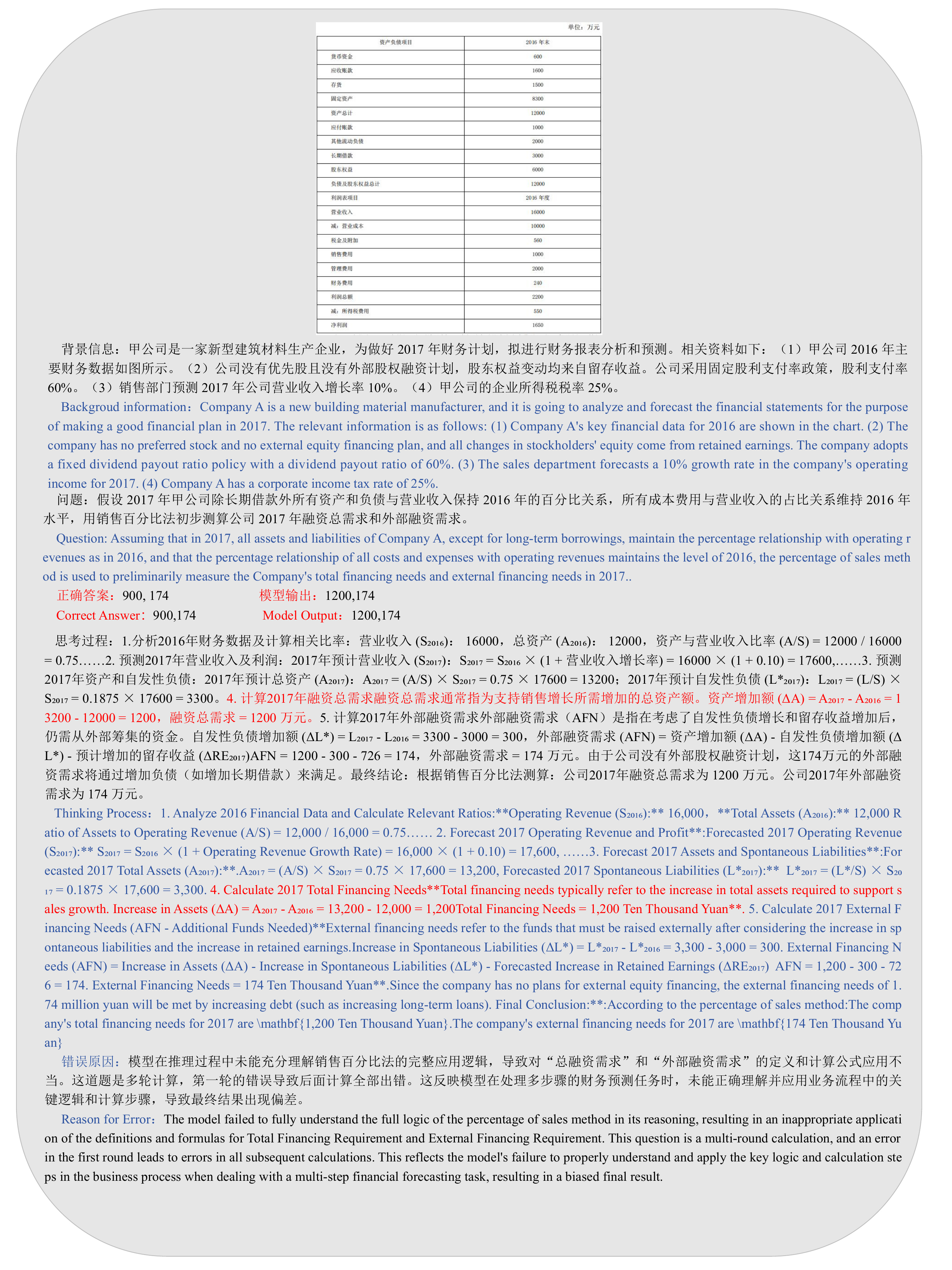}
    \caption{An example of Perceived barriers to financial business processes. This is a multi-round Q\&A, and since the big model has already answered the first round of questions incorrectly, resulting in incorrect answers to the subsequent questions based on this incorrect answer, only the first round of Q\&A is shown here as an example. Here, the model has successfully simulated the finance staff to identify the subjects and corresponding data to be calculated, but the model failed to fully understand the complete application logic of the Percentage of Sales method in the reasoning process, resulting in the improper application of the definitions and formulas of Total Financing Requirement and External Financing Requirement. This leads to improper application of the definitions and formulas of “total financing needs” and “external financing needs”. This error reflects the model's inadequate understanding of the dependencies between the steps and the logic of calculation when dealing with complex financial business processes. }
    \label{errorexamples:PB}
\end{figure*}

\noindent\textbf{Perceived barriers to financial business processes} The model may “make up” facts or make illogical inferences when information is missing or ambiguous, and the output does not match the reality. An example of this can be seen in Figure \ref{errorexamples:PB}.

\section{Prompts Used in This Study}
\label{prompt}
We provide representative prompt examples for question generation, image or question quality verification. Specifically, the prompt examples for question generation are shown in Table~\ref{promptquestion1}, Table~\ref{promptquestion2}, Table~\ref{promptquestion3}, Table~\ref{promptquestion4}. 
Prompt examples for quality verification are shown in Table~\ref{promptquality1}, Table~\ref{promptquality2}. 

Prompts for financial scenario classification are shown in Table~\ref{promptl1}, Table~\ref{promptl2}, Table~\ref{promptl3}.

\begin{table*}[ht]
\caption{Financial Scenario Data Distribution. This table systematically presents the distribution of financial scenarios across the three progressive depths of the VisFinEval dataset, as follows: financial literacy and data analytics covering 8,700 questions, financial analytics and decision support covering 4,650 questions, and financial risk control and asset optimization covering 2,498 questions, culminating in 15,848 questions that have been rigorously manually annotated. This structured presentation accurately assesses the benchmark's ability to simulate real-world financial complexity through increasing difficulty.}
\label{tab:visfineval_dist}
\vspace{10pt}
\resizebox{0.9\textwidth}{!}{
    \begin{tabular}{llr}
    \toprule[2pt]
    \textbf{Scenario Depth} & \textbf{Financial Scenario} & \textbf{Questions} \\
    \midrule[1pt]
    \begin{tabular}[t]{@{}l@{}}\raggedright\textbf{Financial Knowledge and Data Analysis}\end{tabular} 
    & Financial Data Statistics & 3655 \\ 
    & Candlestick Chart Analysis & 1124 \\
    & Financial Indicator Assessment & 1160 \\ 
    & Financial Entity Relationships Interpretation & 919 \\
    & Stock Selection Strategies Backtesting  & 719 \\
    & Financial Information Extraction & 924 \\
    & Financial Seal Recognition & 199 \\
    & All & \textbf{8700} \\
    \midrule[1pt]
    \begin{tabular}[t]{@{}l@{}}\raggedright\textbf{Financial Analysis and Decision Support}\end{tabular} 
    & Financial Scenario Analysis & 2040 \\ 
    & Industry Analysis and Inference & 1361 \\
    & Investment Analysis & 933 \\
    & Financial Market Sentiment Analysis & 316 \\
    & All & \textbf{4650} \\
    \midrule[1pt]
    \begin{tabular}[t]{@{}l@{}}\raggedright\textbf{Financial Risk Control and Asset Optimization}\end{tabular} 
    & Financial Strategy Optimization & 111 \\
    & Financial Risk and Policy Analysis & 181 \\
    & Financial Data Reasoning and Interpretation & 1839 \\
    & Asset Allocation Analysis & 367 \\
    & All & \textbf{2498} \\
    \midrule[1pt]
    \textbf{VisFinEval} & All& \textbf{15848} \\
    \bottomrule[2pt]
    \end{tabular}
}
\centering
\end{table*}

\begin{table*}[ht]
\caption{Evaluation Results of Financial Knowledge and Data Analysis. This table presents comparative evaluation results of various LLMs in Financial Knowledge and Data Analysis scenario. “Human” refers to the scores of human experts in the test, and the last column shows the average scores for each respective model.}
\label{tab:FKDA}
\vspace{10pt}
\resizebox{0.9\textwidth}{!}{
    \begin{tabular}{lcccccccccc}
    \toprule[2pt]
     & & & \multicolumn{7}{c}{\textbf{Financial Knowledge and Data Analysis}} \\
    \cmidrule(l){4-10}
    \textbf{Model} & \textbf{Size} & \textbf{Limit} &\textbf{FDS} & \textbf{CCA} & \textbf{FIA} & \textbf{FERI} &\textbf{SSSB} & \textbf{FIE} & \textbf{FSR} & \textbf{WA}\\
    \midrule[1pt]

    Qwen-VL-max & Unknown & / & \textbf{78.8} & \textbf{90.5} & \textbf{87.4} & \textbf{89.2} & \textbf{86.2} & 90.6 & 77.9 & \textbf{85.8} \\
    Qwen-VL-max-latest & Unknown & / & 76.0 & 84.5 & 86.1 & 87.1 & 79.3 & 88.6 & 84.4 & 83.7 \\
    InternVL3-78B & 78B & / & 71.2 & 83.5 & 71.4 & 86.7 & 79.5 & 87.8 & 87.4 & 81.1 \\
    Doubao-1.5-vision-pro-32k & Unknown & / & 75.6 & 79.0 & 84.2 & 85.5 & 76.8 & \textbf{91.7} & 74.4 & 81.0 \\
    InternVL2.5-78B & 78B & / & 73.3 & 77.9 & 72.3 & 84.2 & 84.0 & 88.4 & 82.9 & 80.4 \\
    Qwen2.5-VL-72B & 72B & / & 75.9 & 77.0 & 72.8 & 85.4 & 81.5 & 88.3 & 80.4 & 80.2 \\
    GPT-4o-2024-11-20  & Unknown & / & 72.0 & 76.8 & 74.9 & 81.7 & 71.8 & 83.8 & 83.9 & 77.8 \\
    Step-1o-vision-32k & Unknown & / & 48.9 & 78.4 & 80.2 & 84.1 & 75.3 & 88.2 & \textbf{98.0} & 79.0 \\
    Moonshot-V1-32k-vision-preview & Unknown & / & 56.2 & 82.8 & 73.4 & 80.5 & 73.9 & 87.6 & 68.3 & 74.7 \\
    Qwen2.5-VL-7B & 7B & / & 71.4 & 75.9 & 69.2 & 80.9 & 74.0 & 85.5 & 69.9& 75.3 \\
    InternVL3-8B & 8B & / & 68.2 & 78.0 & 62.8 & 87.0 & 74.1 & 84.0 & 77.4 & 75.9 \\
    Gemini-2.5-pro-exp-03-25 & Unknown & / & 73.6 & 76.7 & 72.6 & 81.0 & 73.0 & 89.4 & 87.4 & 79.1 \\
    Claude-3-7-Sonnet-20250219 & Unknown & / & 70.5 & 73.4 & 80.3 & 71.1 & 77.5 & 83.2 & 34.7  & 70.1 \\
    Qwen2.5-VL-3B & 3B & / & 69.5 & 81.1 & 65.9 & 76.6 & 73.6 & 83.4 & 72.4 & 74.6 \\
    MiniCPM-V-2.6 & 8B & / & 61.3 & 83.5 & 56.9 & 76.7 & 75.2 & 73.4 & 80.9 & 72.5 \\
    Llama-3.2-11B-Vision-Instruct & 11B & / & 56.9 & 40.8 & 59.3 & 63.9 & 62.9 & 73.1 & 70.4 & 61.0 \\
    Molmo-7B-D-0924 & 7B & / & 60.1 & 74.8 & 54.5 & 62.2 & 59.1 & 60.5 & 42.2 & 59.1 \\
    \midrule 
    GLM-4v-Plus-20250111 & Unknown & Multi-image Limit & 73.8 & 86.6 & 87.9 & 87.5 & 81.2 & 89.3 & 72.7 & 82.7 \\
    LLaVA-NEXT-34B & 34B & Context Window Limit & 55.3 & 79.8 & 92.3 & 63.2 & 87.8 & 55.0 & 58.8 & 70.3 \\
    LLaVA-v1.6-Mistral-7B & 7B & Context Window Limit & 54.6 & 73.4 & 65.9 & 62.1 & 47.4 & 47.0 & 62.3 & 58.9 \\
    LLaVA-NEXT-13B & 13B & Context Window Limit & 50.2 & 64.8 & 43.9 & 57.2 & 62.5 & 50.2 & 38.7 & 52.5 \\
    \midrule 
    Human & / & / & 95.8 & 92.5 & 83.7 & 96.8 & 91.8 & 92.4 & 100 & 93.3 \\
    \bottomrule[2pt]
    \end{tabular}
}
\centering
\end{table*}

\begin{table*}[ht]
\caption{Evaluation Results of Financial Analysis and Decision Support. This table presents comparative evaluation results of various LLMs in Financial Analysis and Decision Support scenario. “Human” refers to the scores of human experts in the test, and the last column shows the average scores for each respective model.}
\label{tab:FADS}
\vspace{10pt}
\resizebox{0.9\textwidth}{!}{
    \begin{tabular}{lccc@{\hspace{3.2em}}c@{\hspace{3.2em}}c@{\hspace{2em}}cc}
    \toprule[2pt]
     & & & \multicolumn{4}{c}{\textbf{Financial Analysis and Decision Support}} \\
    \cmidrule(l){4-7}
    \textbf{Model} & \textbf{Size} & \textbf{Limit} &\textbf{FSA} & \textbf{IAI} & \textbf{IA} & \textbf{FMASA} & \textbf{WA}\\
    \midrule[1pt]

    Qwen-VL-max & Unknown & / & 65.3 & \textbf{83.1} & 82.3 & 76.8 & \textbf{76.9} \\
    Qwen-VL-max-latest & Unknown & / & 59.6 & 82.6 & \textbf{82.8} & \textbf{79.3} & 76.1 \\
    InternVL3-78B & 78B & / & \textbf{64.3} & 82.1 & 80.4 & 78.7 & 76.4 \\
    Doubao-1.5-vision-pro-32k & Unknown & / & 56.7 & 80.2 & 79.8 & 77.3 & 73.5 \\
    InternVL2.5-78B & 78B & / & 63.3 & 81.5 & 80.1 & 75.2 & 75.0 \\
    Qwen2.5-VL-72B & 72B & / & 57.4 & 82.4 & 80.3 & 74.5 & 73.6 \\
    GPT-4o-2024-11-20  & Unknown & / & 61.9 & 77.9 & 78.5 & 73.2 & 72.9 \\
    Step-1o-vision-32k & Unknown & / & 40.3 & 78.8 & 78.6 & 76.1 & 68.4 \\
    Moonshot-V1-32k-vision-preview & Unknown & / & 61.9 & 77.7 & 77.0 & 72.3 & 72.2 \\
    Qwen2.5-VL-7B & 7B & / & 53.4 & 79.7 & 76.5 & 70.7 & 70.1 \\
    InternVL3-8B & 8B  & / & 56.5 & 76.1 & 76.8 & 71.7 & 70.3 \\
    Gemini-2.5-pro-exp-03-25 & Unknown & / & 53.2 & 72.4 & 70.8 & 75.5 & 68.0 \\
    Claude-3-7-Sonnet-20250219 & Unknown & / & 48.0 & 76.1 & 75.5 & 64.0 & 65.9 \\
    Qwen2.5-VL-3B & 3B & / & 50.0 & 75.4 & 74.7 & 66.6 & 66.7 \\
    MiniCPM-V-2.6 & 8B & / & 48.3 & 69.7 & 70.7 & 69.1 & 64.5 \\
    Llama-3.2-11B-Vision-Instruct & 11B & / & 45.3 & 69.7 & 67.1 & 63.4 & 61.4 \\
    Molmo-7B-D-0924 & 7B & / & 39.7 & 64.4 & 62.8 & 63.4 & 57.5 \\
    \midrule 
    GLM-4v-Plus-20250111 & Unknown & Multi-image Limit & 56.5 & 78.1 & 74.9 & 74.6 & 71.0 \\
    LLaVA-NEXT-34B & 34B & Context Window Limit & 54.3 & 88.2 & 88.1 & 66.9 & 74.4 \\
    LLaVA-v1.6-Mistral-7B & 7B & Context Window Limit & 42.3 & 58.3 & 56.4 & 63.7 & 55.2 \\
    LLaVA-NEXT-13B & 13B & Context Window Limit & 34.7 & 59.2 & 59.0 & 52.9 & 51.4 \\
    \midrule 
    Human & / & / & 88.8 & 90.6 & 87.1 & 85.3 & 88.0 \\
    \bottomrule[2pt]
    \end{tabular}
}
\centering
\end{table*}

\begin{table*}[ht]
\caption{Evaluation Results of Financial Risk Control and Asset Optimization. This table presents comparative evaluation results of various LLMs in Financial Risk Control and Asset Optimization. “Human” refers to the scores of human experts in the test, and the last column shows the average scores for each respective model.}
\label{tab:FRCAO}
\vspace{10pt}
\resizebox{0.9\textwidth}{!}{
    \begin{tabular}{lccc@{\hspace{3.6em}}c@{\hspace{3.6em}}c@{\hspace{2em}}cc}
    \toprule[2pt]
     & & & \multicolumn{4}{c}{\textbf{Financial Risk Control and Asset Optimization}} \\
    \cmidrule(l){4-7}
    \textbf{Model} & \textbf{Size} & \textbf{Limit} &\textbf{FSO} & \textbf{FRPA} & \textbf{FDRI} & \textbf{AAA} & \textbf{WA}\\
    \midrule[1pt]

    Qwen-VL-max & Unknown & / & \textbf{49.1} & \textbf{58.2} & \textbf{58.2} & 71.0 & \textbf{59.1} \\
    Qwen-VL-max-lastest & Unknown & / & 44.0 & 52.2 & 48.9 & 71.8 & 54.2 \\
    InternVL3-78B & 78B & / & \textbf{49.1} & 52.8 & 46.6 & 66.5 & 53.7 \\
    Doubao-1.5-vision-pro-32k & Unknown & / & 30.0 & 54.5 & 54.5 & \textbf{75.6} & 53.7 \\
    InternVL2.5-78B & 78B & / & 41 & 53.1 & 47.6 & 68.4 & 52.5 \\
    Qwen2.5-VL-72B & 72B & / & 41.4 & 53.4 & 42.6 & 71.9 & 52.3 \\
    GPT-4o-2024-11-20  & Unknown & / & 41 & 40.5 & 41.6 & 67.9 & 47.7 \\
    Step-1o-vision-32k & Unknown & / & 39.2 & 45.2 & 49 & 65.8 & 49.8 \\
    Moonshot-V1-32k-vision-preview & Unknown & / & 39.2 & 55.8 & 53.6 & 64.0 & 53.1\\
    Qwen2.5-VL-7B & 7B & / & 37.2 & 37.6 & 35.4 & 63.2 & 43.4 \\
    InternVL3-8B & 8B & / & 29.7 & 46.2 & 36.8 & 55.3 & 42.0 \\
    Gemini-2.5-pro-exp-03-25 & Unknown & / & 28.4 & 38 & 41.5 & 37.7 & 36.4 \\
    Claude-3-7-Sonnet-20250219 & Unknown & / & 26.8 & 50.3 & 48.6 & 64.4 & 47.5 \\
    Qwen2.5-VL-3B & 3B & / & 22.9 & 34.8 & 35.9 & 53.8 & 36.9 \\
    MiniCPM-V-2.6 & 8B & / & 20.6 & 35.5 & 26.8 & 52.7 & 33.9 \\
    Llama-3.2-11B-Vision-Instruct & 11B & / & 18.0 & 22.1 & 19.9 & 31.1 & 22.8 \\
    Molmo-7B-D-0924 & 7B & / & 23.4 & 31.7 & 21.9 & 26.5 & 25.9\\
    \midrule 
    GLM-4v-Plus-20250111 & Unknown & Multi-image Limit & 45.1 & 54.1 & 45.3 & 73.2 & 54.4 \\
    LLaVA-NEXT-34B & 34B & Context Window Limit & 13.1 & 17.5 & 12.7 & 7.7 & 12.7 \\
    LLaVA-v1.6-Mistral-7B & 7B & Context Window Limit & 10.2 & 16.3 & 35.9 & 21.1 & 20.9 \\
    LLaVA-NEXT-13B & 13B & Context Window Limit & 14.7 & 10.8 & 15.8 & 31.1 & 18.1 \\
    \midrule 
    Human & / & / & 84.4 & 80.4 & 81.1 & 85.2 & 82.8 \\
    \bottomrule[2pt]
    \end{tabular}
}
\centering
\end{table*}

\begin{table*}[ht]
\centering
\caption{Models evaluated in this paper. The "Access" column shows whether we have full access to the model weights or we can only access through API. The “Version Date” column shows the release date of the corresponding version of the model we evaluated.}
\label{tab:model-overview}
\vspace{10pt}
\resizebox{0.9 \textwidth}{!}{
\begin{tabular}{llcccc}
    \toprule[2pt]
    \textbf{Category} & \textbf{Model} & \textbf{Creator} & \textbf{Parameter} & \textbf{Access} & \textbf{Version Date}\\
    \midrule[1pt]
    Close-Source & Qwen-VL-max-latest & Alibaba Cloud & Undisclosed & API & 2025.1 \\
    & Qwen-VL-max & Alibaba Cloud & Undisclosed & API & 2025.1 \\
    & Step-1o-vision-32k & StepStar & Undisclosed & API & 2025.1 \\
    & Gemini-2.5-pro-exp-03-25 & Google & Undisclosed & API & 2025.3\\
    & GPT-4o-2024-11-20 & OpenAI & Undisclosed & API & 2024.11\\
    & Moonshot-V1-32k-vision-preview & MoonshotAI & Undisclosed & API & 2025.1 \\
    & Claude-3-7-Sonnet-20250219 & Anthropic & Undisclosed & API & 2024.10\\
    & Doubao-1.5-vision-pro-32k & ByteDance & Undisclosed & API & 2025.1 \\
    & GLM-4v-Plus-20250111 & Zhipu.AI & Undisclosed & API & 2025.1\\
    \midrule[1pt]    
    Open-Source & Qwen2.5-VL-3B & Alibaba Cloud & 3B & Weights & 2025.1\\
    & Qwen2.5-VL-7B & Alibaba Cloud & 7B & Weights & 2025.1 \\
    & Qwen2.5-VL-72B & Alibaba Cloud & 72B & Weights & 2025.1\\
    & MiniCPM-V-2.6 & OpenBMB & 8B & Weights & 2025.1 \\
    & InternVL3-8B & Shanghai AI Lab & 8B & Weights & 2025.4\\
    & InternVL2.5-78B & Shanghai AI Lab & 78B & Weights & 2024.12 \\
    & InternVL3-78B & Shanghai AI Lab& 78B & Weights & 2025.4\\
    & Molmo-7B-D-0924 & Allen Institute for AI & 7B & Weights & 2024.9 \\
    & Llama-3.2-11B-Vision-Instruct & Meta AI & 11B & Weights & 2024.9\\
    & LLaVA-v1.6-Mistral-7B & Liu et.al & 7B & Weights & 2024.5 \\
    & LLaVA-NeXT-13B & LLaVA-VL & 13B & Weights & 2024.1 \\
    & LLLaVA-NeXT-34B & LLaVA-VL & 34B & Weights & 2024.1 \\
    \bottomrule[2pt]
\end{tabular}}

\end{table*}

\label{sec:appendix}


\begin{table*}[htbp]
\centering
\caption{Prompt Template for Constructing Four-Option Multiple-Choice Questions Based on Line Charts (Chinese and English Versions)}
\label{promptquestion1}
\vspace{2mm}
\begin{subtable}[t]{0.95\textwidth}
\centering
\begin{CJK}{UTF8}{gbsn}
\begin{tabular}{@{}p{13cm}@{}}
\toprule
你是一名金融分析师，请根据提供的折线图，生成三道四选题。\\
题目应基于折线图中的数据趋势、关键点或特征。\\
要求：\\
\hspace*{1em}1. 每道问题必须清晰明确，选项应具有区分度。\\
\hspace*{1em}2. 每道题的选项 A、B、C、D 应涵盖不同的可能性，避免过于简单或明显。\\
\hspace*{1em}3. 每道题的答案必须是 A、B、C 或 D 中的一个。\\
\hspace*{1em}4. 每道题的问题长度不少于 10 字。\\
\hspace*{1em}5. 三道题目必须完全不同，且每道题需要编号为 1、2、3。\\
\hspace*{1em}6. **只输出 JSON 格式的内容，不要包含任何额外的描述性文本。**\\
你可以参考的示例:\\
\{random\_few\_shots\} \\
输出格式为：\\
\texttt{[} \\
\hspace*{1em} \{"id": "1", "q": "问题1", "A": "选项A", "B": "选项B", "C": "选项C", "Answer": "正确答案"\},\\
\hspace*{1em} \{"id": "2", "q": "问题2", "A": "选项A", "B": "选项B", "C": "选项C", "Answer": "正确答案"\},\\
\hspace*{1em} \{"id": "3", "q": "问题3", "A": "选项A", "B": "选项B", "C": "选项C", "Answer": "正确答案"\}\\
\texttt{]} \\
\bottomrule
\end{tabular}
\end{CJK}
\caption{Chinese Version}
\end{subtable}

\vspace{1mm}

\begin{subtable}[t]{0.95\textwidth}
\centering
\begin{tabular}{@{}p{13cm}@{}}
\toprule
You are a financial analyst. Based on the provided line chart, generate three four-option multiple-choice questions.\\
The questions should be grounded in the data trends, key points, or features shown in the chart.\\
Requirements: \\
\hspace*{1em}1. Each question must be clearly stated, and the options should be meaningfully differentiated.\\
\hspace*{1em}2. Options A, B, C, and D for each question should represent distinct possibilities and avoid being overly obvious or simplistic.\\
\hspace*{1em}3. The answer to each question must be one of A, B, C, or D.\\
\hspace*{1em}4. Each question should be no fewer than 10 Chinese characters in length.\\
\hspace*{1em}5. All three questions must be entirely different, and each should be labeled as 1, 2, and 3.\\
\hspace*{1em}6. **Only output the content in JSON format. Do not include any additional descriptive text.**\\
You may refer to the following examples:\\
\{random\_few\_shots\} \\
Output format: \\
\texttt{[} \\
\hspace*{1em} \{"id": "1", "q": "Question 1", "A": "Option A", "B": "Option B", "C": "Option C", "Answer": "Correct Answer"\},\\
\hspace*{1em} \{"id": "2", "q": "Question 2", "A": "Option A", "B": "Option B", "C": "Option C", "Answer": "Correct Answer"\},\\
\hspace*{1em} \{"id": "3", "q": "Question 3", "A": "Option A", "B": "Option B", "C": "Option C", "Answer": "Correct Answer"\}\\
\texttt{]} \\
\bottomrule
\end{tabular}
\caption{English Version}
\end{subtable}
\end{table*}

\begin{table*}[htbp]
\centering
\caption{Prompt Template for Constructing Counterfactual Inference Questions Based on Histograms (Chinese and English Versions)}
\label{promptquestion2}
\vspace{1mm}
\begin{subtable}[t]{0.95\textwidth}
\centering
\begin{CJK}{UTF8}{gbsn}
\begin{tabular}{@{}p{13cm}@{}}
\toprule
你是一名资深数据分析师，请根据提供的直方图，生成三道反事实推断的单选题，一般形式就是如果某个不是事实的事情发生，会有什么结果。\\
题目应基于你对直方图的理解，参考直方图的分布、峰值、偏态、异常值等特征。要求：\\
\hspace*{1em}1. 每道问题必须使用中文语言，清晰明确，选项应具有区分度。\\
\hspace*{1em}2. 每道题的选项 A、B、C、D 应涵盖不同的可能性，避免过于简单或明显。\\
\hspace*{1em}3. 每道题的答案必须是 A、B、C 或 D 中的一个，不能是其他内容。\\
\hspace*{1em}4. 三道题目必须完全不同，且每道题需要编号为 1、2、3。\\
\hspace*{1em}5. **只输出 JSON 格式的内容，不要包含任何额外的描述性文本。**\\
你可以参考的示例:\\
\{random\_few\_shots\} \\
输出格式为：\\
\texttt{[} \\
\hspace*{1em} \{"id": "1", "q": "问题1", "A": "选项A", "B": "选项B", "C": "选项C", "D": "选项D", "Answer": "A/B/C/D"\},\\
\hspace*{1em} \{"id": "2", "q": "问题1", "A": "选项A", "B": "选项B", "C": "选项C", "D": "选项D", "Answer": "A/B/C/D"\},\\
\hspace*{1em} \{"id": "3", "q": "问题1", "A": "选项A", "B": "选项B", "C": "选项C", "D": "选项D", "Answer": "A/B/C/D"\}\\
\texttt{]} \\
\bottomrule
\end{tabular}
\end{CJK}
\caption{Chinese Version}
\end{subtable}

\vspace{1mm}

\begin{subtable}[t]{0.95\textwidth}
\centering
\begin{CJK}{UTF8}{gbsn}
\begin{tabular}{@{}p{13cm}@{}}
\toprule
You are a senior data analyst. Based on the provided histogram, generate three single-choice counterfactual inference questions. These questions should generally take the form: if something that did not actually happen were to occur, what would be the result?\\
Questions should be based on your understanding of the histogram, referencing features such as distribution, peaks, skewness, and outliers. Requirements: \\
\hspace*{1em}1. Each question must be written in Chinese, clearly stated, and the options should be distinguishable.\\
\hspace*{1em}2. Each question should have four options A, B, C, and D representing different possibilities. Avoid options that are too simple or obvious.\\
\hspace*{1em}3. The answer to each question must be one of A, B, C, or D and nothing else.\\
\hspace*{1em}4. The three questions must be completely different and should be numbered as 1, 2, and 3.\\
\hspace*{1em}5. **Only output the content in JSON format. Do not include any additional descriptive text.**\\
You may refer to the following examples:\\
\{random\_few\_shots\} \\
Output format: \\
\texttt{[} \\
\hspace*{1em} \{"id": "1", "q": "Question 1", "A": "Option A", "B": "Option B", "C": "Option C", "D": "Option D", "Answer": "A/B/C/D"\},\\
\hspace*{1em} \{"id": "2", "q": "Question 2", "A": "Option A", "B": "Option B", "C": "Option C", "D": "Option D", "Answer": "A/B/C/D"\},\\
\hspace*{1em} \{"id": "3", "q": "Question 3", "A": "Option A", "B": "Option B", "C": "Option C", "D": "Option D", "Answer": "A/B/C/D"\}\\
\texttt{]} \\
\bottomrule
\end{tabular}
\end{CJK}
\caption{English Version}
\end{subtable}
\end{table*}

\begin{table*}[htbp]
\centering
\caption{Prompt Template for Constructing Multi-turn Dialogue Tasks Based on Candlestick Charts (Chinese Versions)}
\label{promptquestion3}
\vspace{2mm}

\begin{subtable}[t]{0.95\textwidth}
\centering
\begin{CJK}{UTF8}{gbsn}
\begin{tabular}{@{}p{13cm}@{}}
\toprule
你是一名专业的金融分析师，擅长分析K线图。\\
现在给你的是不同的股票的几张含有不同参数的K线图，请根据提供的多张K线图生成三道专业金融题目，要求：\\
\hspace*{1em}1. 题目类型包括趋势分析、数据比较、计算题等，尽量是客观题，尽量丰富题型，保证正确答案是客观的。\\
\hspace*{1em}2. 每道题四个选项并标注正确答案，正确的答案只能有一个，即单选题。\\
\hspace*{1em}3. 必须基于所有图片内容，即每个题目都必须用到图对的所有图片的内容，每个题目前面可以用一两句话描述下图片和题目，最后再次重申：使用到所有图片的内容，题目尽量长一点。\\
你可以参考的示例:\\
\{random\_few\_shots\} \\
输出格式（每个问题一个 JSON 对象）：\\
\texttt{[} \\
\hspace*{1em} \{ \\
\hspace*{2em} "q": "问题描述",\\
\hspace*{2em} "A": "选项A",\\ 
\hspace*{2em} "B": "选项B",\\
\hspace*{2em} "C": "选项C",\\
\hspace*{2em} "D": "选项D",\\
\hspace*{2em} "Answer": "正确答案"\},\\
\hspace*{1em} \}, \\
\hspace*{1em} \{ \\
\hspace*{2em} "q": "问题描述",\\
\hspace*{2em} "A": "选项A",\\ 
\hspace*{2em} "B": "选项B",\\
\hspace*{2em} "C": "选项C",\\
\hspace*{2em} "D": "选项D",\\
\hspace*{2em} "Answer": "正确答案"\},\\
\hspace*{1em} \}, \\
\hspace*{1em} \{ \\
\hspace*{2em} "q": "问题描述",\\
\hspace*{2em} "A": "选项A",\\ 
\hspace*{2em} "B": "选项B",\\
\hspace*{2em} "C": "选项C",\\
\hspace*{2em} "D": "选项D",\\
\hspace*{2em} "Answer": "正确答案"\},\\
\hspace*{1em} \} \\
\texttt{]} \\
\bottomrule
\end{tabular}
\end{CJK}
\end{subtable}
\end{table*}

\begin{table*}[htbp]
\centering
\caption*{(continued)Prompt Template for Constructing Multi-turn Dialogue Tasks Based on Candlestick Charts (English Versions)} 
\vspace{2mm}

\begin{subtable}[t]{0.95\textwidth}
\centering
\begin{CJK}{UTF8}{gbsn}
\begin{tabular}{@{}p{13cm}@{}}
\toprule
You are a professional financial analyst skilled in analyzing candlestick charts. \\
You are provided with several candlestick charts of different stocks with varying parameters. Please generate three professional financial questions based on these charts, with the following requirements: \\
\hspace*{1em}1. Question types should include trend analysis, data comparison, calculation problems, etc. Prioritize objective questions with diverse formats, ensuring answers are fact-based. \\
\hspace*{1em}2. Each question must have four options with one clearly marked correct answer (single-choice format). \\
\hspace*{1em}3. All questions must incorporate content from every provided image. Each question may be preceded by 1-2 sentences describing the relevant chart elements. Remember: every question must utilize all images' content, and questions should be sufficiently detailed. \\
Reference examples: \\
\{random\_few\_shots\} \\
Output format (one JSON object per question): \\
\texttt{[} \\
\hspace*{1em} \{ \\
\hspace*{2em} "q": "Question description", \\
\hspace*{2em} "A": "Option A", \\ 
\hspace*{2em} "B": "Option B", \\
\hspace*{2em} "C": "Option C", \\
\hspace*{2em} "D": "Option D", \\
\hspace*{2em} "Answer": "Correct option"\},\\
\hspace*{1em} \}, \\
\hspace*{1em} \{ \\
\hspace*{2em} "q": "Question description", \\
\hspace*{2em} "A": "Option A", \\ 
\hspace*{2em} "B": "Option B", \\
\hspace*{2em} "C": "Option C", \\
\hspace*{2em} "D": "Option D", \\
\hspace*{2em} "Answer": "Correct option"\},\\
\hspace*{1em} \}, \\
\hspace*{1em} \{ \\
\hspace*{2em} "q": "Question description", \\
\hspace*{2em} "A": "Option A", \\ 
\hspace*{2em} "B": "Option B", \\
\hspace*{2em} "C": "Option C", \\
\hspace*{2em} "D": "Option D", \\
\hspace*{2em} "Answer": "Correct option"\},\\
\hspace*{1em} \} \\
\texttt{]} \\
\bottomrule
\end{tabular}
\end{CJK}
\end{subtable}
\end{table*}

\addtocounter{table}{-1}
\begin{table*}[htbp]
\centering
\caption{Prompt Template for Constructing True/False Judgment Tasks Based on Pie Charts (Chinese and English Versions)}
\label{promptquestion4}
\vspace{2mm}

\begin{subtable}[t]{0.95\textwidth}
\centering
\begin{CJK}{UTF8}{gbsn}
\begin{tabular}{@{}p{13cm}@{}}
\toprule
你作为一名专业的金融分析师，擅长分析饼图，请帮我根据我给你提供的饼图及图片的caption，给我出3道专业且带有难度的金融判断题目。\\
要求如下：\\
\hspace*{1em}1. 只给我输出我对应的格式信息，不要给我其他信息。\\
\hspace*{1em}2. 问题尽可能的多样化、复杂化，所有的问题请基于我的图片内容。\\
\hspace*{1em}3. 你最后的正确答案应该同时符合你的问题逻辑和图片内容。\\
\hspace*{1em}4. 请确保问题基于图片内容生成。\\
\hspace*{1em}5. **只输出 JSON 格式的内容，不要包含任何额外的描述性文本。**\\
你可以参考的示例:\\
\{random\_few\_shots\} \\
输出格式为：\\
\texttt{[} \\
\hspace*{1em} \{\{ "Question1":"","Answer":"True" \}\},\\
\hspace*{1em} \{\{ "Question2":"","Answer":"True" \}\},\\
\hspace*{1em} \{\{ "Question3":"","Answer":"True" \}\}\\
\texttt{]} \\
\bottomrule
\end{tabular}
\caption{Chinese Version}
\end{CJK}
\end{subtable}

\vspace{4mm}

\begin{subtable}[t]{0.95\textwidth}
\centering
\begin{tabular}{@{}p{13cm}@{}}
\toprule
As a professional financial analyst with expertise in interpreting pie charts, please generate 3 professional and challenging true/false financial questions based on the pie chart and its caption that I provide.\\
Requirements: \\
\hspace*{1em}1. Only return the formatted information I requested. Do not include any additional content.\\
\hspace*{1em}2. The questions should be as diverse and complex as possible, and must be based entirely on the content of the image.\\
\hspace*{1em}3. The correct answers must logically align with both the question structure and the image content.\\
\hspace*{1em}4. Please ensure that the questions are generated based on the image.\\
\hspace*{1em}5. **Only output content in JSON format. Do not include any descriptive or explanatory text.**\\
You may refer to the following examples:\\
\{random\_few\_shots\} \\
Output format: \\
\texttt{[} \\
\hspace*{1em} \{\{ "Question1":"","Answer":"True" \}\},\\
\hspace*{1em} \{\{ "Question2":"","Answer":"True" \}\},\\
\hspace*{1em} \{\{ "Question3":"","Answer":"True" \}\}\\
\texttt{]} \\
\bottomrule
\end{tabular}
\caption{English Version}
\end{subtable}

\end{table*}

\begin{table*}[htbp]
\centering
\caption{Prompt Template for Verifying the Content Quality of Images (Chinese and English Versions)}
\label{promptquality1}
\vspace{2mm}

\begin{subtable}[t]{0.95\textwidth}
\centering
\begin{CJK}{UTF8}{gbsn}
\begin{tabular}{@{}p{13cm}@{}}
\toprule
你是一个专业的图像分析助手。请根据以下标准筛选高质量的折线图：\\
\hspace*{1em}1. **数据多样性**：折线图中应展示至少2条不同的折线，每条折线代表一个独立的数据类别或指标，且数据变化趋势应具有一定的多样性（如上升、下降、波动等）。\\
\hspace*{1em}2. **数据清晰可辨**：折线图中的数据点、坐标轴、图例等应清晰可辨，避免模糊或难以解读的图表。\\
\hspace*{1em}3. **具备问答意义**：图表中的数据应能够产生有效的问答对，且问题应具有一定的计算意义或挑战性。确保问答可以基于这些数据进行推理、计算或者对比。\\
\hspace*{1em}4. **去除低质量图表**：如果折线图中只展示了一条折线，或者数据变化趋势过于简单（如单调上升或下降），则不符合要求。\\
'''\\
\hspace*{1em} messages=[\{"role": "user", "content": [\{"type": "image\_url", "image\_url": \{"url": f"data:image/jpeg;base64,\{img\_base\}"\}\}, \{"type": "text", "text": f
'''\\
以下是针对这张图片生成的三个问题：\{questions\_text\} \\
你是一名专业的金融分析师，擅长分析折线图。请根据问题的专业性和难度，从中选择一个最好的问题，返回问题的索引（从1开始）。\\
请只返回一个数字，例如：1、2 或 3。\\
\bottomrule
\end{tabular}
\caption{Chinese Version}
\end{CJK}
\end{subtable}

\vspace{4mm}

\begin{subtable}[t]{0.95\textwidth}
\centering
\begin{tabular}{@{}p{13cm}@{}}
\toprule
You are a professional image analysis assistant. Please filter high-quality line charts based on the following criteria: \\
\hspace*{1em}1. **Data Diversity**: The line chart should display at least two distinct lines, each representing an independent data category or metric. The trends should exhibit diversity (e.g., increase, decrease, fluctuation). \\
\hspace*{1em}2. **Clarity of Data**: Data points, axes, legends, and other elements in the chart should be clearly distinguishable, avoiding any blurry or unreadable visuals. \\
\hspace*{1em}3. **Question-Answer Relevance**: The chart should enable the generation of meaningful QA pairs. The questions should involve some degree of calculation or reasoning. Ensure that the data in the chart supports logical inference, computation, or comparison. \\
\hspace*{1em}4. **Exclude Low-Quality Charts**: If the chart contains only one line, or if the data trend is overly simplistic (e.g., strictly increasing or decreasing), it should be excluded. \\
'''\\
\hspace*{1em} messages=[\{"role": "user", "content": [\{"type": "image\_url", "image\_url": \{"url": f"data:image/jpeg;base64,\{img\_base\}"\}\}, \{"type": "text", "text": f
'''\\
Here are three questions generated based on this image: \{questions\_text\} \\
You are a professional financial analyst skilled in interpreting line charts. Based on the professionalism and difficulty of the questions, select the best one and return its index (starting from 1).\\
Only return a single number, such as: 1, 2, or 3.\\
\bottomrule
\end{tabular}
\caption{English Version}
\end{subtable}

\end{table*}

\begin{table*}[htbp]
\centering
\caption{Prompt Template for Validating the Consistency and Correctness of QA Pairs (Chinese and English Versions)}
\label{promptquality2}
\vspace{2mm}

\begin{subtable}[t]{0.95\textwidth}
\centering
\begin{CJK}{UTF8}{gbsn}
\begin{tabular}{@{}p{13cm}@{}}
\toprule
请验证以下问答对的质量：\\
\texttt{[问题]} \{question.get('query', '')\} \\
\texttt{[参考答案]} \{question.get('answer', '')\} \\
验证标准：\\
\hspace*{1em}1. 答案准确性（基于图表数据）\\
\hspace*{1em}2. 问题复杂度（需两步以上推理）\\
\hspace*{1em}3. 问题客观程度，需是客观题或者计算题\\
\hspace*{1em}4. 选项合理性（如为选择题）\\
验证结论格式：通过/不通过|理由\\
\bottomrule
\end{tabular}
\caption{Chinese Version}
\end{CJK}
\end{subtable}

\vspace{4mm}

\begin{subtable}[t]{0.95\textwidth}
\centering
\begin{tabular}{@{}p{13cm}@{}}
\toprule
Please validate the quality of the following QA pair: \\
\texttt{[Question]} \{question.get('query', '')\} \\
\texttt{[Reference Answer]} \{question.get('answer', '')\} \\
Validation Criteria: \\
\hspace*{1em}1. Answer accuracy (based on the chart data) \\
\hspace*{1em}2. Question complexity (requires more than two steps of reasoning) \\
\hspace*{1em}3. Objectivity of the question (must be objective or computational) \\
\hspace*{1em}4. Option quality (if multiple-choice) \\
Validation Output Format: Pass / Fail | Reason \\
\bottomrule
\end{tabular}
\caption{English Version}
\end{subtable}

\end{table*}

\begin{table*}[ht]
\centering
\begin{CJK}{UTF8}{gbsn}
\caption{Prompt Template for Classifying Scenarios in the Financial Knowledge and Data Analysis Category (Chinese Version)}
\label{promptl1}
\vspace{2mm}
\begin{tabular}{@{}p{13cm}@{}}
\toprule
请根据以下内容为问题进行分类：\{combined\_text\} \\
请将问题分类到以下7个金融场景之一：\\
\hspace*{1em}1. 股票K线解读，智能验印，金融信息识别，财务数据统计，金融实体关系解读，金融市场情绪洞察，金融情景分析\\
其中\\
\hspace*{1em}2. 股票K线解读：该场景通过解读股票K线图及相关技术指标（如MACD、RSI、成交量等），分析股价的历史走势、当前状态及未来趋势。\\
\hspace*{1em}3. 智能验印：该场景需要对金融或行政文档中的印章进行识别、验证与比对，以判断其真伪、归属及合规性。\\
\hspace*{1em}4. 金融信息识别：该场景关注金融、经济、投资领域中金融信息的识别和解读任务，识别其所表达的金融含义。\\
财务数据统计：该场景关注对具体财务或经济数据的整理、趋势分析和对比评估，例如地方政府债券发行量、资本项目差额、财政收支变动等。\\
\hspace*{1em}5. 金融实体关系解读：该场景聚焦于经济主体（如公司、政府、部门）之间的关系分析和经济影响链条解读，例如“财政扩张如何影响居民消费”或“资本流入对汇率的影响”。\\
\hspace*{1em}6. 金融市场情绪洞察：该场景侧重从投资者行为、舆情或市场表现中提取市场情绪趋势，例如通过新闻、评论、价格行为等数据推测市场预期。\\
\hspace*{1em}7. 金融情景分析：该场景包含假设性问题和反事实推理，例如“如果2022年债券发行没有增加，将可能发生什么？”此类问题需要基于对金融机制的理解推测可能后果。\\
请仅回答类别名称，不要解释。\\
\bottomrule
\end{tabular}
\end{CJK}
\end{table*}

\begin{table*}[ht]
\centering
\caption*{(continued) Prompt Template for Classifying Scenarios in the Financial Knowledge and Data Analysis Category (English Version)}
\vspace{2mm}
\begin{tabular}{@{}p{13cm}@{}}
\toprule
Please classify the question based on the following content: \{combined\_text\} \\
Assign the question to one of the following seven financial scenarios: \\
\hspace*{1em}1. Candlestick Chart Analysis, Intelligent Seal Recognition, Financial Information Extraction, Statistical Analysis of Financial Data, Interpretation of Financial Entity Relationships, Financial Market Sentiment Analysis, Financial Scenario Analysis \\
Descriptions: \\
\hspace*{1em}2. Candlestick Chart Analysis: This scenario involves interpreting candlestick charts and related technical indicators (e.g., MACD, RSI, trading volume) to analyze historical price trends, current states, and potential future movements.\\
\hspace*{1em}3. Intelligent Seal Recognition: This scenario requires identifying, verifying, and matching seals in financial or administrative documents to determine their authenticity, origin, and compliance.\\
\hspace*{1em}4. Financial Information Extraction: This scenario focuses on identifying and interpreting financial concepts and information in the fields of finance, economics, and investment.\\
Statistical Analysis of Financial Data: This scenario focuses on organizing, analyzing trends, and comparing financial or economic data—such as bond issuance volume by local governments, capital account balance, and fiscal revenue/expenditure changes.\\
\hspace*{1em}5. Interpretation of Financial Entity Relationships: This scenario centers on analyzing the relationships among economic entities (e.g., firms, governments, departments) and tracing economic impact chains, such as "How does fiscal expansion affect household consumption?" or "What is the impact of capital inflows on exchange rates?"\\
\hspace*{1em}6. Financial Market Sentiment Analysis: This scenario emphasizes extracting market sentiment from investor behavior, public opinion, or market movements—e.g., inferring expectations through news, comments, or price behavior.\\
\hspace*{1em}7. Financial Scenario Analysis: This scenario involves hypothetical and counterfactual reasoning, such as "What would have happened if bond issuance had not increased in 2022?" These tasks require understanding financial mechanisms to infer potential outcomes.\\
Please return only the scenario category name. Do not include any explanations.\\
\bottomrule
\end{tabular}
\end{table*}

\begin{table*}[htbp]
\centering
\caption{Prompt Template for Classifying Scenarios in the Financial Analysis and Decision Support Category (Chinese and English Versions)}
\label{promptl2}
\vspace{2mm}

\begin{subtable}[t]{0.95\textwidth}
\centering
\begin{CJK}{UTF8}{gbsn}
\begin{tabular}{@{}p{13cm}@{}}
\toprule
【金融题目智能分类任务】\\
请基于以下信息，选择最合适的分类（仅返回类别名称）：\\
【背景上下文】\\
\{row['background']\} \\
\{problem\_presentation\} \\
【分类标准】（四选一）：\\
\hspace*{1em}1. 产业分析推断 - 行业趋势、政策影响类问题\\
\hspace*{1em}2. 财务指标分析 - 涉及财务比率、指标计算\\
\hspace*{1em}3. 金融报表分析 - 资产负债表/利润表等解读\\
\hspace*{1em}4. 投资分析 - 综合投资决策评估\\
判断要求：\\
\hspace*{1em}1. 单轮问题直接根据问题内容分类\\
\hspace*{1em}2. 多轮问题需综合分析各轮次的关联性\\
\hspace*{1em}3. 背景信息可帮助理解问题场景\\
\hspace*{1em}4. 只需返回最匹配的中文类别名称\\
\bottomrule
\end{tabular}
\caption{Chinese Version}
\end{CJK}
\end{subtable}

\vspace{4mm}

\begin{subtable}[t]{0.95\textwidth}
\centering
\begin{tabular}{@{}p{13cm}@{}}
\toprule
\texttt{[}Financial Question Scenario Classification Task\texttt{]}\\
Based on the following information, select the most appropriate category (return only the category name):\\
\texttt{[}Background Context\texttt{]}\\
\{row\texttt{[}'background'\texttt{]}\} \\
\{problem\_presentation\} \\
\texttt{[}Classification Criteria\texttt{]} (Choose one):\\
\hspace*{1em}1. Industry Analysis and Inference – questions related to industry trends or policy impacts\\
\hspace*{1em}2. Financial Performance Indicator Analysis – questions involving financial ratios or indicator calculations\\
\hspace*{1em}3. Financial Statement Analysis – interpretation of balance sheets, income statements, etc.\\
\hspace*{1em}4. Investment Analysis – comprehensive evaluation of investment decisions\\
Classification Guidelines:\\
\hspace*{1em}1. For single-turn questions, classify based on the question content alone\\
\hspace*{1em}2. For multi-turn questions, consider the relationship between all turns\\
\hspace*{1em}3. Background context may assist in understanding the question\\
\hspace*{1em}4. Only return the best-matching category name in Chinese\\
\bottomrule
\end{tabular}
\caption{English Version}
\end{subtable}

\end{table*}

\begin{table*}[htbp]
\centering
\begin{CJK}{UTF8}{gbsn}
\caption{Prompt Template for Classifying Scenarios in the Financial Risk Control and Asset Optimization Category (Chinese Version)}
\label{promptl3}
\vspace{2mm}
\begin{tabular}{@{}p{13cm}@{}}
\toprule
你现在是一位在金融领域的从业专家，请判断下列问题：\\
题目\\
\{question\} \\
\{options\_text\} \\
属于哪个【金融业务场景】（请从以下给定场景中严格选择一个）。\\
金融业务场景：\\
\hspace*{1em}1.资产配置分析—涉及投资组合结构、资产比例调整、风险收益平衡、企业股权架构设计等等。\\
\hspace*{1em}2.金融策略优化—聚焦企业财务策略调整（如定价/成本/营销策略）及其对盈利的影响等等。\\
\hspace*{1em}3.金融数据推演与解释—依赖数据计算、财务指标预测、数据间逻辑关系推导等等。\\
\hspace*{1em}4.金融风险与政策解读—汇率/利率波动风险识别、政策对金融市场（如股市、债市）或企业的影响分析、市场风险信号判断（如资产价格大幅波动）、政策导向解读（如货币政策调整对信贷的影响）等等。\\
补充说明：\\
以上4个场景仅做了简单的描述，但这些描述不足以囊括该场景的所有情况，因此，如果某一问题并不符合上述描述，此时你可以基于自身对四个场景的理解，自行判断该问题应该属于哪一类金融场景。\\
冲突处理：\\
\hspace*{1em}- 若同时涉及数据+策略，优先选择金融策略优化\\
\hspace*{1em}- 若同时涉及数据+资产配置，优先选择资产配置分析\\
\hspace*{1em}- 若同时涉及数据+风险或政策，优先选择金融风险与政策解读\\ 
输出格式：\\
场景分类：XXX \\
判定依据：YYY \\
禁止行为：\\
添加额外解释\\
脱离给定场景列表分类\\
修改预设输出格式\\
\bottomrule
\end{tabular}
\end{CJK}
\end{table*}

\begin{table*}[htbp]
\centering
\caption*{(continued) Prompt Template for Classifying Scenarios in the Financial Risk Control and Asset Optimization Category (English Version)}
\vspace{2mm}
\begin{tabular}{@{}p{13cm}@{}}
\toprule
You are now a financial domain expert. Please classify the following question: \\
Question: \\
\{question\} \\
\{options\_text\} \\
Determine which of the following **financial business scenarios** it belongs to (please strictly choose one from the list below).\\
Financial Business Scenarios: \\
\hspace*{1em}1. Asset Allocation Analysis — related to portfolio structure, asset proportion adjustment, risk-return balancing, equity structure design, etc.\\
\hspace*{1em}2. Financial Strategy Optimization — focuses on corporate financial strategy adjustments (e.g., pricing/cost/marketing strategies) and their impact on profitability.\\
\hspace*{1em}3. Financial Data Reasoning and Interpretation — relies on numerical computation, financial indicator forecasting, inference of logical relationships between data, etc.\\
\hspace*{1em}4. Financial Risk and Policy Analysis — includes identification of risks from exchange rate/interest rate fluctuations, analysis of policy impacts on financial markets (e.g., stock/bond markets) or firms, judgment of market risk signals (e.g., significant asset price volatility), and interpretation of policy directions (e.g., how monetary policy adjustments affect credit).\\
Supplementary Note: \\
The above descriptions are simplified and do not fully capture all cases under each scenario. If a question does not clearly match any description, you may rely on your own understanding of the four scenarios to make a reasoned judgment.\\
Conflict Resolution: \\
\hspace*{1em}- If the question involves both data and strategy, prioritize Financial Strategy Optimization.\\
\hspace*{1em}- If it involves both data and asset allocation, prioritize Asset Allocation Analysis.\\
\hspace*{1em}- If it involves both data and risk or policy, prioritize Financial Risk and Policy Analysis.\\
Output Format: \\
Scenario Classification: XXX \\
Justification: YYY \\
Prohibited Actions: \\
Adding extra explanation \\
Classifying outside the given list \\
Modifying the preset output format \\
\bottomrule
\end{tabular}
\end{table*}

\end{document}